%% file: strong2l_paper.tex
\pdfoutput=1
\newcommand*{\ATLASLATEXPATH}{}
\documentclass[cernpreprint,texlive=2016,txfonts,UKenglish]{\ATLASLATEXPATH atlasdoc} 
\usepackage{\ATLASLATEXPATH atlaspackage}
\usepackage{multirow}
\usepackage{bm}
\usepackage{\ATLASLATEXPATH atlasbiblatex}
\usepackage{\ATLASLATEXPATH atlascontribute}
\usepackage{\ATLASLATEXPATH atlasphysics}
\usepackage{strong2l_paper-defs}

\newcommand{\zstar}{\ensuremath{Z^{(*)}}\xspace}

\DeclareOldFontCommand{\it}{\normalfont\itshape}{\mathit}
\DeclareOldFontCommand{\sc}{\normalfont\scshape}{\@nomath\sc}
\DeclareOldFontCommand{\bf}{\normalfont\bfseries}{\mathbf}
\DeclareOldFontCommand{\rm}{\normalfont\rmfamily}{\mathrm}
\usepackage[cernpreprint]{\ATLASLATEXPATH atlascover}

\AtlasTitle{Search for new phenomena in events containing a same-flavour opposite-sign dilepton pair, jets, and large missing transverse momentum in $\boldmath\sqrt{s}=13$~\TeV\ $\boldmath pp$ collisions with the ATLAS detector}

\author{The ATLAS Collaboration}
\AtlasRefCode{SUSY-2016-05}
\PreprintIdNumber{CERN-PH-2016-260}
\AtlasDOI{10.1140/epjc/s10052-017-4700-5}
\AtlasJournalRef{Eur.Phys.J. C77 (2017) no.3, 144} 
\AtlasAbstract{
Two searches for new phenomena in final states containing a \SF\ \OS\ lepton (electron or muon) pair,
jets, and large missing transverse momentum are presented.
These searches make use of proton--proton collision data, collected during 2015 and 2016 at a centre-of-mass energy $\sqrt{s}=13$~\TeV\ by the ATLAS detector
at the Large Hadron Collider, which correspond to an integrated luminosity of \lumi. 
Both searches target the pair production of supersymmetric particles, squarks or gluinos, which decay to final states containing a \SF\ \OS\ lepton
pair via one of two mechanisms: a leptonically decaying $Z$ boson in the final state, 
leading to a peak in the dilepton invariant-mass distribution around the $Z$ boson mass; 
and decays of neutralinos (e.g. $\tilde{\chi}_2^0 \rightarrow \ell^+\ell^- \tilde{\chi}_1^0$), 
yielding a kinematic endpoint in the dilepton invariant-mass spectrum. 
The data are found to be consistent with the Standard Model expectation.
Results are interpreted in simplified models of gluino-pair (squark-pair) production, and 
provide sensitivity to gluinos (squarks) with masses as large as~1.70~\TeV\ (980~\GeV).
}

\hypersetup{pdftitle={ATLAS draft},pdfauthor={The ATLAS Collaboration}}

\begin{document}
\maketitle
\tableofcontents

\section{Introduction}
\label{sec:intro}
\input{sec-intro}

%-------------------------------------------------------------------------------
\section{ATLAS detector}
\label{sec:detector}
\input{sec-detector}

%-------------------------------------------------------------------------------
\section{SUSY signal models}
\label{sec:susy}
\input{sec-susy}

%-------------------------------------------------------------------------------
\section{Data and Monte Carlo samples}
\label{sec:data}
\input{sec-data}

\label{sec:mc}
\input{sec-mc}

\clearpage
%-------------------------------------------------------------------------------
\section{Analysis object identification and selection}
\label{sec:objects}
\input{sec-objects}
%-------------------------------------------------------------------------------
\section{Event selection}
\label{sec:selection}
\input{sec-selection}

%-------------------------------------------------------------------------------
\section{Background estimation}
\label{sec:bg}
\input{sec-bg}
%-------------------------------------------------------------------------------
\section{Systematic uncertainties}
\label{sec:syst}
\input{sec-syst}

%-------------------------------------------------------------------------------
\section{Results}
\label{sec:result}
\input{sec-results}

%-------------------------------------------------------------------------------
\clearpage
\section{Interpretation}
\label{sec:interpretation}
\input{sec-interpretation}
%-------------------------------------------------------------------------------
\FloatBarrier
%-------------------------------------------------------------------------------
\section{Conclusion}
\label{sec:conclusion}
\input{sec-conclusion}
%-------------------------------------------------------------------------------

%-------------------------------------------------------------------------------
\section*{Acknowledgements}
%-------------------------------------------------------------------------------

\input{Acknowledgements}

\printbibliography
%-------------------------------------------------------------------------------
\newpage 
\input{atlas_authlist}

\end{document}

%% file: sec-intro.tex
Supersymmetry (SUSY)~\cite{Golfand:1971iw,Volkov:1973ix,Wess:1974tw,Wess:1974jb,Ferrara:1974pu,Salam:1974ig}
is an extension of the Standard Model (SM) that introduces partner particles (called {\it sparticles}) that differ by half a unit of spin from their SM counterparts. 
The squarks ($\tilde{q}$) and sleptons ($\tilde{\ell}$) are the scalar partners of the quarks and leptons, respectively, 
and the gluinos ($\tilde{g}$) are the fermionic partners of the gluons.
The charginos (${\tilde{\chi}}_{i}^{\pm}$) and 
neutralinos (${\tilde{\chi}}_{i}^{0}$) are the mass eigenstates (where the index $i$ is ordered from the lightest to the heaviest) formed from 
the linear superpositions of the SUSY partners of the Higgs bosons (higgsinos) and electroweak gauge bosons.  

If the masses of the gluino, higgsinos, and top squarks are close to the \TeV~scale, SUSY may offer a solution to the SM hierarchy problem~\cite{Sakai:1981gr,Dimopoulos:1981yj,Ibanez:1981yh,Dimopoulos:1981zb}.
In this case, strongly interacting sparticles should be produced at a high enough rate to be detected by the experiments at the Large Hadron Collider (LHC). 
For models with R-parity conservation~\cite{Farrar:1978xj}, such sparticles would be pair-produced and are expected to decay into jets, 
perhaps leptons, and the lightest stable SUSY particle (LSP). 
The LSP is assumed to be only weakly interacting and therefore escapes the detector, 
resulting in events with potentially large missing transverse momentum (${\boldsymbol p}_{\mathrm{T}}^\mathrm{miss}$, with magnitude \met).
In such a scenario the LSP could be a dark-matter candidate~\cite{Goldberg:1983nd,Ellis:1983ew}.

Final states containing pairs of leptons may arise from the cascade decays of squarks and gluinos via several mechanisms. 
In this paper, two search channels are considered that target scenarios with same-flavour (SF) opposite-sign (OS) lepton 
(electron or muon) pairs. 
The first channel requires a lepton pair with an invariant mass \mll\ that is consistent with the $Z$ boson mass $m_Z$ (``on-shell $Z$'' channel), while the second channel considers all SFOS lepton pairs (``edge'' channel).
The presence of two leptons in the final state suppresses large SM backgrounds from, e.g., QCD multijet and $W+\mathrm{jets}$ production,
providing a clean environment in which to search for new physics. As discussed further below, in such events the distribution
of dilepton mass \mll\ may be used to characterise the nature of the SUSY particle decay and constrain mass differences between
SUSY particles.

The SFOS lepton pairs may be produced in the decay $\tilde{\chi}_{2}^{0} \to \ell^{+}\ell^{-} \tilde{\chi}_{1}^{0}$
(or, in models of generalised gauge mediation with a gravitino LSP~\cite{Dine:1981gu,AlvarezGaume:1981wy,Nappi:1982hm}, via $\tilde{\chi}_{1}^{0} \to \ell^{+}\ell^{-} \tilde{G}$).
The properties of the \chitwozero decay depend on the mass difference
$\Delta m_\chi \equiv m_{\tilde{\chi}_{2}^{0}} - m_{\tilde{\chi}_{1}^{0}}$, the mixing of the charginos and neutralinos, and on whether there are additional sparticles
with masses less than $m_{\tilde{\chi}_{2}^{0}}$ that may be produced in the decay of the \chitwozero particle.
For $\Delta m_\chi>m_Z$, SFOS lepton pairs may be produced in the decay 
$\tilde{\chi}_{2}^{0} \to Z \tilde{\chi}_{1}^{0} \to \ell^{+}\ell^{-} \tilde{\chi}_{1}^{0}$, leading
to a peak in the invariant-mass distribution near $m_{\ell\ell} \approx m_Z$. Such models are the target of the on-shell $Z$ search.
For $\Delta m_\chi < m_Z$, the decay $\tilde{\chi}_{2}^{0} \to Z^* \tilde{\chi}_{1}^{0} \to \ell^{+}\ell^{-} \tilde{\chi}_{1}^{0}$
leads to a rising \mll\ distribution that is truncated at a kinematic endpoint, whose position is
given by $m_{\ell\ell}^{\text{max}}=\Delta m_\chi < m_Z$, below
the $Z$ boson mass peak.
If there are sleptons with masses less than $m_{\tilde{\chi}_{2}^{0}}$, the \chitwozero particle may decay
as $\tilde{\chi}_{2}^{0} \to \tilde{\ell}^{\pm}\ell^{\mp} \to \ell^{+}\ell^{-} \tilde{\chi}_{1}^{0}$,
also leading to a kinematic endpoint but with a different shape and \mll\ endpoint position, given by
$m_{\ell\ell}^{\mathrm{max}} = \sqrt{ (m^2_{\chitwozero}-m^2_{\tilde{\ell}})(m^2_{\tilde{\ell}}-m^2_{\chionezero}) / m^2_{\tilde{\ell}}}$,
which may occur below, on, or above the $Z$ boson mass peak.
The latter two scenarios are targeted by the ``edge'' search channel, which considers the full \mll\ range.

This paper reports on a search for SUSY in the same-flavour dilepton final state with \lumi\ of $pp$ collision data at $\sqrt{s}=13$~\TeV\ recorded in 2015 and 2016 by the ATLAS detector at the LHC.
Searches for SUSY in the $Z+\mathrm{jets}+\met$ final state have previously been performed at $\sqrt{s}=8$~\TeV\ by the
CMS~\cite{Chatrchyan:2012qka,CMS2} and ATLAS~\cite{SUSY-2014-10} collaborations using Run-1 LHC data.
In the ATLAS analysis performed with 20.3~fb$^{-1}$ of $\sqrt{s}=8$~\TeV\ data reported in Ref.~\cite{SUSY-2014-10}, 
an excess of events above the SM background with a significance of 3.0 standard deviations was observed. 
The event selection criteria for the on-shell $Z$ search in this paper are almost identical,
differing only in the details of the analysis object definitions and missing transverse momentum.
CMS performed a search with $\sqrt{s}=13$~\TeV\ data in a similar kinematic region but did not observe evidence to corroborate this excess~\cite{CMS:2015bsf}.

Searches for an edge in the \mll\ distribution in events with $2\ell+\mathrm{jets}+\met$ have been performed
by the CMS~\cite{CMS2,Chatrchyan:2012te} and ATLAS~\cite{SUSY-2014-10} collaborations. 
In Ref.~\cite{CMS2}, CMS reported an excess above the SM prediction with a significance of 2.6 standard deviations.
In a similar search region, however, the Run-1 ATLAS analysis~\cite{SUSY-2014-10} and Run-2 CMS analysis~\cite{CMS:2015bsf} observed results consistent with the SM prediction.

%% file: sec-detector.tex
The ATLAS detector~\cite{PERF-2007-01} is a general-purpose detector with almost $4\pi$ coverage in solid angle.\footnote{ATLAS uses a right-handed coordinate system with its origin at the nominal interaction point (IP) in the centre of the detector and the $z$-axis along the beam pipe. The $x$-axis points from the IP to the centre of the LHC ring, and the $y$-axis points upward. Cylindrical coordinates $(r,\phi)$ are used in the transverse plane, $\phi$ being the azimuthal angle around the $z$-axis. The pseudorapidity is defined in terms of the polar angle $\theta$ as $\eta=-\ln\tan(\theta/2)$ and the rapidity is defined as $y=1/2 \cdot\ln[(E+p_{z})/(E-p_{z})])$, where $E$ is the energy and $p_{z}$ the longitudinal momentum of the object of interest. The opening angle between two analysis objects in the detector is defined as $\Delta R=\sqrt{(\Delta y)^2+(\Delta \phi)^2}$.}
The detector comprises an inner tracking detector, a system of calorimeters, and a muon spectrometer.

The inner tracking detector (ID) is immersed in a 2~T magnetic field provided by a superconducting solenoid and allows charged-particle tracking out to $|\eta|=2.5$.
It includes silicon-pixel and silicon-strip tracking detectors inside a straw-tube tracking detector.
In 2015 the detector received a new innermost layer of silicon pixels, 
which improves the track impact parameter resolution by almost a factor of two in both the transverse and longitudinal directions~\cite{ATL-PHYS-PUB-2015-051}.

High-granularity electromagnetic and hadronic calorimeters cover the region $|\eta|<4.9$.
All the electromagnetic calorimeters, as well as the endcap and forward hadronic calorimeters, 
are sampling calorimeters with liquid argon as the active medium and lead, copper, or tungsten as the absorber.
The central hadronic calorimeter is a sampling calorimeter with scintillator tiles as the active medium and steel as the absorber.

The muon spectrometer uses several detector technologies to provide precision tracking out to $|\eta|=2.7$ and triggering in $|\eta|<2.4$, 
making use of a system of three toroidal magnets.

The ATLAS detector incorporates a two-level trigger system, with the first level implemented in custom hardware and the second level implemented in software.
This trigger system selects events of interest at an output rate of about 1~kHz.

%% file: sec-susy.tex
\begin{figure*}[htb]
\centering
\includegraphics[width=0.3\textwidth]{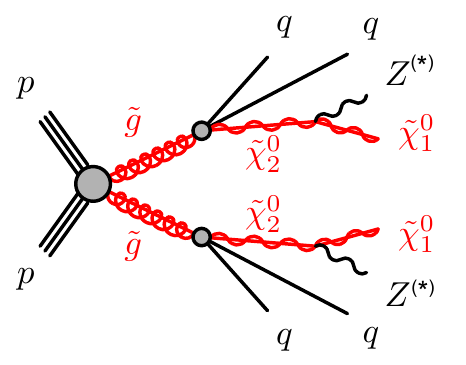}
\hspace{2cm}
\includegraphics[width=0.3\textwidth]{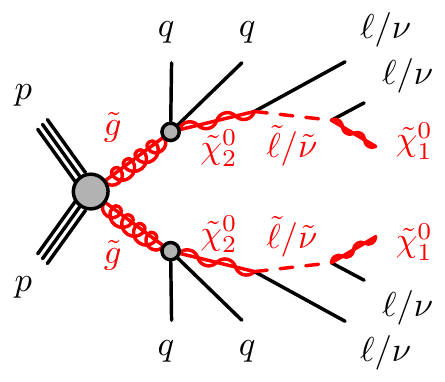}
\caption{Example decay topologies for two of the simplified models considered, 
involving gluino-pair production, with the gluinos following an effective three-body decay for $\tilde{g}\to q \bar{q} \tilde{\chi}_2^0$, 
with $\tilde{\chi}_2^0\rightarrow \zstar \tilde{\chi}_1^0$ (left) and $\tilde{\chi}_{2}^{0} \to \tilde{\ell}^{\mp}\ell^{\pm} / \tilde{\nu}\nu$ (right). 
For simplicity, no distinction is made between particles and antiparticles.}
\label{fig:models}
\end{figure*}

SUSY-inspired simplified models are considered as signal scenarios for these analyses. 
In all of these models, squarks or gluinos are directly pair-pro\-duced, 
decaying via an intermediate neutralino, $\tilde{\chi}_2^0$, into the LSP ($\tilde{\chi}_1^0$).
All sparticles not directly involved in the decay chains considered are effectively decoupled. 
Two example decay topologies are shown in Figure~\ref{fig:models}.  
For all models with gluino-pair production, 
a three-body decay for  $\tilde{g}\to q \bar{q} \tilde{\chi}_2^0$  is used.
Signal models are generated in a grid over a two-dimensional space, 
varying the gluino or squark mass and the mass of either the $\tilde{\chi}_2^0$ or the $\tilde{\chi}_1^0$. 

Three models, one with squark-pair production and two with gluino-pair production, which result exclusively in events with
two on-shell $Z$ bosons in the final state are considered for the on-shell search.
For two of these models, signal mass points are generated across the $\tilde{g}$--$\tilde{\chi}_2^0$ (or $\tilde{q}$--$\tilde{\chi}_2^0$) plane. 
These models are produced following the decays $\tilde{g}\to q \bar{q} \tilde{\chi}_2^0$ or $\tilde{q}\to q \tilde{\chi}_2^0$, 
with the $\tilde{\chi}_1^0$ (LSP) mass set to 1~\GeV, 
inspired by SUSY scenarios with a low-mass LSP (e.g. generalised gauge mediation).
These two models are referred to here as the $\tilde{g}$--$\tilde{\chi}_2^0$ on-shell and $\tilde{q}$--$\tilde{\chi}_2^0$ on-shell grids, respectively,
and are summarised in Table~\ref{tab:models}. 
The third model is based on MSSM-inspired topologies~\cite{Fayet:1976et,Fayet:1977yc,Cahill-Rowley:2015cha} with potentially higher mass LSPs. 
Signal points are generated across the $\tilde{g}$--$\tilde{\chi}_1^0$ plane, and this model is thus referred to as the $\tilde{g}$--$\tilde{\chi}_1^0$ on-shell grid. 
In this case the $\tilde{\chi}_2^0$ mass is set to be 100~\GeV\ above the $\tilde{\chi}_1^0$ mass,
which in many models maximises the branching fraction of the $\tilde{\chi}_2^0$ decay to $Z$ bosons.
For the two models with gluino-pair production, 
since the gluino coupling to $q\tilde{q}$ is flavour independent and the corresponding flavours of squarks are assumed to be mass degenerate, 
the branching fractions for $q=u,d,c,s$ are each 25\%. 
Other ATLAS searches are dedicated to final states with two leptons and heavy flavour jets~\cite{SUSY-2013-08,SUSY-2013-19}.
For the model involving squark-pair production, the superpartners of the $u$, $d$, $c$ and $s$ quarks have the same mass, 
with the superpartners of the $b$ and $t$ quarks being decoupled. 

\begin{table}[htbp!]
\caption{
Summary of the simplified signal model topologies used in this paper.
Here $x$ and $y$ denote the $x$--$y$ plane across which the signal model masses are varied to construct the signal grid.
For the slepton model, the masses of the superpartners of the left-handed leptons are given by $[m(\chitwozero)+m(\chionezero)]/2$,
while the superpartners of the right-handed leptons are decoupled.
}
\begin{center}
    \begin{tabular}{lcccccc}
      \hline
      Model                               & Production mode        & Quark flavours            & $m(\tilde{g})/m(\tilde{q})$ & $m(\tilde{\chi}^{0}_{2})$  & $m(\tilde{\chi}^{0}_{1})$ \\
      \hline\hline
      $\tilde{g}$--$\chitwozero$ on-shell & $\tilde{g}\tilde{g}$   &  $u$, $d$, $c$, $s$       & $x$                         & $y$                              & 1~\GeV        \\
      $\tilde{g}$--$\chionezero$ on-shell & $\tilde{g}\tilde{g}$   &  $u$, $d$, $c$, $s$       & $x$                         & $m(\tilde{\chi}^0_1)+100$~\GeV   & $y$           \\
      $\tilde{q}$--$\chitwozero$ on-shell & $\tilde{q}\tilde{q}$   &  $u$, $d$, $c$, $s$       & $x$                         & $y$                              & 1~\GeV        \\
      \zstar\                             & $\tilde{g}\tilde{g}$   &  $u$, $d$, $c$, $s$, $b$  & $x$                         & $[m(\tilde{g})+m(\chionezero)]/2$        & $y$           \\
      slepton                             & $\tilde{g}\tilde{g}$   &  $u$, $d$, $c$, $s$, $b$  & $x$                         & $[m(\tilde{g})+m(\chionezero)]/2$        & $y$           \\
      \hline
\end{tabular}
\end{center}
\label{tab:models}
\end{table}

The edge search considers two scenarios, 
both of which involve the direct pair production of gluinos and differ by the decay mode of the $\tilde{\chi}_{2}^{0}$. 
These signal models are also summarised in Table~\ref{tab:models}.
In the \zstar\ model the $\tilde{\chi}_{2}^{0}$ decays as $\tilde{\chi}_{2}^{0} \to Z^{(*)}\tilde{\chi}_{1}^{0}$.
For $\Delta m_\chi = m(\tilde{\chi}_2^0) - m(\tilde{\chi}_1^0) > m_Z$, the $Z$ boson is on-shell, leading to a peak in the \mll\ distribution at $m_Z$,
while for $\Delta m_\chi < m_Z$, the $Z$ boson is off-shell, leading to an edge in the dilepton mass distribution 
with a position below $m_Z$. 
The slepton model assumes that the sleptons are lighter than the $\tilde{\chi}_{2}^{0}$, 
which decays as $\tilde{\chi}_{2}^{0} \to \tilde{\ell}^{\mp}\ell^{\pm}$ with $\tilde{\ell} \to \ell\tilde{\chi}_{1}^{0}$ 
or as $\tilde{\chi}_{2}^{0} \to \tilde{\nu}\nu$ with $\tilde{\nu} \to \nu\tilde{\chi}_{1}^{0}$, each with a branching fraction of 50\%,
where $\tilde{\ell}=\tilde{e},\tilde{\mu},\tilde{\tau}$ and $\tilde{\nu}=\tilde{\nu_e},\tilde{\nu_\mu},\tilde{\nu_\tau}$.
The endpoint position can occur at any mass, highlighting the need to search over the full dilepton mass distribution.
The gluino decays as $\tilde{g}\to q \bar{q} \tilde{\chi}_2^0$, and both models have equal branching fractions for $q=u,d,c,s,b$. 
The \chitwozero\ mass is set to the average of the gluino and \chionezero\ masses.
For the slepton model, the masses of the superpartners of the left-handed leptons are set as the average of the \chitwozero\ and \chionezero\ masses, 
while the superpartners of the right-handed leptons are decoupled. 
The three slepton flavours are mass-degenerate. 
In both these models the $\tilde{g}$ and \chionezero\ masses are free parameters that are varied to produce the two-dimensional signal grid. 
The mass splittings are chosen to maximise the differences between these simplified models and other models with only one intermediate particle between the gluino and the LSP~\cite{wacker}.

%% file: sec-data.tex
The data used in this analysis were collected by ATLAS during 2015 and 2016, 
with a mean number of additional $pp$ interactions per bunch crossing ({\it pile-up}) of approximately 14 in 2015 and 21 in 2016,
and a centre-of-mass collision energy of 13~\TeV.  
Following requirements based on beam and detector conditions and data quality, the data set corresponds to an integrated luminosity of \lumi.
The uncertainty in the combined 2015 and 2016 integrated luminosity is $\pm2.9$\%. It is derived, following a methodology similar to that detailed in Refs.~\cite{2011lumi} and~\cite{2012lumi}, from a preliminary calibration of the luminosity scale using $x$--$y$ beam-separation scans performed in August 2015 and May 2016. 

Data events are collected using a combination of single-lepton and dilepton triggers~\cite{ATL-DAQ-PUB-2016-001}, 
in order to maximise the signal acceptance. 
The dielectron, dimuon, and electron--muon triggers have leading-lepton $\pt$ thresholds in the range 12--24~\GeV.
Additional single-electron (single-muon) triggers are also used, with trigger $\pt$ thresholds of 60 (50)~\GeV, to increase the trigger efficiency for models with high-\pt leptons. 
Events are required to contain at least two selected leptons with $\pt>25$~\GeV, 
making the selection fully efficient with respect to the trigger \pt\ thresholds.

An additional control sample of events containing photons is collected using a set of single-photon triggers with
\pt thresholds in the range 20--140~\GeV. 
All triggers except for the one with threshold $\pt=120$~\GeV\ in 2015, or the one with $\pt=140$~\GeV\ in 2016, are prescaled.
Events are required to contain a selected photon with $\pt>37$~\GeV, so that they are selected efficiently by the lowest available \pt\ trigger in 2015, 
which had a threshold of $\pt^{\gamma}=35$~\GeV.

%% file: sec-mc.tex
Simulated event samples are used to aid in the estimation of SM backgrounds, validate the analysis techniques, optimise the event selection, and provide predictions for SUSY signal processes.  
All SM background samples 
used are listed in Table~\ref{tab:MC}, 
along with the parton distribution function (PDF) set, the configuration of underlying-event and hadronisation parameters (underlying-event tune) and the cross-section calculation order in $\alpha_{\text{S}}$ used to normalise the event yields for these samples. 
      
Samples simulated using {\sc MG5\_aMC@NLO} v2.2.2~\cite{Alwall:2014hca}, interfaced with {\sc Pythia} 8.186~\cite{Sjostrand:2007gs} with the {\sc A14} underlying-event tune~\cite{ATL-PHYS-PUB-2014-021} to simulate the parton shower and hadronisation, are generated at leading order in $\alpha_{\text{S}}$ (LO) with the {\sc NNPDF23LO} PDF set~\cite{Ball:2012cx}.
For samples generated using {\sc Powheg Box V2}~\cite{PowhegBOX1,PowhegBOX2,PowhegBOX3}, 
{\sc Pythia} 6.428~\cite{Sjostrand:2006za} is used to simulate the parton shower, hadronisation, and the underlying event. 
The {\sc CTEQ6L1} PDF set is used with the corresponding {\sc Perugia2012}~\cite{pythiaperugia} tune. 
In the case of both the {\sc MG5\_aMC@NLO} and {\sc Powheg} samples, 
the {\sc EvtGen} v1.2.0 program~\cite{EvtGen} is used for properties of the bottom and charm hadron decays. 
{\sc Sherpa} $2.1.1$~\cite{sherpa} simulated samples use the CT10 PDF set with {\sc Sherpa}'s own internal parton shower~\cite{Schumann:2007mg} and hadronisation methods, 
as well as the {\sc Sherpa} default underlying-event tune. 
Diboson processes with four charged leptons, three charged leptons and a neutrino or two charged leptons and two neutrinos are simulated using the {\sc Sherpa} $2.1.1$ generator. 
Matrix elements contain all diagrams with four electroweak vertices. 
They are calculated for up to one ($4\ell$, $2\ell+2\nu$) or zero ($3\ell+1\nu$) partons at next-to-leading order in $\alpha_{\text{S}}$ (NLO) and up to three partons at LO using the Comix~\cite{Gleisberg:2008fv} and 
OpenLoops~\cite{Cascioli:2011va} matrix element generators and merged with the {\sc Sherpa} parton shower 
using the ME+PS@NLO prescription~\cite{Hoeche:2012yf}.
For the \dyjets\ background, \sherpa\ $2.1.1$ is used to generate a sample with up to two additional partons at NLO and up to four at LO.
For Monte Carlo (MC) closure studies, \gjets\ events are generated at LO with up to four additional partons using \sherpa\ $2.1.1$.
Additional MC simulation samples of events with a leptonically decaying vector boson and photon ($V\gamma$, where $V=W,Z$) are generated at LO using \sherpa\ $2.1.1$. 
Matrix elements including all diagrams with three electroweak couplings are calculated with up to three partons.
These samples are used to estimate backgrounds with real \met\ in $\gjets$ event samples.

The SUSY signal samples are produced at LO using {\sc MG5\_aMC@NLO} with the {\sc NNPDF2.3LO} PDF set, interfaced with {\sc Pythia} 8.186.
The scale parameter for CKKW-L matching~\cite{CKKW,CKKWL} is set at a quarter of the mass of the gluino.
Up to one additional parton is included in the matrix element calculation. 
The underlying event is modelled using the {\sc A14} tune for all signal samples, and {\sc EvtGen} is adopted to describe the properties of bottom and charm hadron decays.
Signal cross sections are calculated at NLO in $\alpha_{\text{S}}$.
This includes the resummation of soft gluon emission at next-to-leading-logarithm accuracy (NLO+NLL)~\cite{Beenakker:1996ch,Kulesza:2008jb,Kulesza:2009kq,Beenakker:2009ha,Beenakker:2011fu}. 

All of the SM background MC samples are subject to a full ATLAS detector simulation~\cite{:2010wqa} using {\sc GEANT4}~\cite{Agostinelli:2002hh}. 
A fast simulation~\cite{:2010wqa}, which uses a combination of a parameterisation of the response of the ATLAS electromagnetic and hadronic calorimeters and {\sc GEANT4},
is used in the case of signal MC samples. 
This fast simulation is validated by comparing a few chosen signal samples to some fully simulated points.
Minimum-bias interactions are generated and overlaid on the hard-scattering process to simulate the effect of multiple $pp$ interactions occurring during the same (in-time) or a nearby (out-of-time) bunch-crossing (pile-up).
These are produced using {\sc Pythia} 8.186 with the {\sc A2} tune~\cite{ATLAS:2012uec} and {\sc MSTW 2008} PDF set~\cite{MSTW}.
The pile-up distribution in MC samples is simulated to match that in data during 2015 and 2016 $pp$ data-taking.

\begin{table*}[ht]
\begin{center}
\caption{Simulated background event samples used in this analysis with the corresponding matrix element and parton shower generators,
cross-section order in $\alpha_{\text{S}}$ used to normalise the event yield, underlying-event tune and PDF set.
}
\scriptsize
\begin{tabular}{l c c c c c }
\hline
Physics process &  Generator  & Parton & Cross section & Tune & PDF set\\
                &             & Shower &              &      & \\
%\hline\hline
\noalign{\smallskip}\hline\noalign{\smallskip}
$t\bar{t}+W$ and $t\bar{t}+Z$~\cite{ATL-PHYS-PUB-2016-005,Garzelli:2012bn}& {\sc MG5\_aMC@NLO}        & {\sc Pythia} 8.186 & NLO \cite{Campbell:2012,Lazopoulos:2008} & {\sc A14} & NNPDF23LO\\
$t\bar{t}+WW$~\cite{ATL-PHYS-PUB-2016-005}      & {\sc MG5\_aMC@NLO}          & {\sc Pythia} 8.186 & LO \cite{Alwall:2014hca} & {\sc A14}  &  NNPDF23LO\\
$t\bar{t}$~\cite{ATL-PHYS-PUB-2016-004}         & {\sc Powheg Box v2} r3026   & {\sc Pythia} 6.428 & NNLO+NNLL \cite{ttbarxsec1,ttbarxsec2}          &\sc{Perugia2012}     &NLO CT10\\
Single-top ($Wt$)~\cite{ATL-PHYS-PUB-2016-004}  & {\sc Powheg Box v2} r2856   & {\sc Pythia} 6.428 & Approx. NNLO \cite{Kidonakis:2010b}& \sc{Perugia2012}    &NLO CT10 \\ 
$WW$, $WZ$ and $ZZ$~\cite{ATL-PHYS-PUB-2016-002} & \sherpa\ 2.1.1 & \sherpa\ 2.1.1 & NLO \cite{diboson1,diboson2} & \sherpa\ default & NLO CT10 \\
$Z/\gamma^{*}(\rightarrow \ell \ell)$ + jets~\cite{ATL-PHYS-PUB-2016-003}& \sherpa\ 2.1.1           & \sherpa\ 2.1.1  &NNLO \cite{DYNNLO1,DYNNLO2}       & \sherpa\ default     &NLO CT10\\
\gjets & \sherpa\ 2.1.1 & \sherpa\ 2.1.1 & LO~\cite{sherpa} & \sherpa\ default & NLO CT10 \\
$V(=W,Z)\gamma$ & \sherpa\ 2.1.1 & \sherpa\ 2.1.1 & LO~\cite{sherpa} & \sherpa\ default & NLO CT10 \\
\noalign{\smallskip}\hline\noalign{\smallskip}
\end{tabular}
\label{tab:MC}
\end{center}
\end{table*}

%% file: sec-objects.tex
All analysis objects are categorised as either ``baseline'' or ``signal'' based on various quality and kinematic requirements.
{\it Baseline} objects are used in the calculation of missing transverse momentum and to disambiguate between the analysis objects in the event,
while the jets and leptons entering the final analysis selection must pass more stringent {\it signal} requirements.
The selection criteria for both the baseline and signal objects differ from the requirements used in
the Run-1 ATLAS $Z+\mathrm{jets}+\MET$ search reported in Ref.~\cite{SUSY-2014-10},
owing to the new silicon-pixel tracking layer and significant changes to the reconstruction software since $2012$ data-taking.
In particular, improvements in the lepton identification criteria have reduced the background due to hadrons
misidentified as electrons.
The primary vertex in each event is defined as the reconstructed vertex~\cite{ATL-PHYS-PUB-2015-026} with the highest $\sum p_{\text{T}}^2$,
where the summation includes all particle tracks with $\pt>400$~\MeV\ associated with the vertex.

Electron candidates are reconstructed from energy clusters in the electromagnetic calorimeter matched to ID tracks.
Baseline electrons are required to have transverse energy $E_{\text{T}}>10$~\GeV,
satisfy the ``loose likelihood'' criteria described in Ref.~\cite{ATLAS-CONF-2014-032} and reside within the region $|\eta|<2.47$.
Signal electrons are further required to have $\pT>25$~\GeV, satisfy the ``medium likelihood'' criteria of Ref.~\cite{ATLAS-CONF-2014-032},
and be consistent with originating from the primary vertex.
The signal electrons must originate from within $|z_0\sin\theta| = 0.5$~mm of the primary vertex along the direction of the beamline.\footnote{The distance of closest approach between a particle object and the primary vertex (beamline) in the longitudinal (transverse) plane is denoted by $z_0$ ($d_0$).}
The transverse-plane distance of closest approach of the electron to the beamline, divided by the corresponding uncertainty, must be $|d_0/\sigma_{d_0}|<5$.
These electrons must also be isolated with respect to other objects in the event,
according to a \pt-dependent isolation requirement.
The isolation uses calorimeter- and track-based information to obtain 95\% efficiency at $\pt=25$~\GeV, rising to 99\% efficiency at $\pt=60$~\GeV. 

Baseline muons are reconstructed from either ID tracks matched to muon segments (collections of hits in a single muon spectrometer layer) or combined tracks formed from the ID and muon spectrometer~\cite{PERF-2015-10}. 
They must satisfy the ``medium'' selection criteria described in Ref.~\cite{PERF-2015-10}, and to satisfy $p_{\text{T}}>10$~\GeV\ and $|\eta|<2.5$.
Signal muon candidates are further required to have $\pT>25$~\GeV, be isolated, and have $|z_0\sin\theta| < 0.5$~mm and $|d_0/\sigma_{d_0}|<3$.
Calorimeter- and track-based isolation criteria are used to obtain 95\% efficiency at $\pt=25$~\GeV, rising to 99\% efficiency at $\pt=80$~\GeV~\cite{PERF-2015-10}.
Further, the relative uncertainties in the $q/p$ of each of the ID track alone and muon spectrometer track alone are required to be less than 80\% 
of the uncertainty in the $q/p$ of the combined track.   
This reduces the already low rate of grossly mismeasured muons.
The combined isolation and identification efficiency for single leptons, after the trigger requirements,
is about 70\% (80\%) for electrons (muons) with $\pT \sim 25$~\GeV, rising to about 90\% for $\pT>200$~\GeV.

Jets are reconstructed from topological clusters of energy~\cite{PERF-2014-07} in the calorimeter using the anti-$k_{t}$ algorithm~\cite{Cacciari:2008gp,N3Myth} with a radius parameter of 0.4.
Calibration corrections are applied to the jets based on a comparison to jets made of stable particles
(those with lifetimes $\tau > 0.3 \times 10^{-10}$ s) in the MC simulation.
A residual correction is applied to jets in data, based on studies of \pt\ balance between jets and well-calibrated objects in the MC simulation and data~\cite{PERF-2012-01,ATL-PHYS-PUB-2015-015}.
Baseline jet candidates are required to have $\pt>20$~\GeV\ and reside within the region $|\eta|<4.5$.
Signal jets are further required to satisfy $\pt>30$~\GeV\ and reside within the region $|\eta|<2.5$.
Jets with $\pt<60$~\GeV\ and $|\eta|<2.4$ must meet additional criteria designed to select jets from the hard-scatter interaction and reject those originating from pile-up.
This is enforced by using the jet vertex tagger described in Ref.~\cite{ATLAS-CONF-2014-018}.
Finally, events containing a jet that does not pass specific jet quality requirements are vetoed from the analysis selection in order to remove events impacted by detector noise and non-collision backgrounds~\cite{Aad:2013zwa,ATLAS-CONF-2015-029}.
The MV2c10 boosted decision tree algorithm~\cite{PERF-2012-04,ATL-PHYS-PUB-2016-012} identifies jets with $|\eta|<2.5$ containing $b$-hadrons ($b$-jets) based on quantities such as
the impact parameters of associated tracks and any reconstructed secondary vertices.
A selection that provides 77\% efficiency for tagging $b$-jets in simulated \ttbar\ events is used.
These tagged jets are called $b$-tagged jets.

Photon candidates must satisfy ``tight'' selection criteria described in Ref.~\cite{PERF-2013-05}, have $\pt>25$~\GeV\ and reside within the region $|\eta|<2.37$,
excluding the transition region $1.37<|\eta|<1.6$ where there is a discontinuity in the calorimeter.
Signal photons are further required to have $\pt>37$~\GeV\ and to be isolated from other objects in the event, 
using \pt-dependent requirements on both track- and calorimeter-based isolation. 

To avoid the duplication of analysis objects in more than one baseline selection, an overlap removal procedure is applied.
Any baseline jet within $\Delta R=0.2$ of a baseline electron is removed, unless the jet is $b$-tagged,
in which case the electron is identified as originating from a heavy-flavour decay and is removed instead.
Remaining electrons residing within $\Delta R=0.4$ of a baseline jet are then removed from the event.
Subsequently, any baseline muon residing within $\Delta R=0.2$ of a remaining baseline $b$-tagged jet is discarded.
If such a jet is not $b$-tagged then the jet is removed instead.
Any remaining muon found within $\mathrm{ min } (0.04+(10~\GeV)/\pt ,0.4)$ of a jet is also discarded.
This stage of the overlap removal procedure differs from that used in Ref.~\cite{SUSY-2014-10}.
It was improved to retain muons near jet candidates mostly containing calorimeter energy from final-state radiation from muons,
while still rejecting muons from heavy-flavour decays.
Finally, to remove electron candidates originating from muon bremsstrahlung, any baseline electron within $\Delta R=0.01$ of any remaining baseline muon is removed from the event.
Photons are removed if they reside within $\Delta R=0.4$ of a baseline electron, and any jet within $\Delta R=0.4$ of any remaining photon is discarded.

The \met\ is defined as the magnitude of the negative vector sum, ${\boldsymbol p}_{\mathrm{T}}^\mathrm{miss}$, of the transverse momenta of all baseline electrons,
muons, jets, and photons~\cite{ATL-PHYS-PUB-2015-023,ATL-PHYS-PUB-2015-027}.
Low-momentum contributions from particle tracks from the primary vertex that are not associated with reconstructed analysis objects are included in the calculation of \met.
This contribution to the \met\ is referred to as the ``soft term''.

Models with large hadronic activity are targeted by placing additional requirements on the quantity $H_\text{T}$, defined as the scalar sum of the $\pt$ values
of all signal jets, or on $\htincl$, the scalar sum of the $\pt$ values of all signal jets and the two leptons with largest $\pt$.

All MC samples have correction factors applied to take into account small differences between data and MC simulation in identification, reconstruction and trigger efficiencies for leptons.
The $\pt$ values of leptons in MC samples are additionally smeared to match the momentum resolution in data.

%% file: sec-selection.tex
For each search channel, signal regions (SRs) are designed to target events from specific SUSY signal models.
Control regions (CRs) are defined to be depleted in SUSY signal events and enriched in specific SM backgrounds, 
and they are used to assist in estimating these backgrounds in the SRs.
To validate the background estimation procedures, various validation regions (VRs) are defined to be analogous to the CRs and SRs,
but with less stringent requirements than the SRs on \met, \htincl\ or \HT.
Other VRs with additional requirements on the number of leptons are used to validate the modelling of 
backgrounds in which more than two leptons are expected.

Events in SRs are required to contain at least two signal leptons (electrons or muons).
If more than two signal leptons are present in a given event, the selection process continues based on the two leptons with the highest \pt\ values in the event.

The selected events must pass at least one of the leptonic triggers.
If an event is selected by a dilepton trigger,
the two leading, highest \pT, leptons must be matched to one of the objects that triggered the event. 
These leptons must also have \pt\ higher than the threshold of the trigger in question.
For events selected by a single-lepton trigger, at least one of the two leading leptons must be matched to the trigger object in the same way.
The leading two leptons in the event must have $\pt>25$~\GeV, and form an SFOS pair.

As at least two jets are expected in all signal models studied,
selected events are further required to contain at least two signal jets.
Furthermore, events in which the azimuthal opening angle between either of the leading two jets and the \met\ satisfies 
$\Delta\phi(\text{jet}_{12},{\boldsymbol p}_{\mathrm{T}}^\mathrm{miss})<0.4$ are rejected so as to remove events with \met\ from jet mismeasurements.
This requirement also suppresses \ttbar\ events in which the top quark, the anti-top quark, or the entire \ttbar\ system has a large Lorentz boost.

\begin{figure}[hbtp]
\centering
\includegraphics[width=.8\textwidth]{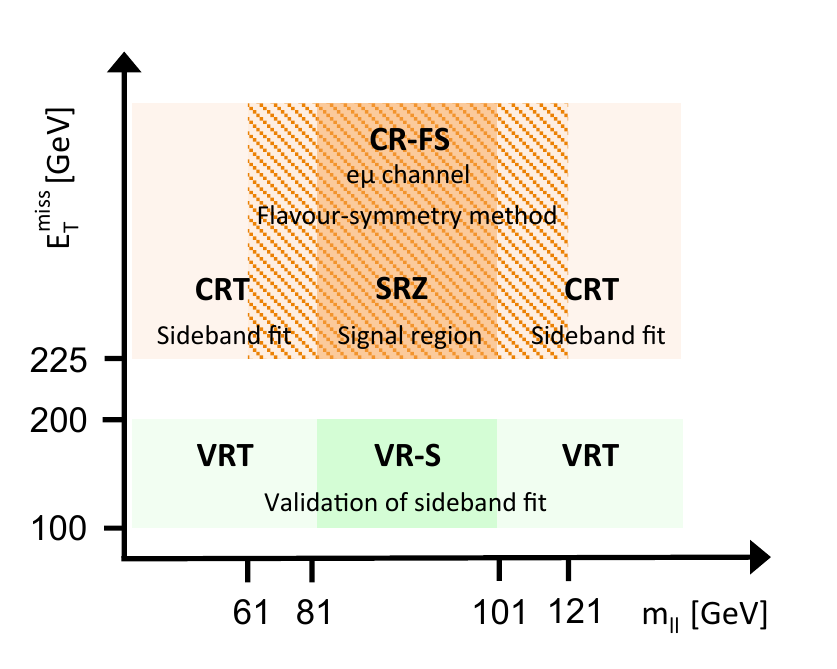}\\
\includegraphics[width=.8\textwidth]{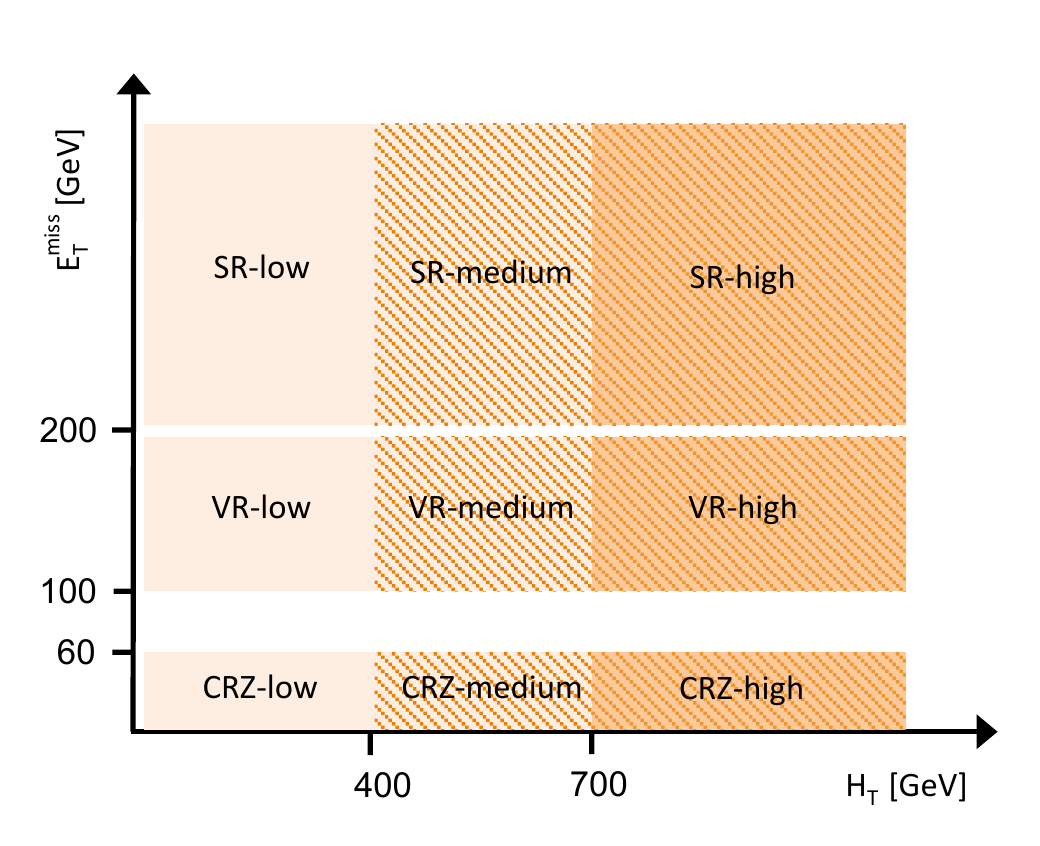}
\caption{
Schematic diagrams of the control (CR), validation (VR) and signal regions (SR) for the on-shell $Z$ (top) and edge (bottom) searches.
For the on-shell $Z$ search the various regions are shown in the $\mll$--$\met$ plane, whereas in the case of the edge search the
signal and validation regions are depicted in the $\HT$--$\met$ plane.  The flavour-symmetry and sideband-fit background estimation methods are described further in Section~\ref{sec:FS}.
\label{fig:region_diagrams}
}
\end{figure}

The various methods used predict the background in the SRs are discussed in Section~\ref{sec:bg}.
The selection criteria for the CRs, VRs, and SRs are summarised in Tables~\ref{tab:regions-z}~and~\ref{tab:regions-edge}.
The most important of these regions are shown graphically in Figure~\ref{fig:region_diagrams}.

\begin{table}[htbp]
\begin{center}
 \caption{Overview of all signal (SR), control (CR) and validation regions (VR) used in the on-shell $Z$ search.
 The flavour combination of the dilepton pair is denoted as either ``SF'' for same-flavour or ``DF'' for different-flavour.
 All regions require at least two leptons, unless otherwise indicated.
 In the case of CR$\gamma$, VR-WZ, VR-ZZ, and VR-3L the number of leptons, rather than a specific flavour configuration, is indicated.
 More details are given in the text.
The main requirements that distinguish the control and validation regions from the signal region are indicated in bold.
The kinematic quantities used to define these regions are discussed in the text. The quantity $m_{\text{T}}(\ell_{3},\met)$
indicates the transverse mass formed by the \met and the lepton which is not assigned to either of the $Z$-decay leptons. }
\resizebox{\textwidth}{!}{
 \begin{tabular}{lcccccccc}
   \noalign{\smallskip}\hline\noalign{\smallskip}
     {\bf On-shell $Z$} &  {\bf \met}   &  {\bf $\htincl$}  &  {\bf $n_{\text{jets}}$}  & {\bf $m_{\ell\ell} $}  &  {\bf SF/DF}  &  {\bf $\Delta\phi(\text{jet}_{12},{\boldsymbol p}_{\mathrm{T}}^\mathrm{miss})$ }  &  {\bf $m_{\text{T}}(\ell_{3},\met)$} &  {\bf $n_{b\text{-jets}}$} \\
     {\bf regions}       &  {\bf [\GeV]} &  {\bf [\GeV]} &                           &    {\bf [\GeV]}        &               &                                             &  [\GeV\
]                        &   \\
   \noalign{\smallskip}\hline\noalign{\smallskip}
   \multicolumn{2}{l}{Signal region} &&&&&& \\
   \noalign{\smallskip}\hline\noalign{\smallskip}
   SRZ  &  $> 225$  &  $> 600$  &  $\geq 2$  & $81 < m_{\ell\ell} < 101$  &  SF  &  $>0.4$ & $-$& $-$\\
   \noalign{\smallskip}\hline\noalign{\smallskip}
   \multicolumn{2}{l}{Control regions} &&&&&&  &  \\
   \noalign{\smallskip}\hline\noalign{\smallskip}
   CRZ              &  $\bm{< 60}$   &  $> 600$  &  $\geq 2$   &  $81 < m_{\ell\ell} < 101$       &  SF  & $>0.4$ & $-$  &  $-$ \\
   CR-FS            &  $> 225$  &  $> 600$  &  $\geq 2$   &  $\bm{61 < m_{\ell\ell} < 121}$       &  {\bf DF}  & $>0.4$ & $-$  &  $-$ \\
   CRT              &  $> 225$  &  $> 600$  &  $\geq 2$   &  $\bm{>45}$, $\bm{m_{\ell\ell} \notin [81,101]}$  &  SF  & $>0.4$ & $-$  &  $-$ \\
   CR$\gamma$       &  $-$        &  $> 600$  &  $\geq 2$   &  $-$                                                        &  {\bf $0\ell$, $1\gamma$}  & $-$ & $-$  &  $-$ \\
   \noalign{\smallskip}\hline\noalign{\smallskip}
   \multicolumn{2}{l}{Validation regions} &&&&&& \\
   \noalign{\smallskip}\hline\noalign{\smallskip}
   VRZ  &   $\bm{<225}$      &  $> 600$   &  $\geq 2$  &    $81 < m_{\ell\ell} < 101$       &  SF        & $>0.4$  & $-$ & $-$ \\
   VRT  &  {\bf 100--200}     &  $> 600 $  &  $\geq 2$  &    $\bm{>45}$, $\bm{m_{\ell\ell} \notin [81,101]}$  &  SF        & $>0.4$  & $-$ & $-$ \\
   VR-S  &  {\bf 100--200}     &  $> 600 $  &  $\geq 2$  &    $81 < m_{\ell\ell} < 101$       &  SF        & $>0.4$  & $-$ & $-$ \\
   VR-FS & {\bf 100--200}     &  $> 600 $  &  $\geq 2$  &    $\bm{61 < m_{\ell\ell} < 121}$  &  {\bf DF}        & $>0.4$  & $-$ & $-$ \\
   VR-WZ  &  {\bf 100--200}   &     $-$      &   $-$        &         $-$                          &  $\bm{3\ell}$   &    $-$    & $<100$  &  $0$  \\
   VR-ZZ  &  {\bf $<100$}     &     $-$      &   $-$        &         $-$                          &  $\bm{4\ell}$   &    $-$    &  $-$      & $0$   \\
   VR-3L  &  {\bf 60--100}    &  $\bm{> 200}$  &  $\geq 2$  &   $81 < m_{\ell\ell} < 101$        &  $\bm{3\ell}$   & $>0.4$  & $-$ & $-$ \\
   \noalign{\smallskip}\hline\noalign{\smallskip}
\end{tabular}
}
\label{tab:regions-z}
\end{center}
\end{table}

\begin{table}[htbp]
\begin{center}
 \caption{Overview of all signal (SR), control (CR) and validation regions (VR) used in the edge search.
 The flavour combination of the dilepton pair is denoted as either ``SF'' for same-flavour or ``DF'' for different-flavour.
The charge combination of the leading lepton pairs are given as ``SS'' for same-sign or ``OS'' for opposite-sign.
All regions require {\it at least} two leptons, with the exception of CR-real, which requires {\it exactly} two leptons,
and the three $\gamma$ CRs, which require no leptons and one photon.
More details are given in the text.
The main requirements that distinguish the control and validation regions from the signal regions are indicated in bold.
The kinematic quantities used to define these regions are discussed in the text.
}
\resizebox{1\textwidth}{!}{
 \begin{tabular}{lcccccccc}
   \noalign{\smallskip}\hline\noalign{\smallskip}
     {\bf Edge }   &  {\bf \met}   & {\bf $\HT$}  &  {\bf $n_{\text{jets}}$}  & {\bf $m_{\ell\ell} $} &  {\bf SF/DF}  & {\bf OS/SS}  &  {\bf $\Delta\phi(\text{jet}_{12},{\boldsymbol p}_{\mathrm{T}}^\mathrm{miss})$ }  & {\bf \mll\ ranges} \\
     {\bf regions} &  {\bf [\GeV]} & {\bf [\GeV]} &                           & {\bf [\GeV]}          &               &              &                &                           \\
   \noalign{\smallskip}\hline\noalign{\smallskip}
   \multicolumn{2}{l}{Signal regions} &&&&&& \\
   \noalign{\smallskip}\hline\noalign{\smallskip}
   SR-low     &  $> 200$  &  $-$      &  $\geq 2$  & $>12$         &  SF  & OS  & $>0.4$ & 9 \\
   SR-medium  &  $> 200$  &  $> 400$  &  $\geq 2$  & $>12$         &  SF  & OS  & $>0.4$ & 8 \\
   SR-high    &  $> 200$  &  $> 700$  &  $\geq 2$  & $>12$         &  SF  & OS  & $>0.4$ & 7 \\
   \noalign{\smallskip}\hline\noalign{\smallskip}
   \multicolumn{2}{l}{Control regions} &&&&&&    \\
   \noalign{\smallskip}\hline\noalign{\smallskip}
   CRZ-low          &  $\bm{< 60}$   &  $-$        &  $\geq 2$ & $>12$     &  SF  & OS  &  $>0.4$ & $-$ \\
   CRZ-medium       &  $\bm{< 60}$   &  $> 400$  &  $\geq 2$ & $>12$     &  SF  & OS  &  $>0.4$ & $-$ \\
   CRZ-high         &  $\bm{< 60}$   &  $> 700$  &  $\geq 2$ & $>12$     &  SF  & OS  &  $>0.4$ & $-$ \\
   CR-FS-low        &  $> 200$  &  $-$        &  $\geq 2$ & $>12$     &  {\bf DF}  & OS  &  $>0.4$ & $-$ \\
   CR-FS-medium     &  $> 200$  &  $> 400$  &  $\geq 2$ & $>12$     &  {\bf DF}  & OS  &  $>0.4$ & $-$ \\
   CR-FS-high       &  $> 200$  &  $> 700$  &  $\geq 2$ & $>12$     &  {\bf DF}  & OS  &  $>0.4$ & $-$ \\
   CR$\gamma$-low        &  $-$  &  $-$        &  $\geq 2$ & $-$    &  {\bf $0\ell$, $1\gamma$ } & $-$  &  $-$ & $-$ \\
   CR$\gamma$-medium     &  $-$  &  $> 400$  &  $\geq 2$   & $-$    &  {\bf $0\ell$, $1\gamma$ } & $-$  &  $-$ & $-$ \\
   CR$\gamma$-high       &  $-$  &  $> 700$  &  $\geq 2$   & $-$    &  {\bf $0\ell$, $1\gamma$ } & $-$  &  $-$ & $-$ \\
   CR-real          &  $-$      &  $\bm{> 200}$  &  $\geq 2$ & {\bf 81--101}    &  $2\ell$ SF  & OS  &  $-$ & $-$\\
   \multirow{2}{*}{CR-fake}          &  \multirow{2}{*}{$\bm{<125}$}   & \multirow{2}{*}{$-$}      &  \multirow{2}{*}{$-$} & {\bf $\in[12,\infty]$,}     & \multirow{2}{*}{ {\bf $2\ell$ SF/DF} } & \multirow{2}{*}{{\bf SS}} &  \multirow{2}{*}{$-$} & \multirow{2}{*}{$-$}\\
    &  &  &  & {\bf $\notin[81,101]$(SF)} &  & & & \\

   \noalign{\smallskip}\hline\noalign{\smallskip}
   \multicolumn{2}{l}{Validation regions} &&&&&& \\
   \noalign{\smallskip}\hline\noalign{\smallskip}
   VR-low          &  {\bf 100--200}  &  $-$      &  $\geq 2$   & $>12$  &  SF  & OS & $>0.4$ & $-$\\
   VR-medium       &  {\bf 100--200}  &  $> 400$  &  $\geq 2$   & $>12$  &  SF  & OS & $>0.4$ & $-$\\
   VR-high         &  {\bf 100--200}  &  $> 700$  &  $\geq 2$   & $>12$  &  SF  & OS & $>0.4$ & $-$\\
   \multirow{2}{*}{VR-fake} & \multirow{2}{*}{$\bm{>50}$} & \multirow{2}{*}{$-$} & \multirow{2}{*}{ $\geq 2$} & {\bf $\in[12,\infty]$,} & \multirow{2}{*}{{\bf SF/DF}} & \multirow{2}{*}{{\bf SS}} & \multirow{2}{*}{$-$} & \multirow{2}{*}{$-$} \\
    &  &  &  & {\bf $\notin[81,101]$(SF)} &  & & & \\
   \noalign{\smallskip}\hline\noalign{\smallskip}
\end{tabular}
}
 \label{tab:regions-edge}
\end{center}
\end{table}

For the on-shell $Z$ search, the leading-lepton \pt\ threshold is raised to $50$~\GeV\ to increase 
the sensitivity to signal models with final-state $Z$ bosons.
This is an increased leading-lepton \pt\ threshold relative to Ref.~\cite{SUSY-2014-10}
and is found to better reject fake-lepton candidates from misidentified jets, photon conversions and $b$-hadron decays,
while retaining high efficiency for signal events, which tend to produce boosted $Z$ bosons.
To select events containing a leptonically decaying $Z$ boson, the invariant mass of the dilepton system is required to be $81<m_{\ell\ell}<101$~\GeV.
In the CRs and VRs that use the $Z$ mass sidebands, only events with $\mll>45$~\GeV\ are used to reject the lower \mll\ region dominated by Drell--Yan (DY) production.
In Ref.~\cite{SUSY-2014-10} an ``on-$Z$'' SR, denoted SRZ, is defined requiring
$\met>225$~\GeV\ and $\htincl  > 600$~\GeV.
The region is motivated by SUSY signals with high gluino or squark mass and high jet activity.
Since $b$-jets are not always expected in the simplified models considered here,
no requirement is placed on $b$-tagged jet multiplicity ($n_{b\mathrm{-jets}}$) so as to be as inclusive as possible and to be consistent with Ref.~\cite{SUSY-2014-10}.
Dedicated CRs are defined, with selection criteria similar to those of SRZ,
to estimate the contribution from the dominant SM backgrounds in SRZ. 
These CRs are discussed in more detail in Section~\ref{sec:bg}.

The edge selection requires at least two leptons with $\pt>25$~\GeV.
The search is performed across the full \mll\ spectrum,
with the exception of the region with $\mll<12$~GeV, which is vetoed to reject 
low-mass DY events and the $J/\psi$ and $\Upsilon$ resonances.
Three regions are defined to target signal models with low, medium and high values of $\Delta m_{\tilde{g}} = m(\tilde{g}) - m(\tilde{\chi}^0_1)$,
denoted SR-low, SR-medium, and SR-high, respectively.
All these regions require $\met>200$~\GeV.
SR-medium and SR-high also include the requirements $\HT >400$~\GeV\ and $\HT>700$~\GeV, respectively,
to further isolate high-$\Delta m_{\tilde{g}}$ events.
Here the leptons are not included in the \HT\ definition to avoid introducing any bias in the \mll\ distribution.
Events selected in SR-low, SR-medium and SR-high are further grouped into non-orthogonal \mll\ windows, 
which represent the search regions used in the edge analysis.
The dilepton mass ranges of these are chosen to maximise sensitivity to the targeted signal models,
with the window boundaries being motivated by the dilepton mass endpoints of generated signal points. 
In total, 24 \mll\ windows are defined by selecting ranges with the best expected sensitivity to signal models.
Of these windows, nine are in SR-low, eight are in SR-medium and seven are in SR-high. 
Details of the \mll\ definitions in these regions are given along with the results in Section~\ref{sec:result}.
Models without light sleptons are targeted by windows with $\mll<60$~\GeV\ or $\mll<80$~\GeV\ for $\Delta m_\chi < m_Z$ leading to off-shell $Z$ bosons,
and by the window with $81<\mll<101$~\GeV\ for $\Delta m_\chi > m_Z$ leading to on-shell $Z$ bosons.
Models with light sleptons are targeted by the remaining \mll\ windows, which cover the full \mll\ range.
The edge selection and on-shell Z selection are not orthogonal.  In particular, SR-medium in the range $81<\mll<101$~\GeV\ overlaps significantly with SRZ.

For the combined $ee+\mu\mu$ channels, the typical signal acceptance times efficiency values for the signal models considered in 
SRZ are 2--8\%.
They are 8--40\%, 3--35\%, and 1--35\%, inclusively in \mll, for SR-low, SR-medium and SR-high, respectively.
The on-shell $Z$ and edge analyses are each optimised for different signal models.
There are models in which signal contamination in CRs or VRs can become significant. 
For example, CRT in Table~\ref{tab:regions-z} is used to normalise the \ttbar\ MC sample to data as a cross-check in the on-shell $Z$ search, 
but it is a region where the signal contamination from signal models targeted by the edge search can be up to 80\% relative to the expected background.
In addition, the contamination from on-shell $Z$ signals in the region used to validate the \dyjets\ and flavour-symmetric estimates, VR-S, 
is up to 60\% for models with $m(\tilde{g})<1$~\TeV.  
The signal contamination from the slepton models in the DF regions used to estimate the flavour-symmetric backgrounds in the edge search, 
CR-FS-low/medium/high in Table~\ref{tab:regions-edge}, is less than 20\% for models with $m(\tilde{g})>600$~\GeV. 
It is only the contamination in these $e\mu$ CRs that is relevant in terms of the model-dependent interpretation of the results, 
and its impact is further discussed in Section~\ref{sec:interpretation}.  
In general, for models giving substantial contamination in the CRs, 
the signal-to-background ratio in the SRs is found to be large enough for this contamination to have negligible impact on the sensitivity of the search.

%% file: sec-bg.tex
The dominant background processes in the SRs are
``flavour-symmetric'' (FS) backgrounds, where the ratio of $ee$, $\mu\mu$ and $e\mu$ dileptonic branching 
fractions is 1:1:2 because the two leptons originate from independent $W\to\ell\nu$ decays.
This background is dominated by \ttbar ($50$--$70$\%) and also includes $WW$, $Wt$, and $Z\to\tau\tau$ processes.
The FS background constitutes $60$--$90$\% of the expected background in the SRs,
and is estimated using control samples of $e\mu$ events.

As all the SRs have a high-\met\ requirement, 
\dyjets\ events only enter the SRs when there is large \met\ originating from instrumental effects or from neutrinos in jet fragments.
This background is generally small, but it is difficult to model with MC simulation and can mimic signal, particularly for the on-shell $Z$ search.
This background is estimated using a control sample of \gjets\ events in data, which are kinematically similar to \dyjets\
and have similar sources of \met.

The production of $WZ/ZZ$ dibosons contributes approximately 30\% of the SM background in SRZ
and up to $20$\% of the background in the edge SR \mll\ windows. These backgrounds are estimated from MC simulation, after
validation in dedicated $3\ell$ ($WZ$) and $4\ell$ ($ZZ$) VRs.
Rare top backgrounds, which include $t\bar{t}W$, $t\bar{t}Z$ and $t\bar{t}WW$ processes, constitute $<5$\% of the expected SM background in all SRs,
and are estimated from MC simulation.
The contribution from events with fake or misidentified leptons is at most $15$\% (in one of the edge \mll\ ranges in SR-low), 
but is generally $<5$\% of the expected SM background in most SRs.

\subsection{Flavour-symmetric backgrounds}
\label{sec:FS}

The flavour-symmetric background is dominant in all SRs. 
To estimate the contribution of this background to each SR, the so-called ``flavour-symmetry'' method, 
detailed in Ref.~\cite{SUSY-2014-10}, is used. 
In this method, data events from a DF control sample, which is defined with the same kinematic requirements as the SR, 
are used to determine the expected event yields in the SF channels.
In the on-shell $Z$ analysis, the method is used to predict the background yield in the $Z$ mass window, defined as $81<\mll<101$~\GeV.
In the edge analysis, the method is extended to predict the full dilepton mass shape, 
such that a prediction can be extracted in any of the predefined \mll\ windows.

For the edge search, the flavour-symmetric contribution to each $\mll$ bin of the signal regions is predicted using data from the corresponding bin in an $e\mu$ control region. 
All edge CR-FS regions (definitions can be seen in Table~\ref{tab:regions-edge}) are $88$--$97$\% pure in flavour-symmetric processes (this purity is calculated from MC simulation).

For the on-shell search, this method is complicated slightly by a widening of the $\mll$ window used in CR-FS, 
the $e\mu$ control region (defined in Table~\ref{tab:regions-z}). 
The window is enlarged to $61<\mll<121$~\GeV\ to approximately triple the amount of data in the control region and thus increase the statistical precision of the method. 
This results in a region that is $\sim95$\% pure in flavour-symmetric processes 
(the expected composition of this $95$\% is $\sim80$\% \ttbar, $\sim10$\% $Wt$, $\sim10$\% $WW$ and $<1$\% $Z \rightarrow \tau\tau$).

Apart from the $\mll$ widening in CR-FS, the method used is identical for the on-shell and edge regions. 
Events in the control regions are subject to lepton $\pt$- and $\eta$-dependent correction factors measured in data and MC simulation. 
Because the triggers used are not identical in 2015 and 2016, these factors are measured separately for each year
and account for the different identification and reconstruction efficiencies for electrons and muons, 
as well as the different trigger efficiencies for the dielectron, dimuon and electron--muon selections. 
The estimated numbers of events in the SF channels, $N^\text{est}_{ee/\mu\mu}$, are given by:

\begin{eqnarray}
N_{ee}^\text{est} = \frac{1}{2} \cdot  f_{\mathrm{FS}} \cdot f_{Z \mathrm{\text{-}mass}} \cdot\sum^{N_{e\mu}^\text{data}}_{i} k_{e}(\pT^{i,\mu}, \eta^{i,\mu})\cdot \alpha(\pT^{i,\mu}, \eta^{i,\mu}) ,\\
N_{\mu\mu}^\text{est} = \frac{1}{2} \cdot  f_{\mathrm{FS}} \cdot f_{Z \mathrm{\text{-}mass}} \cdot \sum^{N_{e\mu}^\text{data}}_{i} k_{\mu}(\pT^{i,e}, \eta^{i,e})\cdot \alpha(\pT^{i,e}, \eta^{i,e}) ,
\end{eqnarray}

\noindent where $N_{e\mu}^\text{data}$ is the number of data events observed in a given control region, 
$\alpha(\pT^i, \eta^i)$ accounts for the different trigger efficiencies for SF and DF events, 
and $k_{e}(\pT^{i,\mu}, \eta^{i,\mu})$ and $k_{\mu}(\pT^{i,e}, \eta^{i,e})$ are electron and muon selection efficiency factors for the kinematics of the lepton being replaced, in event $i$.
The trigger and selection efficiency correction factors are derived from the events in an inclusive on-$Z$ selection ($81<\mll<101$~\GeV, $\geq2$ jets), 
according to:

\begin{eqnarray}\label{eq:kfac}
k_{e}(\pT, \eta) = \sqrt{\frac{N_{ee}^{\text{meas}(\pT, \eta)}}{N_{\mu\mu}^{\text{meas}(\pT, \eta)}}} \\
k_{\mu}(\pT, \eta) = \sqrt{\frac{N_{\mu\mu}^{\text{meas}(\pT, \eta)}}{N_{ee}^{\text{meas}(\pT, \eta)}}} \\
\alpha(\pT, \eta) = \frac{\sqrt{\epsilon^\text{trig}_{ee}(\pt^{\ell_1},\eta^{\ell_1})\times\epsilon^\text{trig}_{\mu\mu}(\pt^{\ell_1},\eta^{\ell_1})}}{\epsilon^\text{trig}_{e\mu}(\pt^{\ell_1},\eta^{\ell_1})}
\end{eqnarray}

\noindent where $\epsilon^\text{trig}_{ee/\mu\mu}$ is the trigger efficiency and $N_{ee/\mu\mu}^{\text{meas}}$ 
is the number of $ee/\mu\mu$ events in the inclusive on-$Z$ region outlined above. 
Here $k_{e}(\pT, \eta)$ and $k_{\mu}(\pT, \eta)$ are calculated separately for leading and sub-leading leptons, while $\alpha$ is calculated for the leading lepton, $\ell_1$. 
The correction factors are typically within 10\% of unity, except in the region $|\eta|<0.1$ where, 
because of the lack of coverage by the muon spectrometer, they are up to 50\% from unity.
For all background estimates based on the flavour-symmetry method, results are computed separately for $ee$ and $\mu\mu$ and then summed to obtain the combined predictions. 
The resulting estimates from the DF channels are scaled according to the fraction of flavour-symmetric backgrounds
in each $e\mu$ control sample, $f_{\mathrm{FS}}$ ($95\%$ in CR-FS), 
which is determined by subtracting non-flavour-symmetric backgrounds taken from MC simulation from the data observed in the corresponding $e\mu$ region. 
In the on-shell case, the result is also scaled by the fraction of events in CR-FS expected to be contained within $81<\mll<101$~\GeV, $f_{Z \mathrm{-mass}}$ ($38\%$), 
which is otherwise set to 100\% for the edge regions. 
The validity of extrapolating in \mll\ between CR-FS and SRZ 
was checked by comparing the \mll\ shapes in data and MC simulation in a region similar to VR-S, 
but with the \mll\ requirement relaxed and \htincl$>300$~\GeV\ to obtain a sample with a large number of events.
The resulting on-$Z$ fractions in MC simulation were found to agree with data within statistical uncertainties, 
which are summed in quadrature to assign a systematic uncertainty. 
In the case of the edge search the full \mll\ distribution is validated by applying a flavour-symmetry method to \ttbar\ MC evnets in VR-low, 
VR-medium and VR-high. 
This procedure results in good closure, which is further discussed in Section~\ref{sec:VRs}. 
The difference between the prediction and the observed distribution is used to assign an MC non-closure uncertainty to the estimate.

The flavour-symmetry method in SRZ is further cross-checked by performing a profile likelihood fit~\cite{statforumlimits} of MC yields to data in the $Z$-mass sidebands ($\mll \notin [81, 101]$~\GeV), 
the region denoted CRT in Table~\ref{tab:regions-z}, which is dominated by \ttbar\ (with a purity of $>75\%$) and contains $273$ events in data.   
The other flavour-symmetric processes in this region contribute $\sim12$\% ($Wt$), $10$\% ($WW$) and $<1$\% ($Z\rightarrow\tau\tau$).
All SM background processes are taken directly from MC simulation in this cross-check, including backgrounds also estimated using the flavour-symmetry method. 
The normalisation of the dominant \ttbar\ background is a free parameter and is the only parameter affected by the fit.
For this cross-check, the contamination from Beyond Standard Model processes in the $Z$-mass sidebands is assumed to be negligible.
The fit results in a scale factor of $0.64$ for the \ttbar\ yield predicted by simulation. 
This result is extrapolated from the $Z$-mass sidebands to SRZ and gives a prediction of $ 29 \pm 7 $ events, 
which is consistent with the nominal flavour-symmetry background estimate of $33 \pm 4$ in this region.  

The sideband fit is repeated at lower \met\ in VRT, with the results being propagated to VR-S, so as to test the \mll\ extrapolation used in the sideband fit method. 
The normalisation to data in this region, which is at lower \met\ relative to CRT, results in a scale factor of $0.80$ for the \ttbar\ yield predicted by simulation. 
The number of FS events predicted in VR-S using the sideband fit in VRT is compatible with the number estimated by applying the FS method to 
data in VR-FS.
The results of the background estimate in both VR-S and SRZ obtained from the flavour-symmetry method are compared with the values obtained by the sideband fit cross-check in 
Table~\ref{tab:FS}. 
The methods result in consistent estimates in both regions. 
Further results in the edge VRs are discussed in Section~\ref{sec:VRs}.

\begin{table}[h]
\centering
\caption{
Comparison of the predicted yields for the flavour-symmetric backgrounds in SRZ and VR-S as obtained from the nominal data-driven method using CR-FS and the $Z$-mass sideband method.
The quoted uncertainties include statistical and systematic contributions. }
\begin{tabular}{lcc}
\noalign{\smallskip}\hline\noalign{\smallskip}
Region  & Flavour-symmetry  & Sideband fit  \\
\noalign{\smallskip}\hline\noalign{\smallskip}
SRZ & $33 \pm 4$   &  $29 \pm 7$  \\ [+0.05cm]
VR-S & $99\pm8$        &  $\makebox[3ex]{\hfill 92} \pm \makebox[2ex]{25\hfill}$  \\ [+0.05cm]
\noalign{\smallskip}\hline
\end{tabular}
\label{tab:FS}
\end{table}

A potential cause of the low scale factors obtained from the sideband fit at large \HT\ and \met\ is mismodelling of the top-quark \pt\ distribution, 
where measurements of \ttbar\ differential cross sections by the ATLAS and CMS experiments indicate that the top-quark \pt\ distribution predicted by most generators is harder than that observed in data~\cite{Aad:2015hna,Khachatryan:2016gxp}. 
Corrections to the MC predictions according to NNLO calculations provided in Ref.~\cite{TopPtNNLO} indicate an improvement in the top-quark pair modelling at high \HT, which should lead to scale factors closer to unity.  
With the data-driven method used to estimate \ttbar\ contributions in this analysis, the results do not depend on these corrections.
They are therefore not applied to the \ttbar\ MC sample for the sideband-fit cross-check.

\subsection{\dyjets\ background}
\label{sec:zjets}

The \dyjets\ background estimate is based on a data-driven method that uses \gjets\ events in data to model the \met\ distribution of \dyjets.  
The \gjets\ and \dyjets\ processes have similar event topologies, with a well-measured object recoiling against a hadronic system, 
and both tend to have \met\ that stems from jet mismeasurements and neutrinos in hadronic decays. 
In this method, which has been used by CMS in a search in this final state~\cite{Chatrchyan:2012qka}, 
a sample of data events containing at least one photon and no leptons is constructed using the same kinematic selection as each of the SRs, 
without the \met\ and $\Delta\phi(\text{jet}_{12},{\boldsymbol p}_{\mathrm{T}}^\mathrm{miss})$ requirements (the CR$\gamma$ regions defined in Tables~\ref{tab:regions-z} and~\ref{tab:regions-edge}).

The requirement
$\Delta\phi(\text{jet}_{12},{\boldsymbol p}_{\mathrm{T}}^\mathrm{miss}) > 0.4$ applied in the SRs suppresses \MET from jet mismeasurements
and increases the relative contributions to \MET from the photon, electrons, and muons.
The difference in resolution between photons, electrons, and muons can be significant at high $\pt$.  
Therefore, before the $\Delta\phi(\text{jet}_{12},{\boldsymbol p}_{\mathrm{T}}^\mathrm{miss}) > 0.4$ requirement is applied, the photon \pt\ is 
smeared according to a $Z\rightarrow ee$ or $Z\rightarrow\mu\mu$ resolution function.
The smearing function is derived by comparing the \met-projection along the boson momentum in \dyjets\ and \gjets\ MC events in a 1-jet control region with no other event-level kinematic requirements.
A deconvolution is applied to avoid including the photon resolution in the $Z$ resolution.
For each event, a photon \pt\ smearing $\Delta\pt$ is obtained by sampling the smearing function.
The photon \pt\ is shifted by $\Delta\pt$, with the parallel component of the \met\ being correspondingly adjusted by $-\Delta\pt$.

The smeared \gjets\ events are then reweighted to match the boson $\pt$ distribution of the \dyjets\ events. 
This reweighting is applied separately in each region and accounts for small differences between the \gjets\ events and \dyjets\ events, 
which arise mainly from the mass of the $Z$ boson. 
The reweighting is done using \dyjets\ events in data, and is checked using \dyjets\ MC simulation in an MC closure test, as described further below. 
Following this smearing and reweighting procedure, the \met\ of each \gjets\ event is recalculated, 
and the final \met\ distribution is obtained after applying the $\Delta\phi(\text{jet}_{12},{\boldsymbol p}_{\mathrm{T}}^\mathrm{miss} ) > 0.4$ requirement.
For each SR, the resulting \met\ distribution is normalised to data in a CRZ with the same requirements except that the SR \MET requirement
is replaced by $\MET<60$~\GeV. 

The shape of the \dyjets\ \mll\ distribution is extracted from MC simulation
and validated by comparing to data in events with lower \MET requirements and a veto on $b$-tagged jets, to suppress the background from \ttbar. 
The \mll\ distribution is modelled by parameterising the \mll\ in \dyjets\ events as a function of the difference between reconstructed and true $Z$ boson \pt\ in MC simulation. 
This parameterization ensures that the correlation between lepton momentum mismeasurement and observed \mll\ values far from the $Z$ boson mass is preserved.
Each photon event is assigned an \mll\ via a random sampling of the corresponding distribution, equating photon $\Delta\pt$ and the difference between true and reconstructed $Z$ boson \pt.
The resulting \mll\ distribution in \gjets\ MC simulation is compared to that extracted from \dyjets\ MC simulation
and the difference is assessed as a systematic uncertainty in the background prediction for each \mll\ bin.

The full smearing, reweighting, and \mll\ assignment procedure is applied to the $V\gamma$ MC sample in parallel with the \gjets\ data sample. 
After applying all corrections to both samples, the $V\gamma$ contribution to the \gjets\ data sample is subtracted to remove contamination from backgrounds with real \met.
Contamination by events with fake photons in these \gjets\ data samples is small, and this contribution is therefore neglected.
 
In the \HT-inclusive region corresponding to VR-low, 
there is a non-negligible contribution expected from \dyjets\ events with $\pt^{Z}<37$~\GeV. 
Given the photon trigger strategy discussed in Section~\ref{sec:data}, no photons with $\pt<37$~\GeV\ are included in the event selection. 
To account for this photon \pt\ threshold, a boson-\pt\ correction of up to 50\% is applied as a function of \met\ in VR-low. 
This correction uses the fraction of \dyjets\ events in a given \met\ bin expected to have $\pt^{Z}<37$~\GeV, according to MC simulation.
The \gjets\ data are then scaled according to this fraction, as a function of \met, to correct for the missing $\pt^{Z}<37$~\GeV\ contribution.
The correction is found to be negligible in all signal regions.

\begin{figure}[!htb]
\centering
\subfloat[]{
\includegraphics[width=0.47\textwidth]{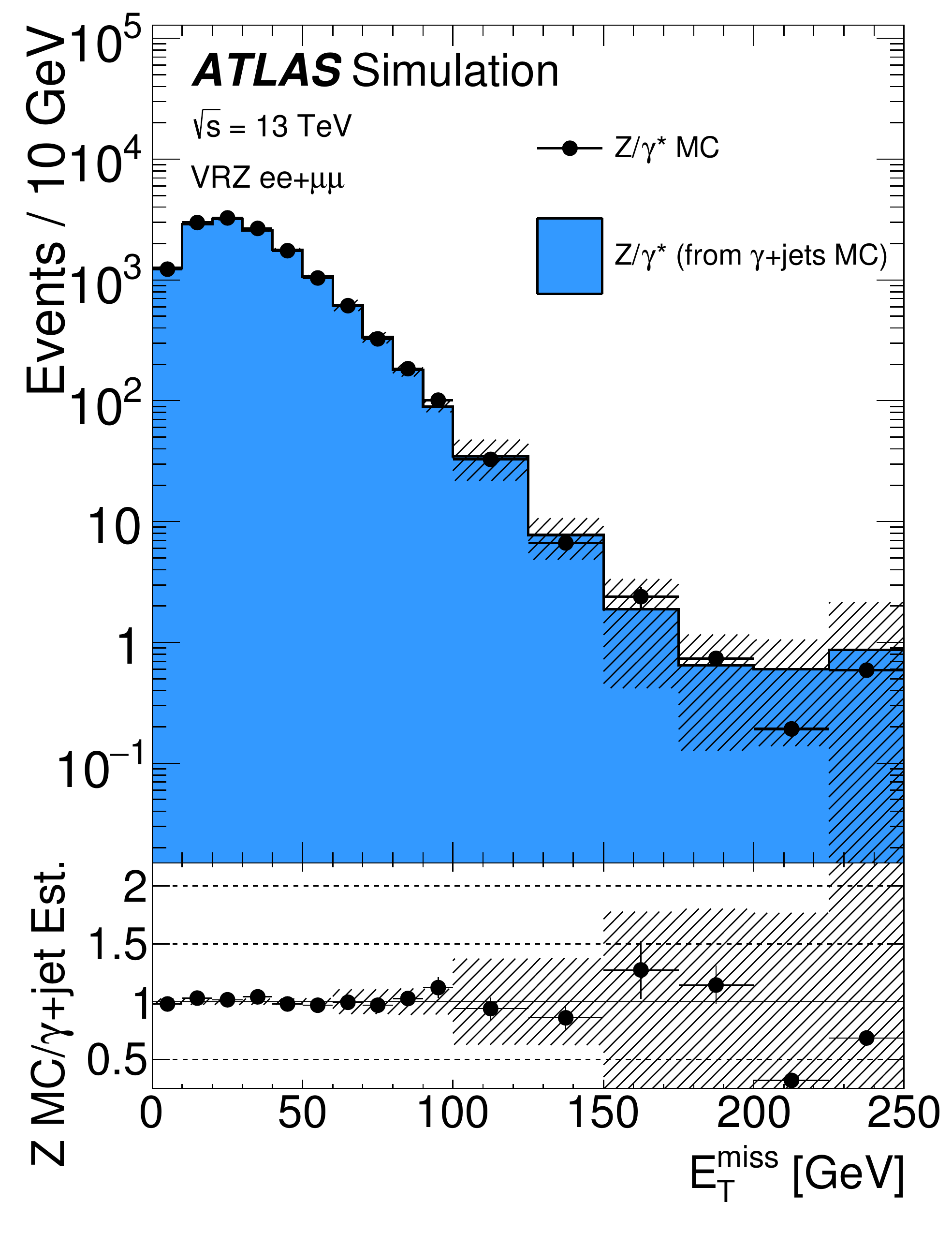}
\label{fig:gammajet_closure_a}
}
\subfloat[]{
\includegraphics[width=0.47\textwidth]{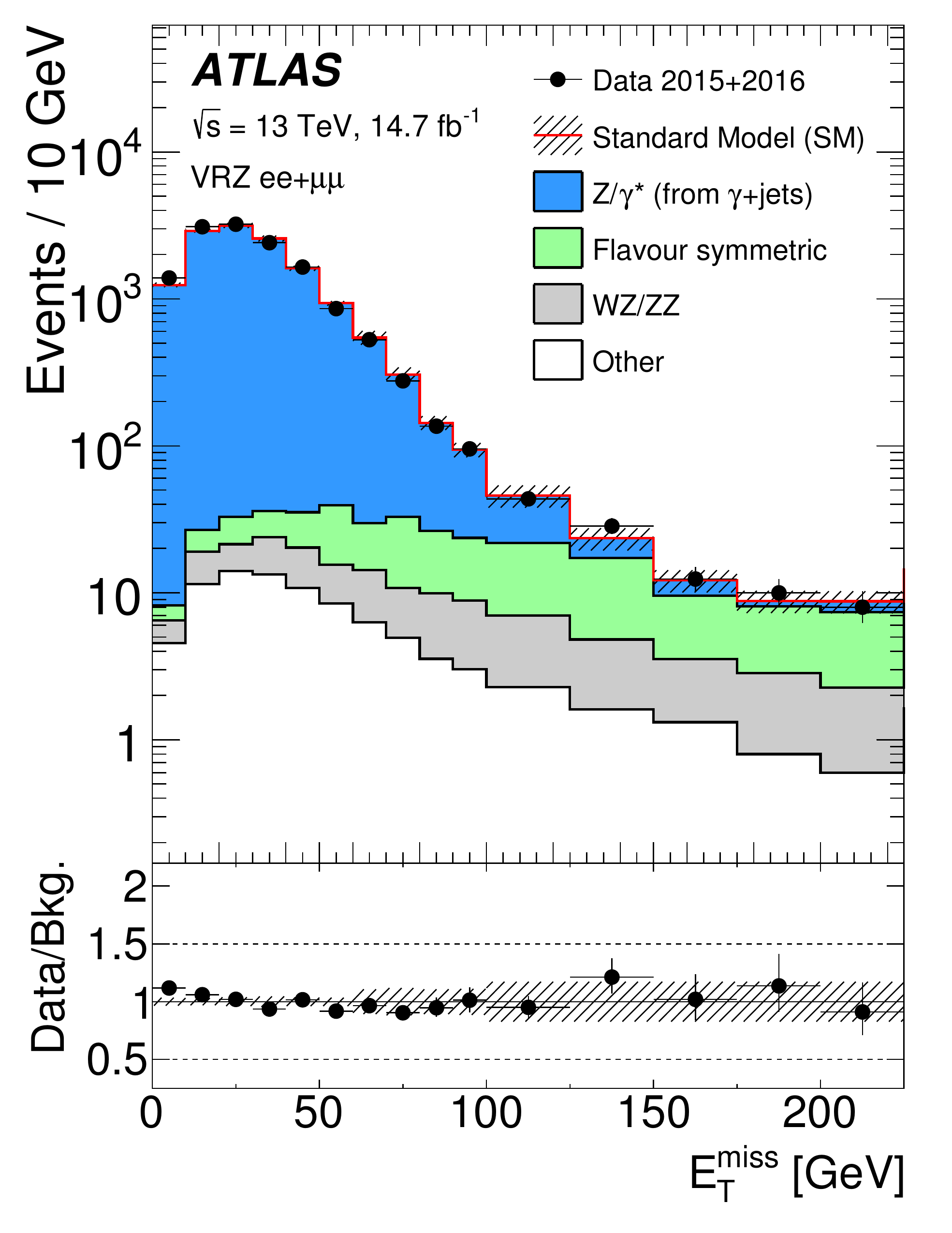}
\label{fig:gammajet_closure_b}
}
\caption{
Left, the \met\ spectrum in {\sc Sherpa} \dyjets\ MC simulation compared to that of the \gjets\ background estimation technique applied to {\sc Sherpa} \gjets\ MC simulation in VRZ.  
The error bars on the points indicate the statistical uncertainty of the \dyjets\ MC simulation, 
and the hashed uncertainty bands indicate the statistical and reweighting systematic uncertainties of the $\gamma+$jet background method.  
For this MC comparison the upper \met\ cut has been removed from VRZ and the overflow is included in the rightmost bin.
Right, the \met\ spectrum when the method is applied to data in VRZ.  
Here the flavour-symmetric background is estimated using the data-driven flavour-symmetry method, 
and the fake-lepton background is estimated using the data-driven method explained in Section~\ref{sec:fakes}.
Rare top and diboson backgrounds are taken from MC simulation. 
The rare top and data-driven fake-lepton backgrounds are grouped under ``other'' backgrounds.
The hashed bands indicate the systematic uncertainty of only the \gjets\ and flavour-symmetric backgrounds below 100~\GeV\ and the full uncertainty of the VR-S prediction above 100~\GeV. 
The bottom panel of each figure shows the ratio of the observation (left, in MC simulation; right, in data) to the prediction.
}
\label{fig:gammajet_closure}
\end{figure}

The distribution of \met\ obtained in {\sc Sherpa} \dyjets\ MC simulation is compared to that obtained by applying this background estimation technique to {\sc Sherpa} \gjets\ MC samples.
In this check the \gjets\ MC simulation is reweighted according to the \pt\ distribution given by the \dyjets\ MC simulation.
The result of this MC closure check is shown in Figure~\ref{fig:gammajet_closure_a} for events in VRZ (without an upper \met\ cut), 
where good agreement between \dyjets\ and corrected \gjets\ MC simulation can be seen across the entire \met\ spectrum.
A comparison between the full \met\ spectrum in data and the \dyjets\ background estimated via the \gjets\ method is also shown in Figure~\ref{fig:gammajet_closure_b} for events in VRZ.
The systematic uncertainties associated with this method are described in Section~\ref{sec:syst}.

\subsection{Fake-lepton background}
\label{sec:fakes}

Semileptonic \ttbar, $W\rightarrow \ell\nu$ and single top ($s$- and $t$-channel) events enter the dilepton channels via ``fake'' leptons.
These can include misidentified hadrons, converted photons or non-prompt leptons from $b$-hadron decays. 
The extent of this background is estimated using the matrix method, detailed in Ref.~\cite{SUSY-2013-20}.
Its contribution in regions with high lepton \pt\ and dilepton invariant mass is negligible, 
but in the edge search, where lower-\pt\ leptons are selected and events can have low \mll,
the fake-lepton background can make up to 15\% of the total background.
In this method a control sample is constructed using baseline leptons, 
thereby enhancing the probability of selecting a fake lepton due to the looser lepton selection and identification criteria relative to the signal lepton selection. 
For each relevant CR, VR or SR, the region-specific kinematic requirements are placed upon this sample of baseline leptons. 
The number of events in this sample in which the selected leptons subsequently pass ($N_{\text{pass}}$) or fail ($N_{\text{fail}}$) 
the signal lepton requirements in Section~\ref{sec:objects} are then counted. 
In the case of a one-lepton selection, the number of fake-lepton events in a given region is then estimated according to:

\begin{equation}
N_{\text{pass}}^{\text{fake}} = \frac{N_{\text{fail}} - (1/\epsilon^{\text{real}} - 1) \times N_{\text{pass}} }{1/\epsilon^{\text{fake}} - 1/\epsilon^{\text{real}}}.
\end{equation}

\noindent Here $\epsilon^{\text{real}}$ is the relative identification efficiency (from baseline to signal) for 
genuine, prompt (``real'') leptons and $\epsilon^{\text{fake}}$ is the relative identification efficiency (again from baseline to signal) with which non-prompt leptons or jets might be misidentified as prompt leptons. 
This principle is then expanded to a dilepton selection by using a four-by-four matrix to account for the various possible real--fake combinations for the two leading leptons in an event.

The real-lepton efficiency, $\epsilon^{\text{real}}$, is measured in $Z\rightarrow\ell\ell$ data events using a tag-and-probe method in CR-real, 
defined in Table~\ref{tab:regions-edge}. 
In this region the \pt\ of the leading lepton is required to be $>40$~\GeV, and only events with exactly two SFOS leptons are selected. 
The fake-lepton efficiency, $\epsilon^{\text{fake}}$, is measured in CR-fake, a region enriched with fake leptons by requiring same-sign lepton pairs. 
The lepton \pt\ requirements are the same as those in CR-real, with the leading lepton being tagged as the ``real'' lepton and the fake efficiency being evaluated based on the sub-leading lepton in the event. 
An \met\ requirement of $<125~\GeV$ is used to reduce possible contamination from Beyond Standard Model processes.
In this region the background due to prompt-lepton production, estimated from MC simulation, is subtracted from the total data contribution. 
Prompt-lepton production makes up $7$\% ($11$\%) of the baseline electron (muon) sample and $10$\% ($61$\%) of the signal electron (muon) sample in CR-fake.
From the resulting data sample the fraction of events in which the baseline leptons pass a signal-like selection yields the fake efficiency.  
Both the real- and fake-lepton efficiencies are binned as a function of lepton \pt\ and calculated separately for the 2015 and 2016 data sets.

This method is validated by checking the closure in MC simulation and data--background agreement in VR-fake.  

\subsection{Diboson and rare top processes}

The remaining SM background contribution in the SRs is due to $WZ/ZZ$ diboson production and rare top processes 
($\antibar{t}Z$, $\antibar{t}W$ and $\antibar{t}WW$). 
The rare top processes compose $<5\%$ of the expected SM background in the SRs and are taken directly from MC simulation.

Production of $WZ/ZZ$ dibosons constitutes about $30$\% of the expected background in SRZ and up to 20\% in some edge SR \mll\ windows.
In SRZ, this background is composed of roughly $70$\% $WZ$, about $40$\% of which is $WZ\rightarrow\ell\ell\tau\nu$. 
This is the largest background contribution that is estimated from MC simulation, and must be carefully validated,
especially because these backgrounds contain $Z$ bosons and can thus mimic a signal by producing a peak at $\mll \approx m_Z$.
To validate the MC modelling of these backgrounds, VRs with three leptons (VR-WZ) and four leptons (VR-ZZ)
are defined (selection shown in Table~\ref{tab:regions-z}). 
In VR-WZ, from the three selected leptons in an event, the SFOS pair with \mll\ most consistent with the $Z$ mass is indentified as the $Z$ candidate. 
The transverse mass of the remaining lepton and the \met, $m_{\text{T}}(\ell_{3},\met)$, is then required to be $<100$~\GeV, forming the $W$ candidate. 
In VR-ZZ an $\met<100$~\GeV\ requirement is used to suppress $WZ$ and top processes.
The yields and kinematic distributions observed in these regions are well-modelled by MC simulation.
In particular, the \met, $H_\text{T}$, jet multiplicity, and boson $\pt$ distributions show good agreement. 
An additional three-lepton VR (VR-3L) is defined
to provide validation of the diboson background in a region of phase space closer to the SR; good agreement is observed in this region as well.

\subsection{Results in validation regions}
\label{sec:VRs}
The expected background yields in VR-S are shown in Table~\ref{tab:VRresults} and compared with the observed data yield.
Agreement between the data and the expected Standard Model background is observed.
The expected background yields in the three diboson VRs are also shown in Table~\ref{tab:VRresults}.
The data are consistent with the expected background. 
Similar information for the edge VRs is provided in Table~\ref{tab:VRedge}. 
Data and background estimates are in agreement within uncertainties.

\begin{figure*}[!t]
\centering
\includegraphics[width=0.32\textwidth]{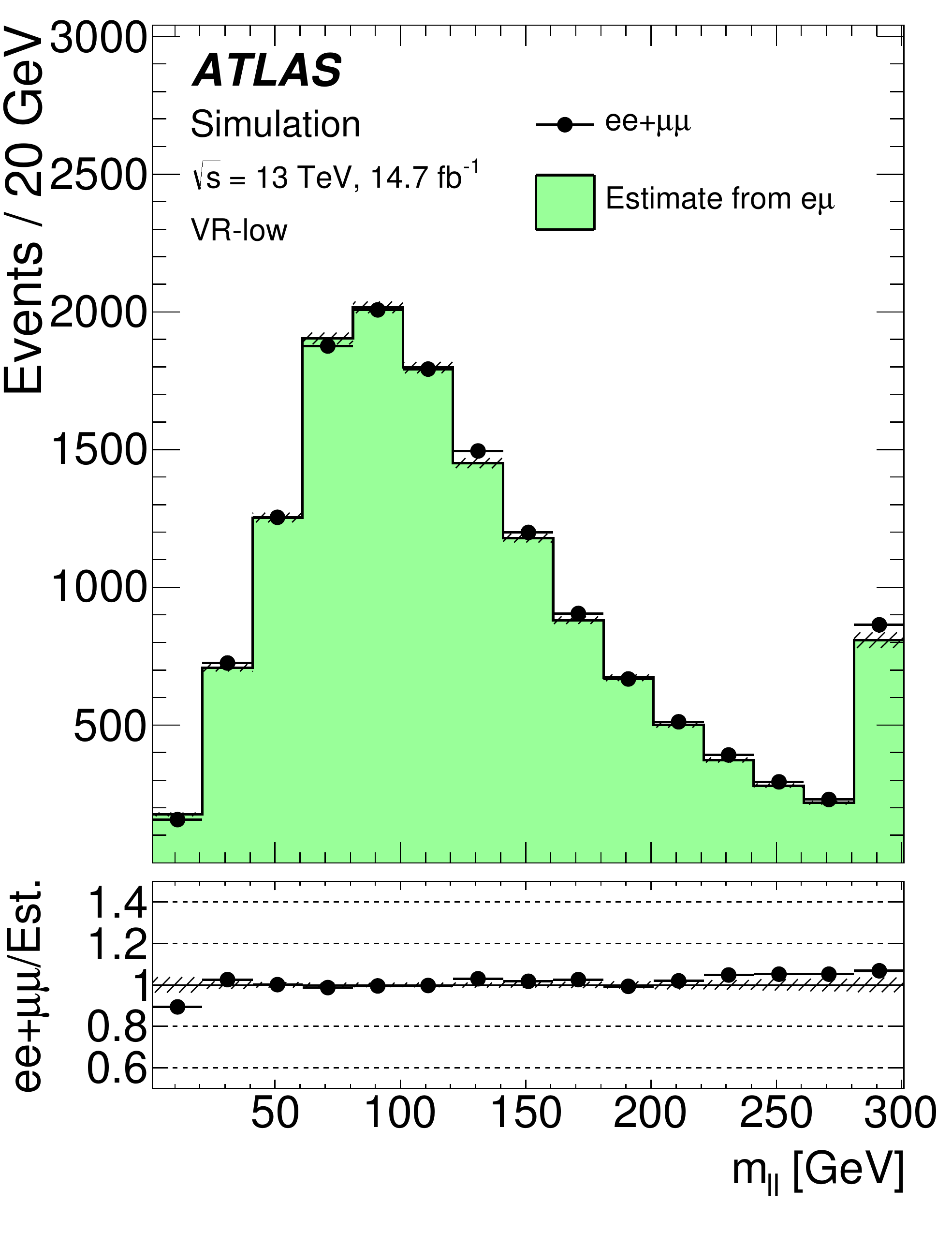}
\includegraphics[width=0.32\textwidth]{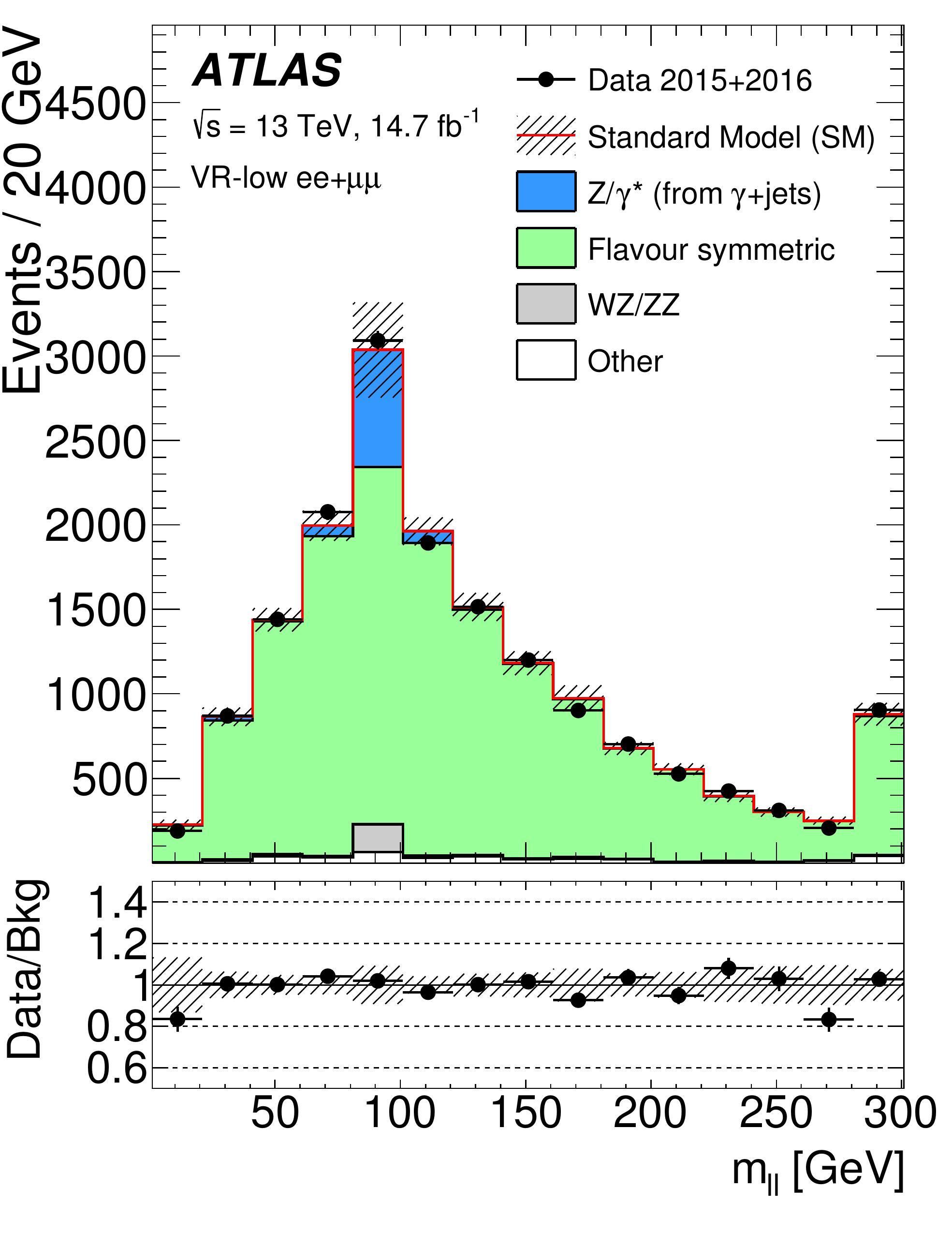} \\
\includegraphics[width=.32\textwidth]{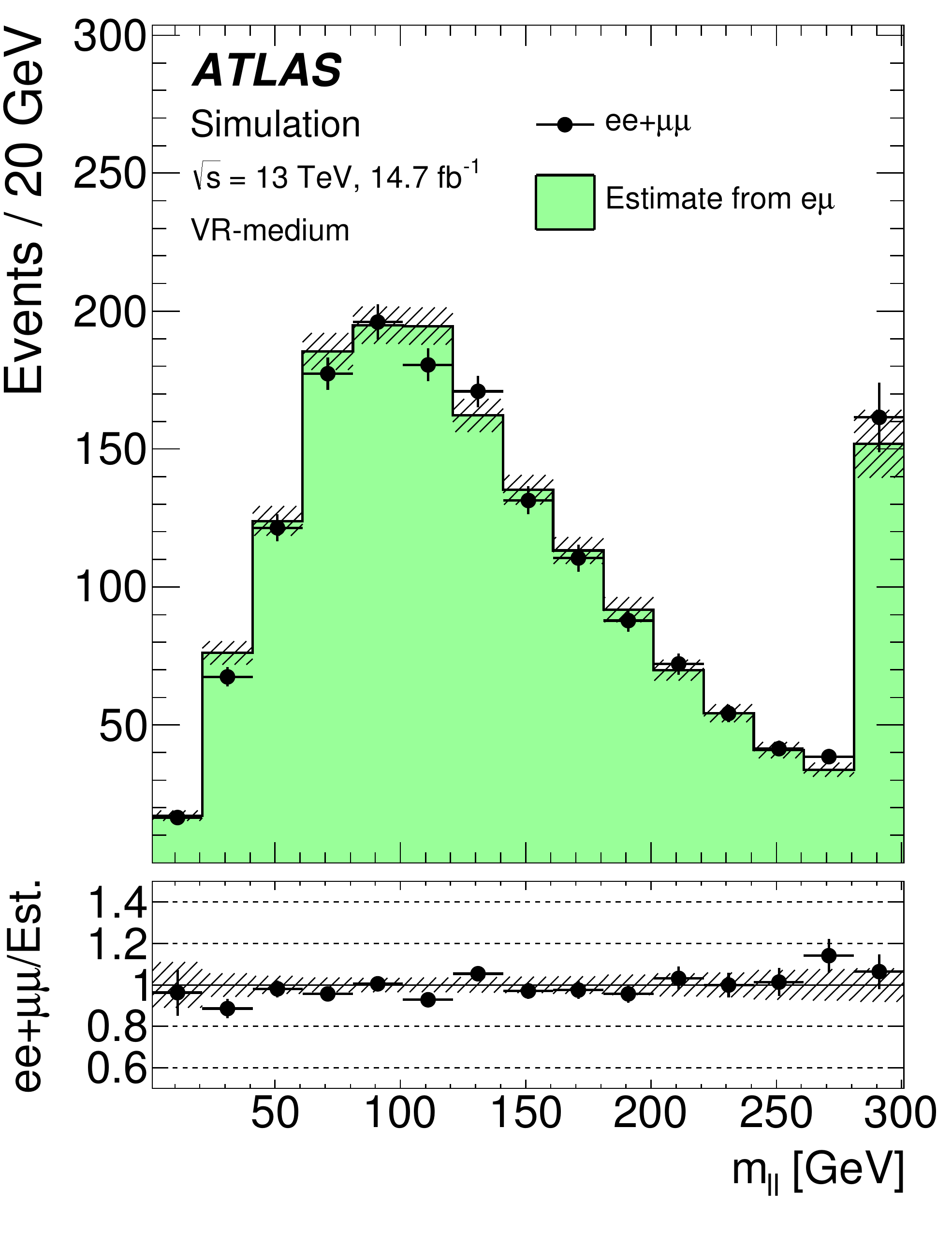}
\includegraphics[width=.32\textwidth]{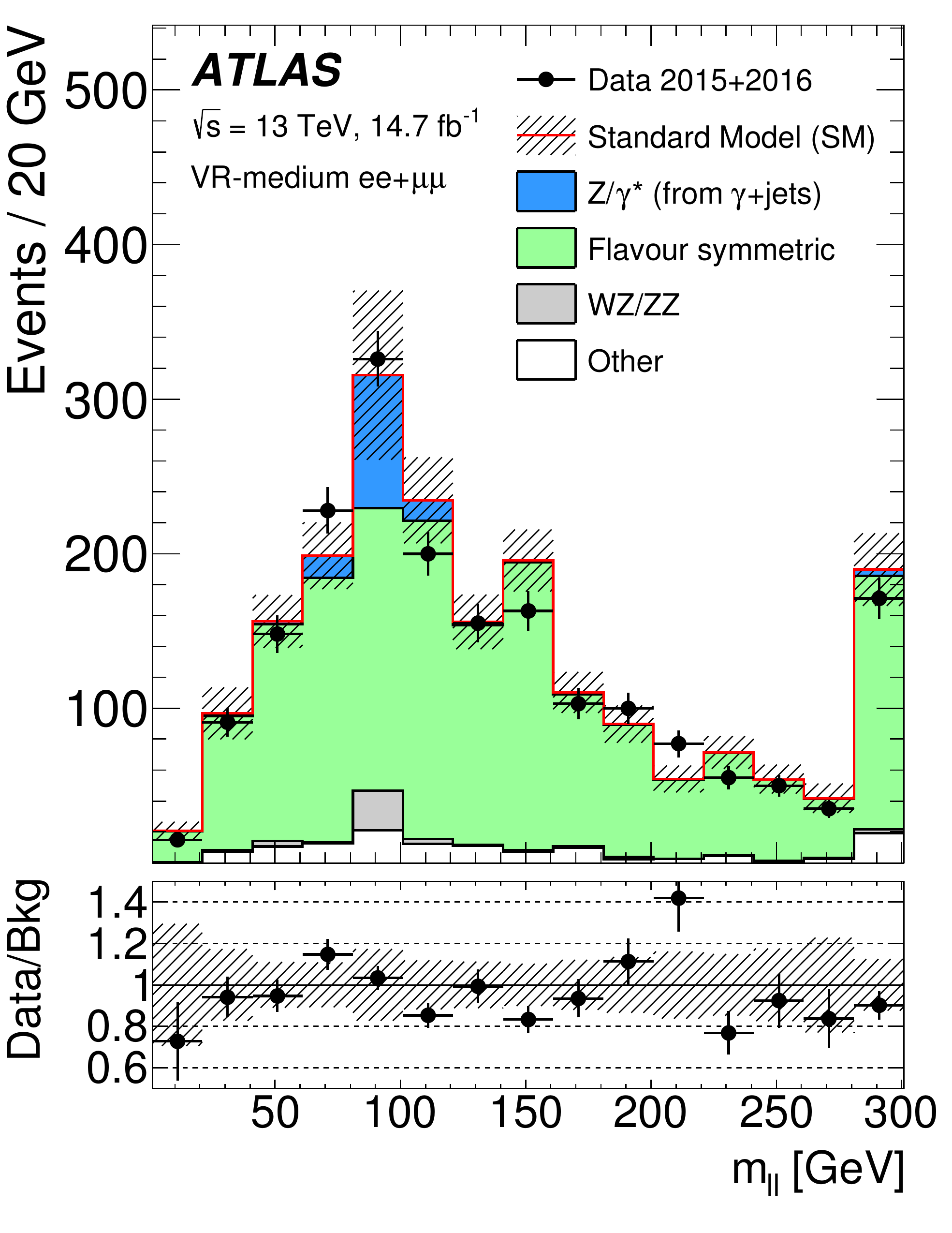} \\
\includegraphics[width=.32\textwidth]{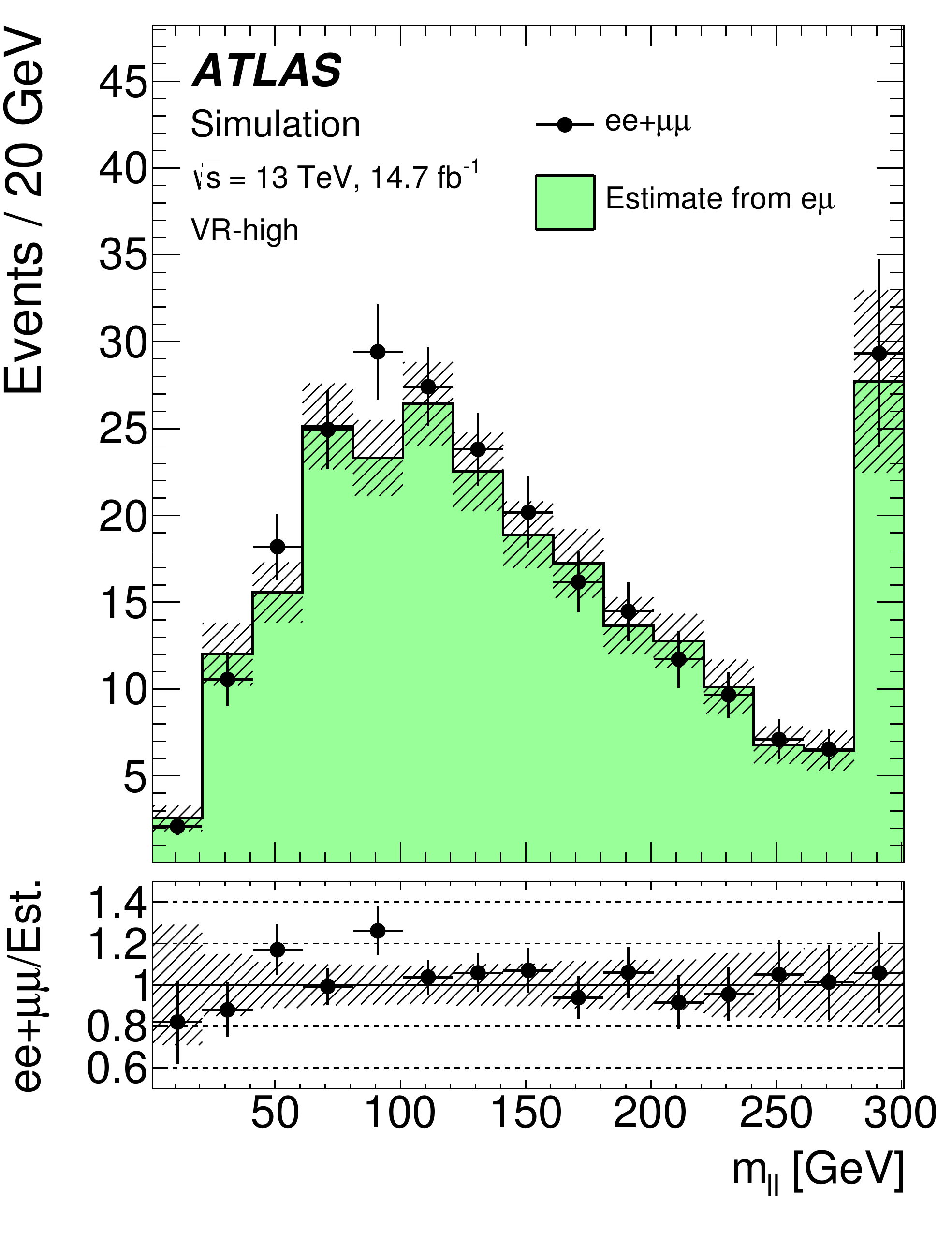}
\includegraphics[width=.32\textwidth]{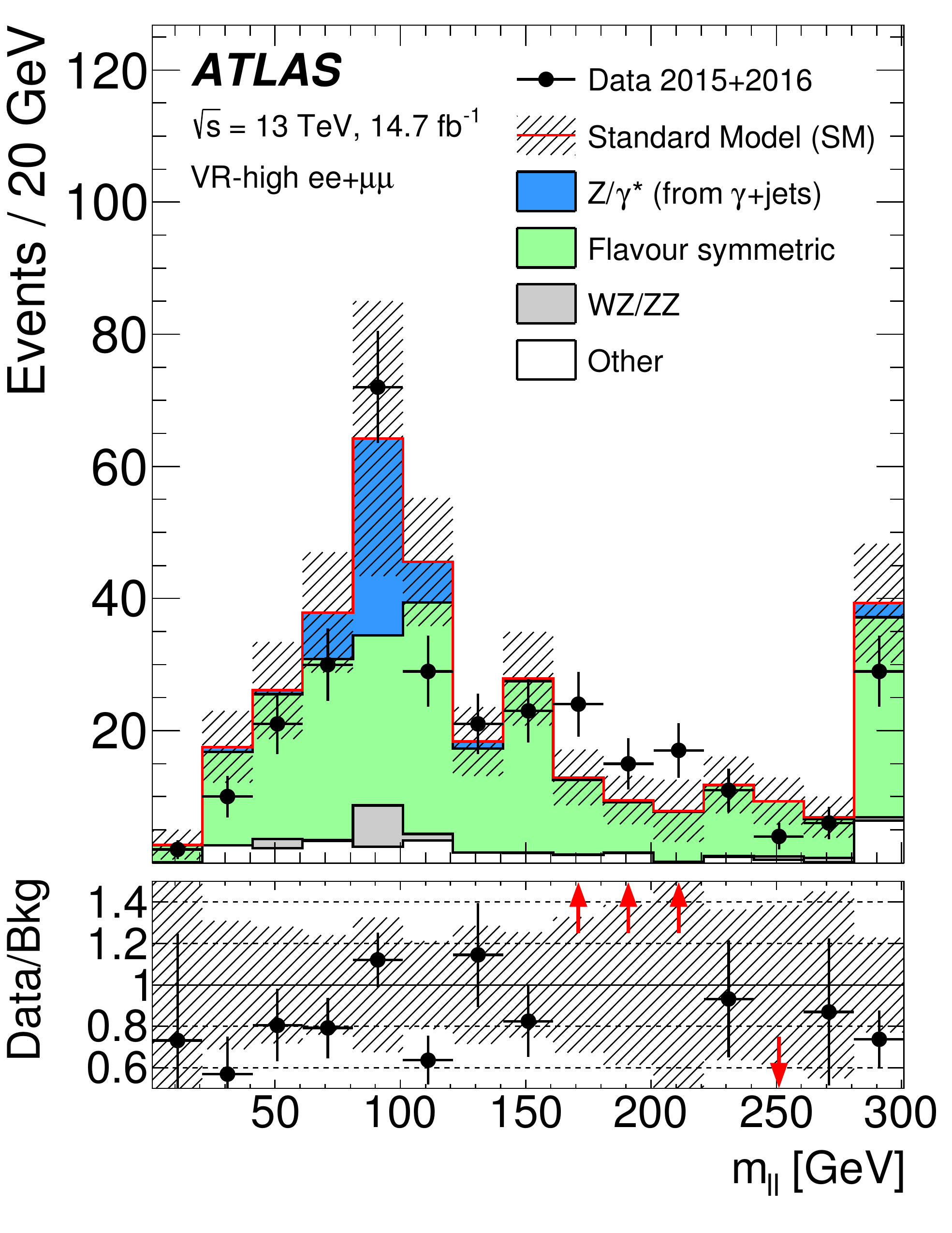}
\caption{Validation of the flavour-symmetry method for the edge search using MC events (left) and data (right), 
in the VR-low (top), VR-medium (middle), and VR-high (bottom) regions.
In the MC plots the flavour-symmetry estimate from $e\mu$ \ttbar\ MC samples
is compared with the observed SF distribution from these MC samples, with the MC statistical uncertainty indicated by the hashed bands. 
In the data plots, all uncertainties in the background prediction are included in the hashed band.  
The rare top and data-driven fake-lepton backgrounds are grouped under ``other'' backgrounds.
The bottom panel of each figure shows the ratio of the observation (left, in MC simulation; right, in data) to the prediction.
In cases where the data point is not accommodated by the scale of this panel, 
a red arrow indicates the direction in which the point is out of range.
The last bin contains the overflow.  }
\label{fig:fs}
\end{figure*}

Figure~\ref{fig:fs} shows the observed and expected \mll\ distributions in the same edge VRs.
The same background estimation methods are applied to both MC simulation and data.
In the MC studies, the flavour-symmetry method of Section~\ref{sec:FS} is applied to \ttbar\ MC simulation, and the observed
SF \mll\ distribution is compared to the prediction based on DF events. 
In the data studies, the observed SF \mll\ distribution is compared to the sum of FS backgrounds from the 
extended flavour-symmetry method,
the \dyjets\ background from the \gjets method, and the $WZ/ZZ$ diboson, rare top, and fake-lepton backgrounds.

The observed MC closure is good in all validation regions.  
The data agree with the expected background in the validation regions as well.
No significant discrepancies or trends are apparent.

\begin{table}
\caption{
Expected and observed event yields in the four validation regions, VR-S, VR-WZ, VR-ZZ, and VR-3L.
The flavour-symmetric, \dyjets, and fake-lepton contributions to VR-S are derived using the data-driven estimates described in Section~\ref{sec:bg}.
All remaining backgrounds, and all backgrounds in the diboson validation regions, are taken from MC simulation.
The quoted uncertainties in VR-S include statistical and all systematic contributions.
In VR-WZ, VR-ZZ, and VR-3L, the rare top and diboson uncertainties include statistical and all theoretical uncertainties described in Section~\ref{sec:syst}.
The fake-lepton contribution in these three regions is predominantly due to \dyjets, and in this case only the statistical uncertainty is given.
The individual uncertainties can be correlated and do not necessarily add up in quadrature to the total systematic uncertainty.
}
\begin{center}
\setlength{\tabcolsep}{0.0pc}
\begin{tabular*}{\textwidth}{@{\extracolsep{\fill}}lcccc@{}}
\noalign{\smallskip}\hline\noalign{\smallskip}
                                                                             & VR-S            & VR-WZ        & VR-ZZ          & VR-3L   \\[-0.05cm]
\noalign{\smallskip}\hline\noalign{\smallskip}
Observed events                                                              & $236$            & $698$        & $132$          & $32$ \\
\noalign{\smallskip}\hline\noalign{\smallskip}
Total expected background events                                             & $224 \pm 41$   &   $622\pm66$   &    $139\pm25$    &    $35\pm10$ \\
\noalign{\smallskip}\hline\noalign{\smallskip}
  Flavour-symmetric (\ttbar, $Wt$, $WW$, $Z\rightarrow\tau\tau$)             & $99\pm8$   &   \makebox[2ex]{\hfill -}                &    \makebox[2ex]{\hfill -}                 &    \makebox[2ex]{\hfill -}              \\
  $WZ/ZZ$ events                                                             & $\makebox[9ex]{\hfill 27}\pm\makebox[8ex]{13\hfill}$  &   $573\pm66$ &    $139\pm25$        &    $25\pm10$ \\
  Rare top events                                                            & $11\pm3$   &   $14\pm3$   &    $\makebox[9ex]{\hfill 0.44}\pm\makebox[8ex]{0.11\hfill}$     &    $9.1\pm2.3$  \\
  \dyjets\ events                                                            & $\makebox[9ex]{\hfill84}\pm\makebox[8ex]{37\hfill}$  &   \makebox[2ex]{\hfill -}                &    \makebox[2ex]{\hfill -}                 &    -              \\
  Fake-lepton events                                                         & $\makebox[9ex]{\hfill4} \pm\makebox[8ex]{4\hfill}$    &   $35\pm6$   &    \makebox[2ex]{\hfill -}                 &    $0.6\pm0.3$  \\
 \noalign{\smallskip}\hline\noalign{\smallskip}
\end{tabular*}
\end{center}
\label{tab:VRresults}
\end{table}

\begin{table}
\caption{
Expected and observed event yields in the three validation regions, VR-low, VR-medium and VR-high.
The quoted uncertainties include statistical and systematic contributions.
The individual uncertainties can be correlated and do not necessarily add up in quadrature to the total systematic uncertainty.
}
\begin{center}
\setlength{\tabcolsep}{0.0pc}
\begin{tabular*}{\textwidth}{@{\extracolsep{\fill}}lccc@{}}
\noalign{\smallskip}\hline\noalign{\smallskip}
                                                                             & VR-low            & VR-medium        & VR-high  \\[-0.05cm]
\noalign{\smallskip}\hline\noalign{\smallskip}
Observed events                                                              & $16253$ & $1917$ & $314$ \\
\noalign{\smallskip}\hline\noalign{\smallskip}                                  
Total expected background events                                              & $16500\pm 700$ & $1990\pm 150$ & $340\pm 60$\\
\noalign{\smallskip}\hline\noalign{\smallskip}
Data-driven flavour-symmetry events                                           & $14700\pm 600$ & $1690\pm 120$ & $250\pm 50$ \\
$WZ/ZZ$ events                                                                & $\makebox[6ex]{\hfill 250} \pm \makebox[4ex]{80\hfill}$ & $\makebox[5ex]{\hfill 40}\pm \makebox[4ex]{19\hfill}$ & $\makebox[5ex]{\hfill 9}\pm \makebox[4ex]{6\hfill}$\\
Data-driven \dyjets (\gjets) events                                           & $\makebox[6ex]{\hfill 1100}\pm \makebox[4ex]{400\hfill}$ & $130\pm 70$ & $\makebox[5ex]{\hfill 50}\pm \makebox[4ex]{29\hfill}$\\
Rare top events                                                               & $\makebox[6ex]{\hfill 87} \pm \makebox[4ex]{23\hfill}$ & $27\pm 7$ & $\makebox[5ex]{\hfill 6.5}\pm \makebox[4ex]{1.8\hfill}$\\
Data-driven fake-lepton events                                                & $\makebox[6ex]{\hfill 270}\pm \makebox[4ex]{100\hfill}$ & $\makebox[5ex]{\hfill 98}\pm \makebox[4ex]{35\hfill}$ & $\makebox[5ex]{\hfill 20}\pm \makebox[4ex]{11\hfill}$\\
 \noalign{\smallskip}\hline\noalign{\smallskip}
\end{tabular*}
\end{center}
\label{tab:VRedge}
\end{table}

\FloatBarrier

%% file: sec-syst.tex
The data-driven background estimates are subject to uncertainties associated with the methods employed and the limited 
number of events used in their estimation. 
The dominant uncertainty (10\%) for the flavour-symmetry-based background estimate in SRZ is due to the limited number of events in CR-FS.  
Other systematic uncertainties assigned to this background estimate include those due to 
MC closure (3\%), the measurement of the efficiency correction factors (3\%) and the extrapolation in \mll\ (1\%).
In the case of the edge SRs the statistical uncertainty is also the dominant uncertainty in the flavour-symmetric background estimate in the case of SR-high, 
but for both SR-medium and SR-low the uncertainties from the MC non-closure and efficiency correction factors are comparable in size, or in some cases larger. 
These uncertainties can contribute up to 5\% in SR-low and SR-medium and 10\% in SR-high. 

Several sources of systematic uncertainty are assessed for the \dyjets\ background. 
The boson $\pt$ reweighting procedure is assigned an uncertainty based on a comparison of the nominal results with those obtained by reweighting using
three other kinematic variables, namely \HT, $Z$-boson $E_{\text{T}}$ and jet multiplicity.
For the smearing function, which is measured using MC events in a 1-jet control region, 
an uncertainty is derived by comparing the results obtained using the nominal smearing function with those obtained using a smearing function from a 2-jet sample of MC events, 
and also using a smearing function measured in a 1-jet data sample. 
An uncertainty of between 40--100\% is assigned to account for different reweighting procedures 
and between 20--100\% for the smearing procedure applied to \gjets\ events. 
The smearing uncertainty dominates in SR-high, while the reweighting uncertainty dominates in SR-low and SR-medium, 
with both being around 60\% in SRZ.
The full reweighting and smearing procedure is carried out using \gjets\ MC events such that an MC non-closure uncertainty can be derived 
by comparing the resulting \gjets\ MC \met\ distribution to that in \dyjets\ MC events. 
The resulting uncertainty of up to 35\% is calculated in the VRs, so as to maximise the number of events that contribute.  
An uncertainty of 16\% is assessed for the $V\gamma$ backgrounds, based on data-to-MC agreement in a $V\gamma$-enriched control region. 
This uncertainty is propagated to the final \dyjets\ estimate following the subtraction of the $V\gamma$ background. 
In VR-low, a correction is applied to the \met\ distribution in \gjets\ events to account for the fraction of \dyjets\ 
events in this \HT-inclusive region expected to have boson \pt\ less than 37~\GeV. 
The full size of this correction (up to 50\% for $\met=150$~\GeV) is applied as a systematic uncertainty.
The \mll\ distribution assigned to \gjets\ MC events is compared to that of \dyjets\ MC events, 
and the relative difference in a given \mll\ bin is assigned as an uncertainty. 
Finally, the statistical precision of the estimate also enters as a systematic uncertainty of $\sim 10\%$ in the final background estimate. 
After applying the correction procedure, differences in the number of $b$-tagged jets between \dyjets\ and \gjets\ are found to be negligible, indicating good agreement in heavy-flavour content.

The uncertainties in the fake-lepton background stem from the number of events in the regions used to measure the real- and fake-lepton efficiencies, 
the limited size of the inclusive loose-lepton sample, 
and from varying the region used to measure the fake-lepton efficiency.  
The nominal fake-lepton efficiency is compared with those measured in a region with $b$-tagged jets and a region with a $b$-jet, 
as well as a region with the prompt-lepton subtraction varied by $20\%$.  
Varying the sample composition via $b$-jet tagging gives the largest uncertainty.  
The uncertainty for the edge SRs from the statistical component of the lepton efficiencies is $30$--$45$\%, 
and from varying the region for the fake-lepton efficiency it is $50$--$75$\%.  
The uncertainties in SRZ are generally larger due to the small number of events contributing to the estimate in this region.  

Theoretical and experimental uncertainties are taken into account for the signal models, as well as background processes that rely on MC simulation.
The estimated uncertainty in the luminosity measurement is $2.9\%$~\cite{2011lumi,2012lumi}.
The jet energy scale is subject to uncertainties associated with the jet flavour composition, 
the pile-up and the jet and event kinematics~\cite{ATL-PHYS-PUB-2015-015}. 
Uncertainties in the jet energy resolution are included to account for differences between data and MC simulation~\cite{ATL-PHYS-PUB-2015-015}.
An uncertainty in the \met\ soft-term resolution and scale is taken into account~\cite{ATL-PHYS-PUB-2015-023}, 
and uncertainties due to the lepton energy scales and resolutions, as well as trigger, reconstruction, and identification efficiencies, are also considered.

The $WZ/ZZ$ processes are assigned a cross-section uncertainty of $6\%$ and an additional uncertainty based on comparisons between {\sc Sherpa} and {\sc Powheg} MC samples, 
which is up to $50\%$ in the SRs. 
Uncertainties due to the choice of factorisation and renormalisation scales are calculated by varying the nominal values up and down by a factor of two and can be up to $23\%$.
For rare top processes, 
a $13\%$ PDF and scale variation uncertainty is applied~\cite{Alwall:2014hca} in addition to a $22\%$ cross-section uncertainty~\cite{Campbell:2012,Lazopoulos:2008,Garzelli:2012bn}.

For signal models, 
the nominal cross section and the uncertainty are taken from an envelope of cross-section predictions using different PDF sets and factorisation and renormalisation scales, 
as described in Refs.~\cite{Kramer:2012bx,Borschensky:2014cia}.
These are calculated at next-to-leading-logarithm accuracy (NLO+NLL)~\cite{Beenakker:1996ch,Kulesza:2008jb,Kulesza:2009kq,Beenakker:2009ha,Beenakker:2011fu}, 
and the resulting uncertainties range from  $16$\% to $30$\%.

A breakdown of the dominant uncertainties in the background prediction in the SRs is provided in Table~\ref{tab:syst}
for the on-shell $Z$ and edge searches.
Here these uncertainties are quoted relative to the total background. 
In the case of the edge regions a range is quoted, 
taking into account the relative contribution of the given uncertainty in each of the \mll\ ranges in SR-low, SR-medium and SR-high. 
The largest uncertainties in the signal regions are due to the size of the $e\mu$ data sample in CR-FS, used to provide the flavour-symmetric background estimate, 
the combined systematic uncertainty in the same background, the systematic uncertainty in \gjets, or, in the case of SRZ, the $WZ/ZZ$ generator uncertainty.  
The statistical component of the uncertainty from the flavour-symmetry estimate is largest for the edge analysis in SR-medium and SR-high in the highest \mll\ regions.
In the edge SRs the uncertainty in the $WZ/ZZ$ background tends to be highest in the \mll\ ranges that include the $Z$ window. 
The uncertainty in the fake-lepton background is largest in SR-high, where fake leptons can compose a larger fraction of the background. 
Experimental uncertainties have a far lower impact on the systematic uncertainty of the total background ($<2\%$).

\begin{table}[h]
\centering
\caption{
Overview of the dominant sources of systematic uncertainty in the total background estimate in the signal regions.
The values shown are relative to the total background estimate, shown in \%.
The systematic uncertainties for the edge search are quoted as a range across the \mll\ regions used for statistical interpretations.
}
\small
\begin{tabular}{lrccc}
\noalign{\smallskip}\hline\noalign{\smallskip}
Source  & Relative systematic uncertainty [\%] \\
\noalign{\smallskip}\hline\noalign{\smallskip}
 & SRZ & SR-low & SR-medium & SR-high\\
\noalign{\smallskip}\hline\noalign{\smallskip}
Total systematic uncertainty & 17    & 8--30                       & 6--34 & 10--45 \\ 
\noalign{\smallskip}\hline
$WZ/ZZ$ generator uncertainty  & 13  & 0--\makebox[2ex]{7\hfill}   & 0--\makebox[2ex]{6\hfill}   & \makebox[2ex]{\hfill 0}--10 \\
Flavour symmetry (statistical) & 7   & 3--16                       & 5--16                       & \makebox[2ex]{\hfill 7}--28           \\
$WZ/ZZ$ scale uncertainty      & 6   & 0--\makebox[2ex]{1\hfill}   & 0--\makebox[2ex]{1\hfill}   & 0--2 \\
\dyjets\ (systematic)          & 4   & 0--15                       & 0--25                       & \makebox[2ex]{\hfill 0}--15 \\
Flavour symmetry (systematic)  & 3   & 2--23                       & 2--15                       & \makebox[2ex]{\hfill 4}--25    \\
\dyjets\ (statistical)         & 2   & 0--\makebox[2ex]{3\hfill}   & 0--\makebox[2ex]{5\hfill}   & 0--1\\
Fake leptons                   & 1   & 0--17                       & 2--18                       & \makebox[2ex]{\hfill 2}--20 \\
\noalign{\smallskip}\hline
\end{tabular}
\label{tab:syst}
\end{table}

%% file: sec-results.tex
\subsection{Results in SRZ}

\begin{figure*}[!htbp]
\centering
\includegraphics[width=0.9\textwidth]{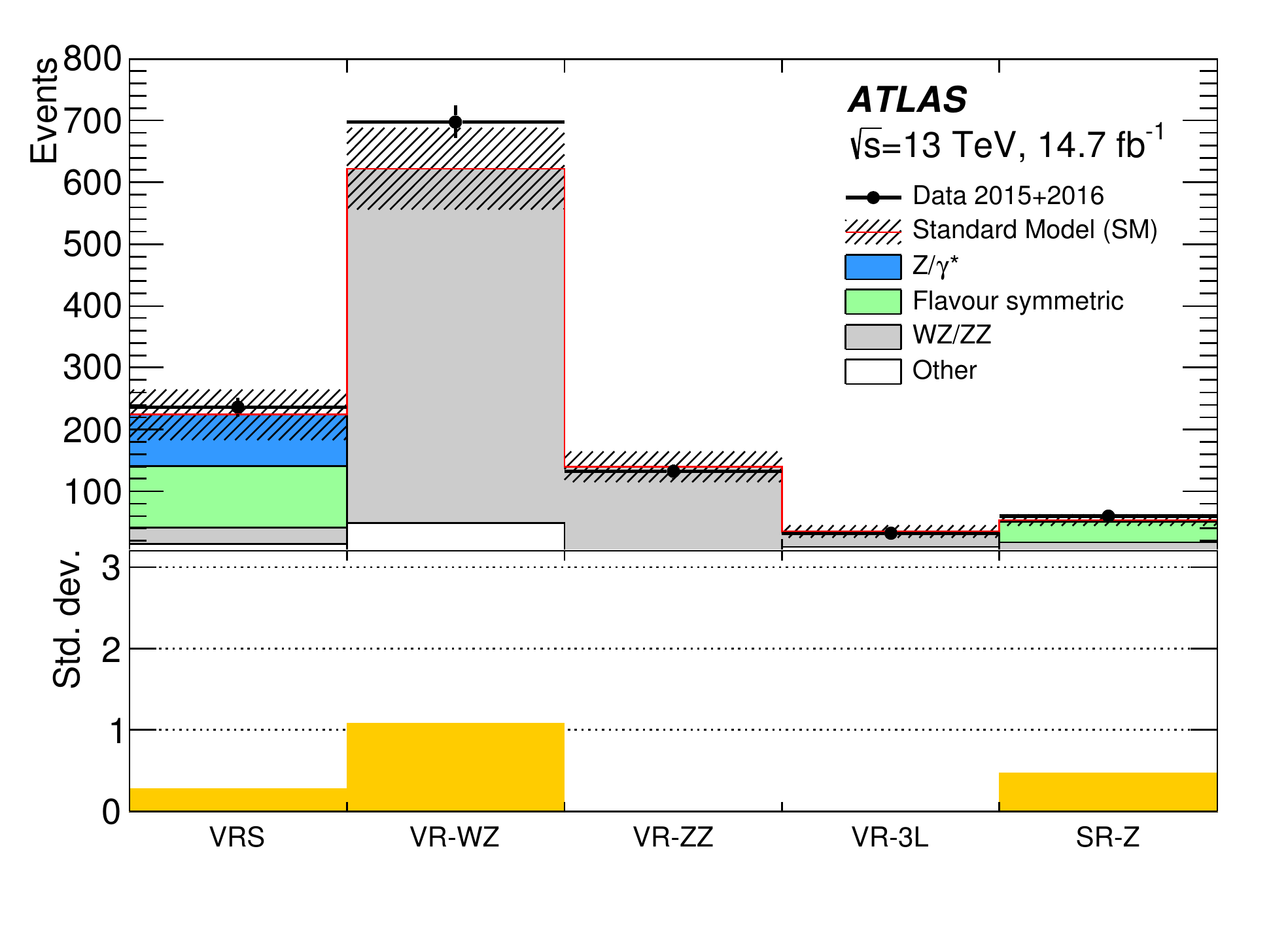}
\caption{
The expected and observed yields in the validation regions and signal region of the on-shell $Z$ search. 
The rare top and data-driven fake-lepton backgrounds are grouped under ``other'' backgrounds.
The significance of the difference between the data and the expected background (see text for details) is shown in the bottom plot; 
for regions in which the data yield is less than expected, the significance is set to zero. 
The hashed uncertainty bands include the statistical and systematic uncertainties in the background prediction.
}
\label{fig:summary}
\end{figure*}

For the on-shell $Z$ search, the expected background and observed yields in the SR are shown in Table~\ref{tab:results}.
A total of 60 events are observed in data with a predicted background of $53.5 \pm 9.3$ events.
There are $35$ events observed in data in the $ee$ channel, and $25$ events observed in the $\mu\mu$ channel.
The probability for the background to produce a fluctuation greater than or equal to that observed in the data, 
called the significance when expressed in terms of the number of standard deviations, corresponds to $0.47\sigma$
(details of the significance calculation are presented in Section~\ref{sec:interpretation}). 
The level of agreement between the observed event yields in data and the background predictions in the VRs, shown previously in Table~\ref{tab:VRresults}, 
is also displayed in Figure~\ref{fig:summary}, along with the results in SRZ.

\begin{table}
\caption{
Expected and observed event yields in SRZ, inclusively, in the $ee$ channel, and in the $\mu\mu$ channel, along with the discovery $p$-value for zero signal strength ($p(s=0)$)~\cite{Baak:2014wma}, Gaussian significance, 95\% confidence level (CL) observed and expected upper limits on the number of signal events ($S^{95}$), and the corresponding observed upper limit on the visible cross section ($\langle\epsilon\sigma\rangle^{95}_\text{obs}$).
For regions in which the data yield is less than expected, the discovery $p$-value is truncated at 0.5 and the significance is set to zero.
The flavour-symmetric, \dyjets\ and fake-lepton components are all derived using data-driven estimates described in Section~\ref{sec:bg}.
All remaining backgrounds are taken from MC simulation.
The quoted uncertainties include statistical and systematic contributions.
The individual uncertainties can be correlated and do not necessarily add up in quadrature to the total systematic uncertainty.
}
\begin{center}
\setlength{\tabcolsep}{0.0pc}

\begin{tabular*}{\textwidth}{@{\extracolsep{\fill}}lccc}
\noalign{\smallskip}\hline\noalign{\smallskip}
                                                                            & SRZ                 & SRZ $ee$      & SRZ $\mu\mu$  \\[-0.05cm]
\noalign{\smallskip}\hline\noalign{\smallskip}

Observed events                                                             & $60$                     & $35$                   &   $25$ \\
\noalign{\smallskip}\hline\noalign{\smallskip}

Total expected background events                                            & $53.5 \pm 9.3$         & $27.1 \pm 5.1$       &   $26.8 \pm 4.4$ \\
\noalign{\smallskip}\hline\noalign{\smallskip}

  Flavour-symmetric (\ttbar, $Wt$, $WW$ and $Z\rightarrow\tau\tau$) events  & $33.2 \pm 3.9$         & $16.5 \pm 2.1$       &   $16.7 \pm 2.0$       \\

  \dyjets\ events                                                           & $\makebox[3ex]{\hfill 3.1} \pm \makebox[2ex]{2.8\hfill}$          & $1.0_{-1.0}^{+1.3}$ &   $\makebox[3ex]{\hfill 2.1} \pm \makebox[2ex]{1.4\hfill}$     \\

  $WZ/ZZ$ events                                                            & $14.2 \pm 7.7$         & $\makebox[3ex]{\hfill 7.8} \pm \makebox[2ex]{4.3\hfill}$        &   $\makebox[3ex]{\hfill 6.4} \pm \makebox[2ex]{3.5\hfill}$        \\

  Rare top events                                                           & $\makebox[3ex]{\hfill 2.9} \pm \makebox[2ex]{0.8\hfill}$          & $\makebox[3ex]{\hfill 1.4} \pm \makebox[2ex]{0.4\hfill}$        &   $\makebox[3ex]{\hfill 1.5} \pm \makebox[2ex]{0.4\hfill}$  \\

 Fake-lepton events                                                         & $\makebox[3ex]{\hfill 0.1}_{-0.1}^{+0.8}$   & $0.5_{-0.5}^{+0.7}$ &   $\makebox[3ex]{\hfill 0}^{+0.2}$    \\
 \noalign{\smallskip}\hline\noalign{\smallskip}
$p(s=0)$                                                                    & 0.32      & 0.15        & 0.5        \\
Significance $(\sigma)$                                                               & 0.47                     & 1.02                   & 0           \\
Observed (Expected) $S^{95}$                                                & 28.2 ($24.5_{-6.7}^{+8.9}$) & 22.0 ($15.8_{-4.5}^{+6.5}$) & 12.9 ($14.0_{-3.9}^{+5.7}$) \\
$\langle\epsilon\sigma\rangle^{95}_\text{obs}$~[fb]                         & 1.9             & 1.5                    & 0.88   \\
 \noalign{\smallskip}\hline\noalign{\smallskip}
\end{tabular*}
\end{center}
\label{tab:results}
\end{table}

\begin{figure*}[!htbp]
\centering
\includegraphics[width=0.47\textwidth]{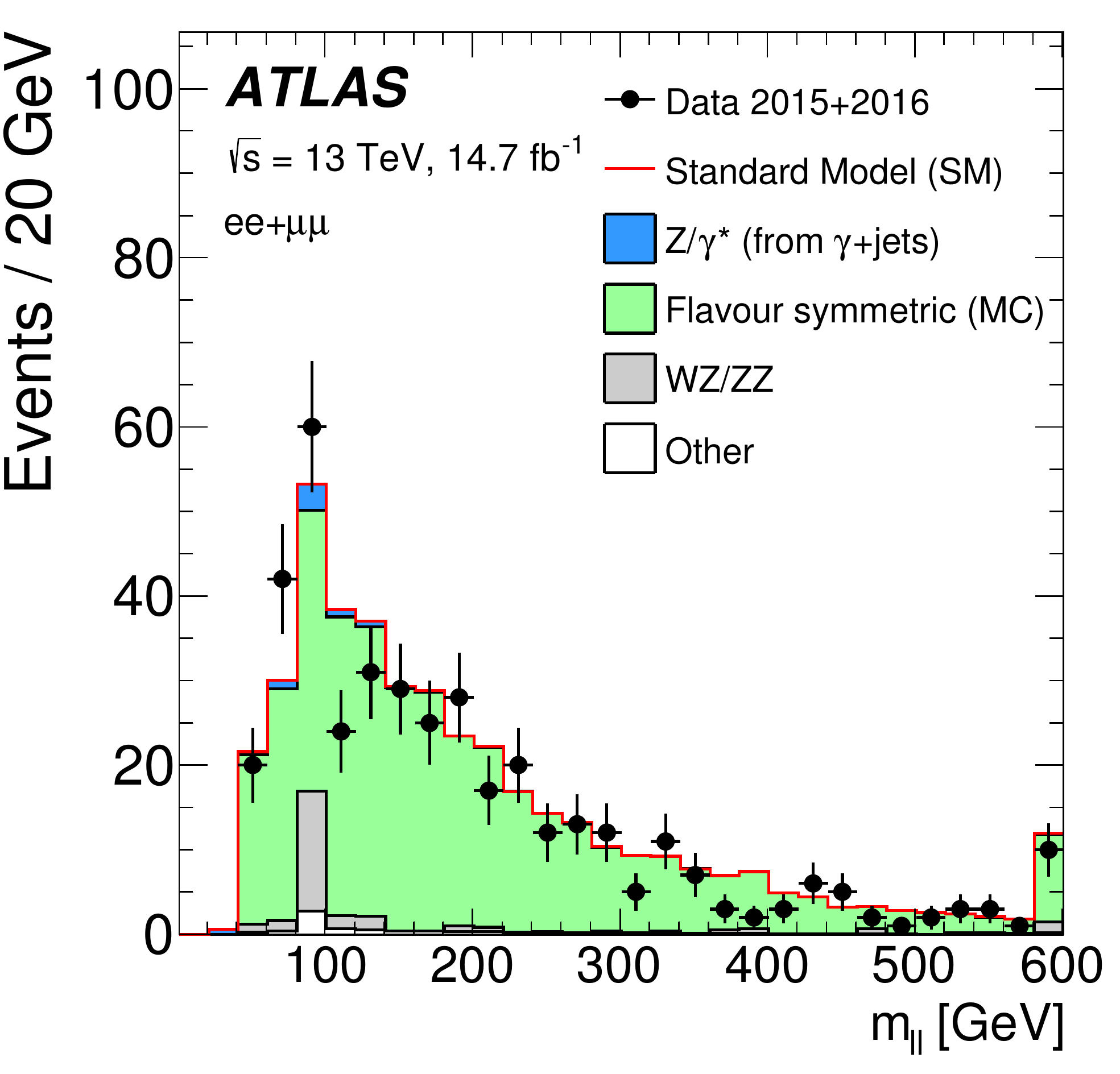}
\includegraphics[width=0.47\textwidth]{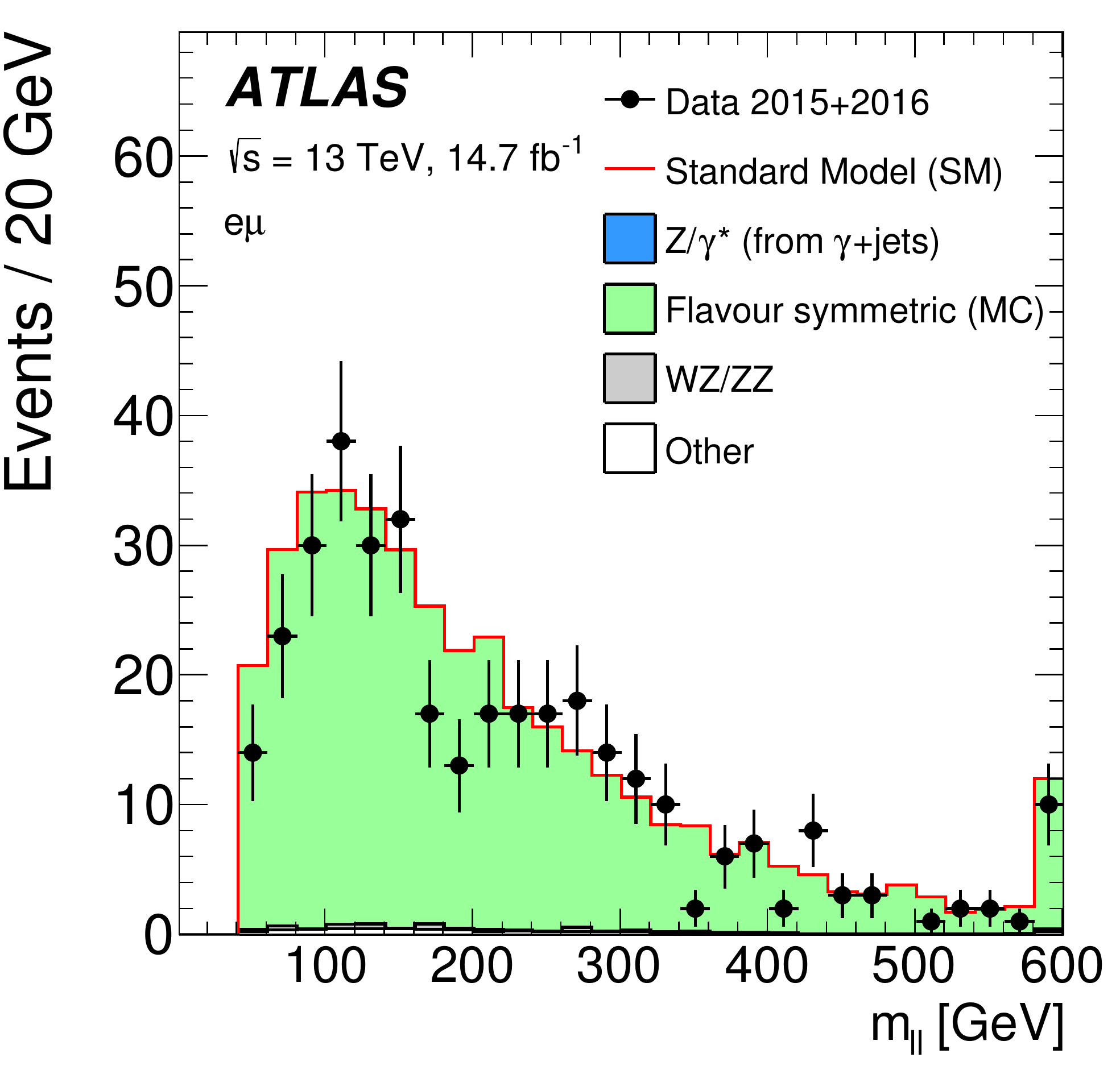}
\caption{
The dilepton invariant-mass distribution for an SRZ-like selection, but with the $Z$ mass requirement removed, in the same-flavour (left) and different-flavour (right) channels.  
With the exception of the \dyjets\ background, MC simulation is used to show the expected shapes of the \mll\ distributions, with the backgrounds being normalised according to their SRZ prediction. 
For the \dyjets\ background, the \mll\ shape is taken from the \gjets\ method.
The rare top and data-driven fake-lepton backgrounds are grouped under ``other'' backgrounds.
The last bin includes the overflow.
}
\label{fig:mll_n-1}
\end{figure*}

\begin{figure*}[!htbp]
\centering
\includegraphics[width=0.44\textwidth]{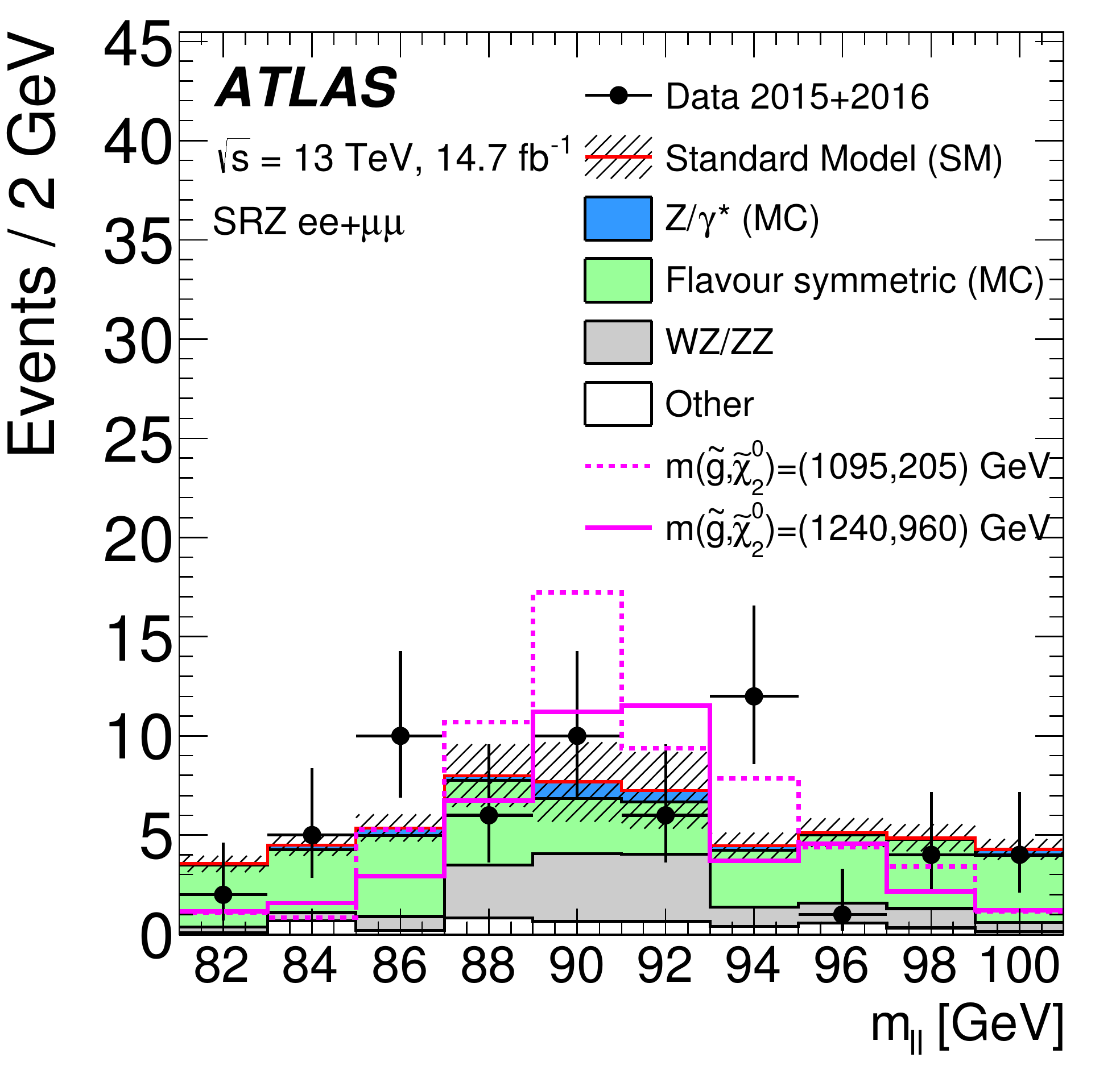}
\includegraphics[width=0.44\textwidth]{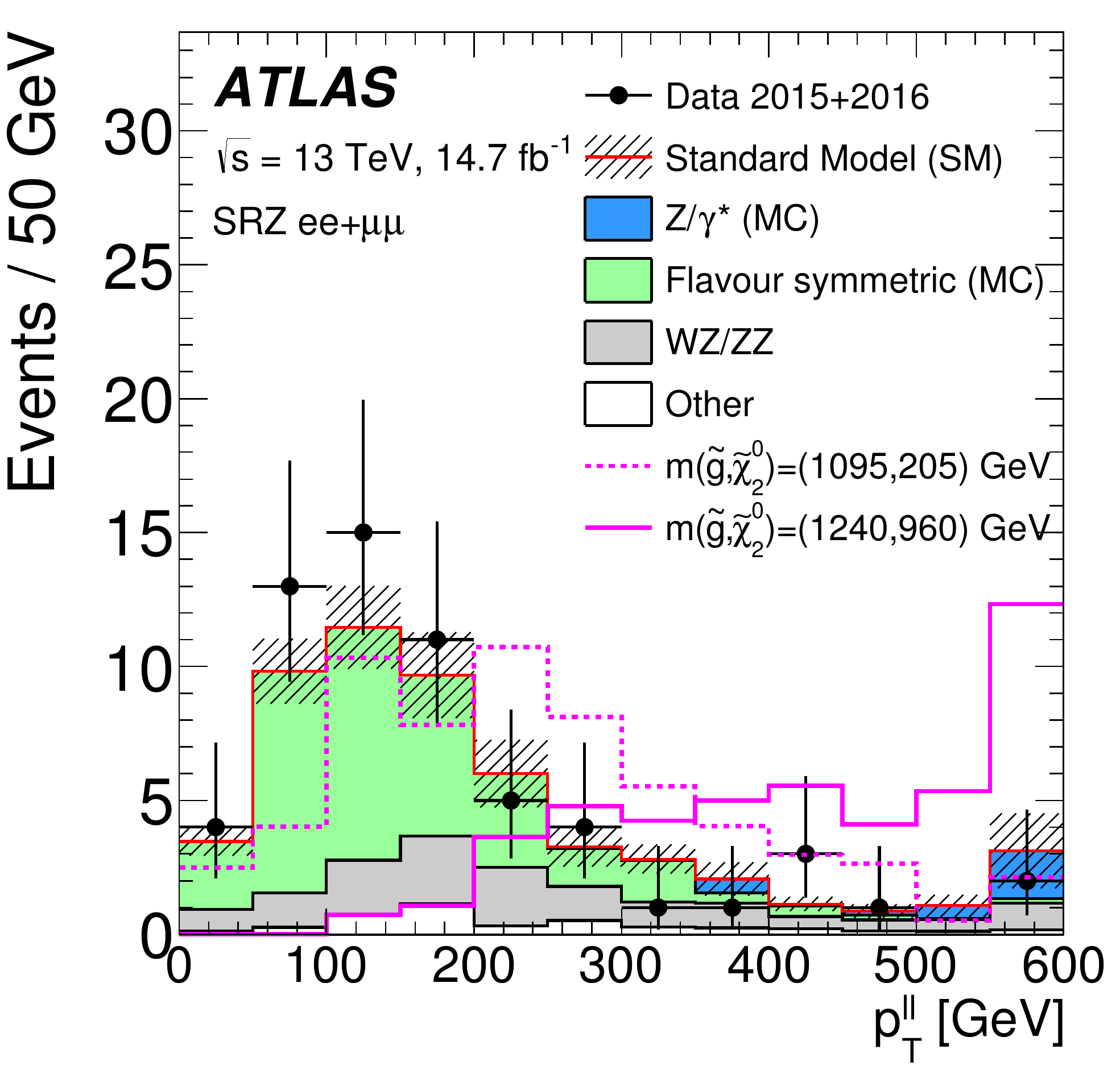}
\includegraphics[width=0.44\textwidth]{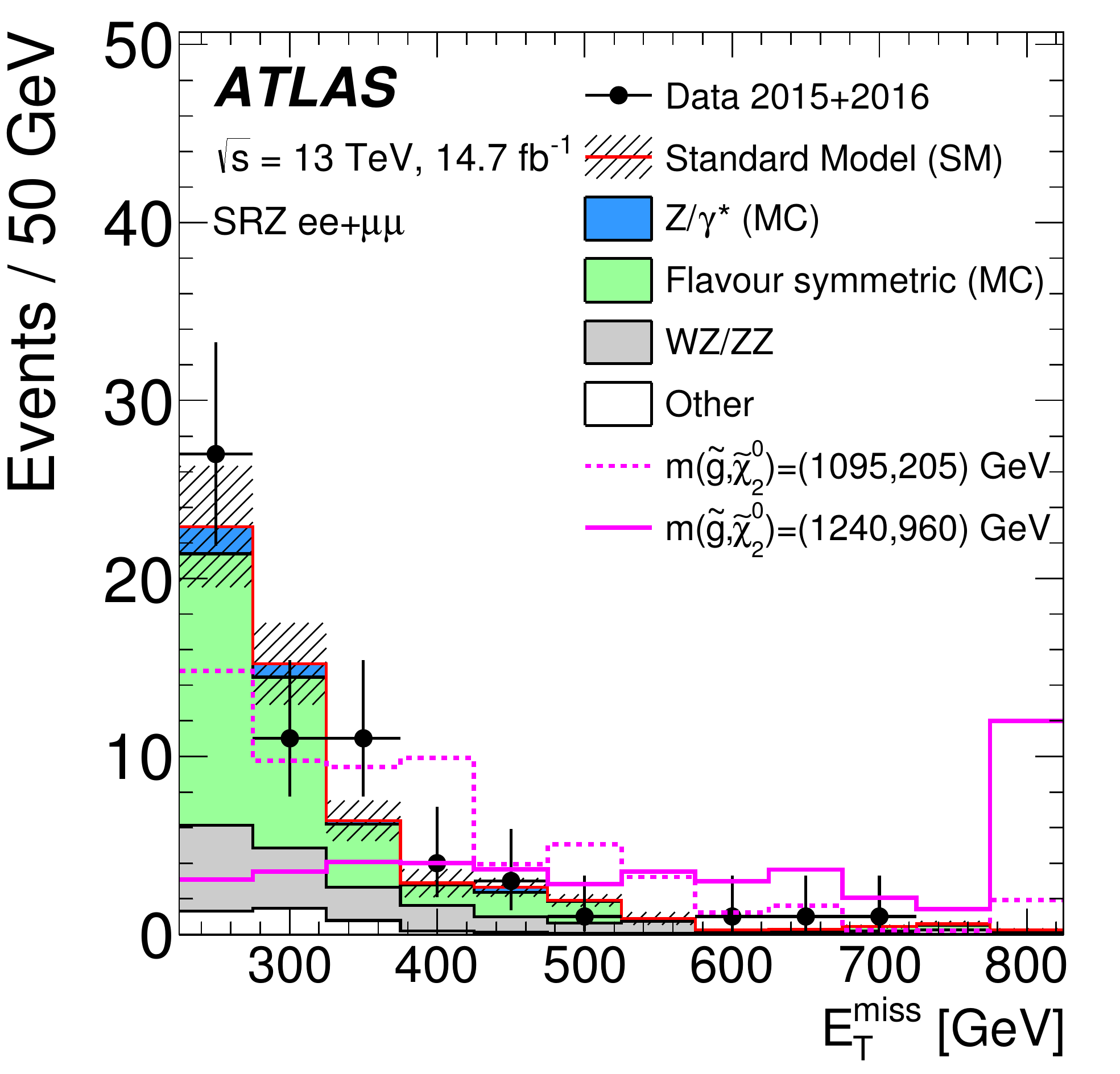}
\includegraphics[width=0.44\textwidth]{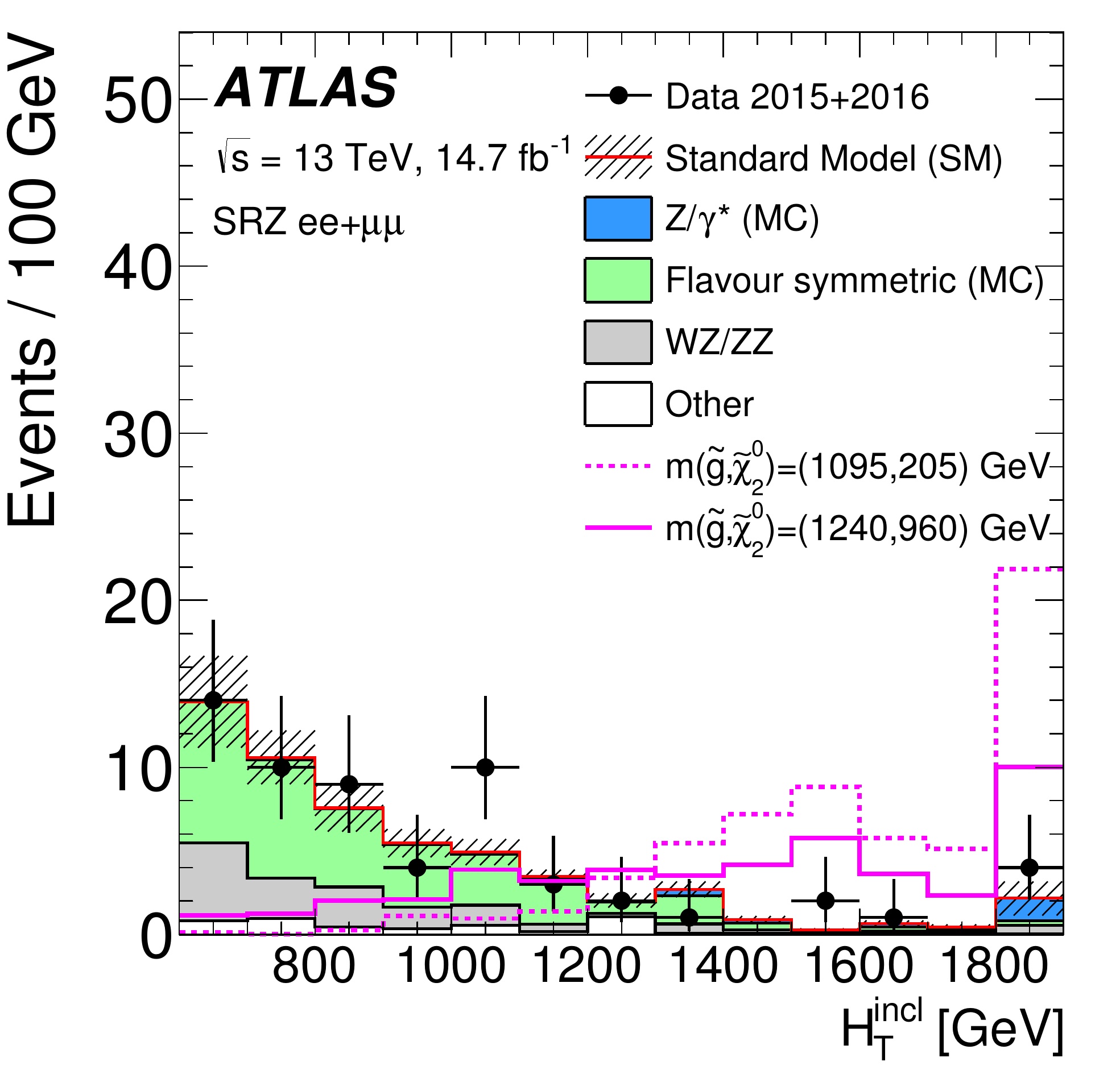}
\includegraphics[width=0.44\textwidth]{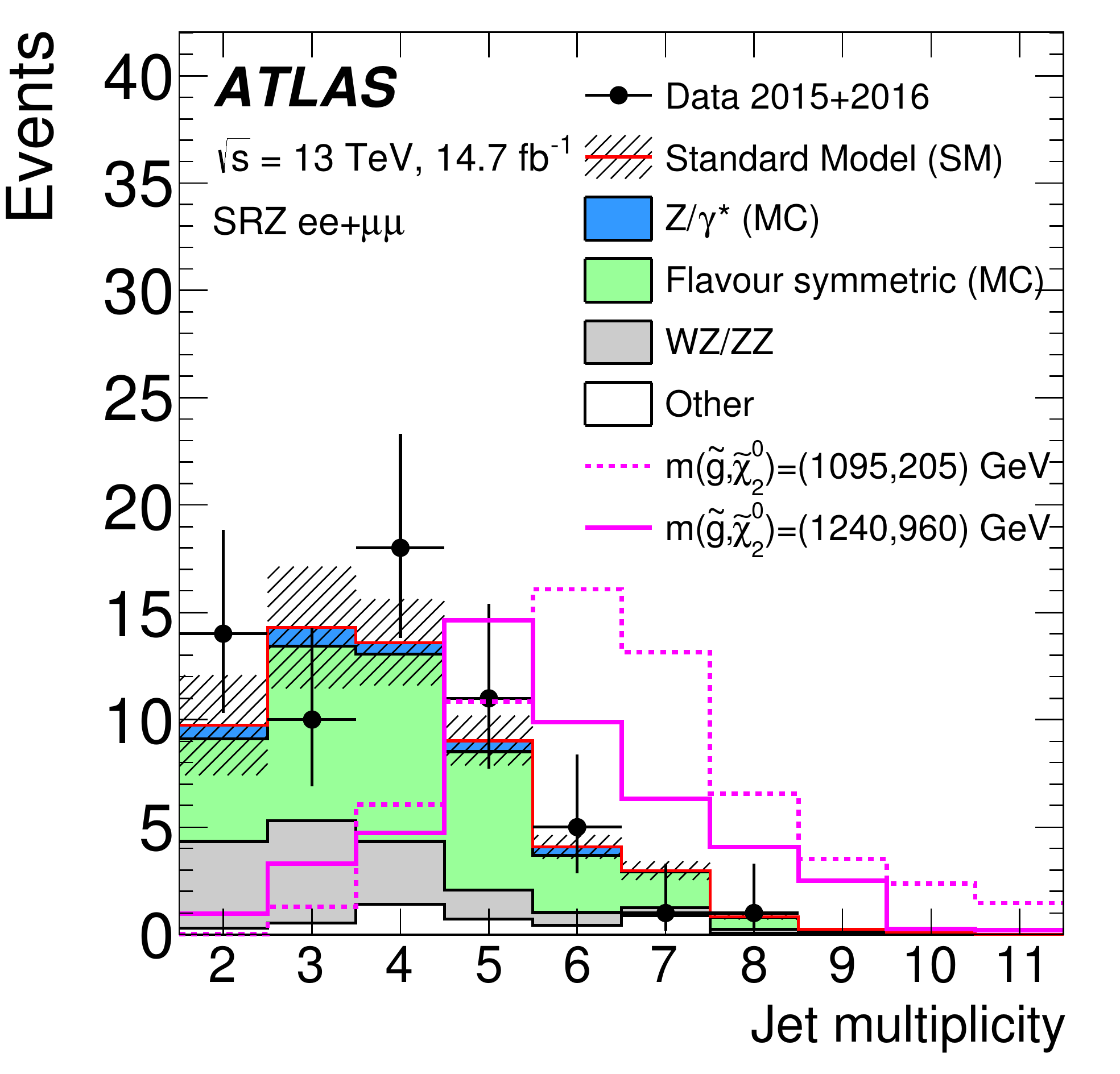}
\includegraphics[width=0.44\textwidth]{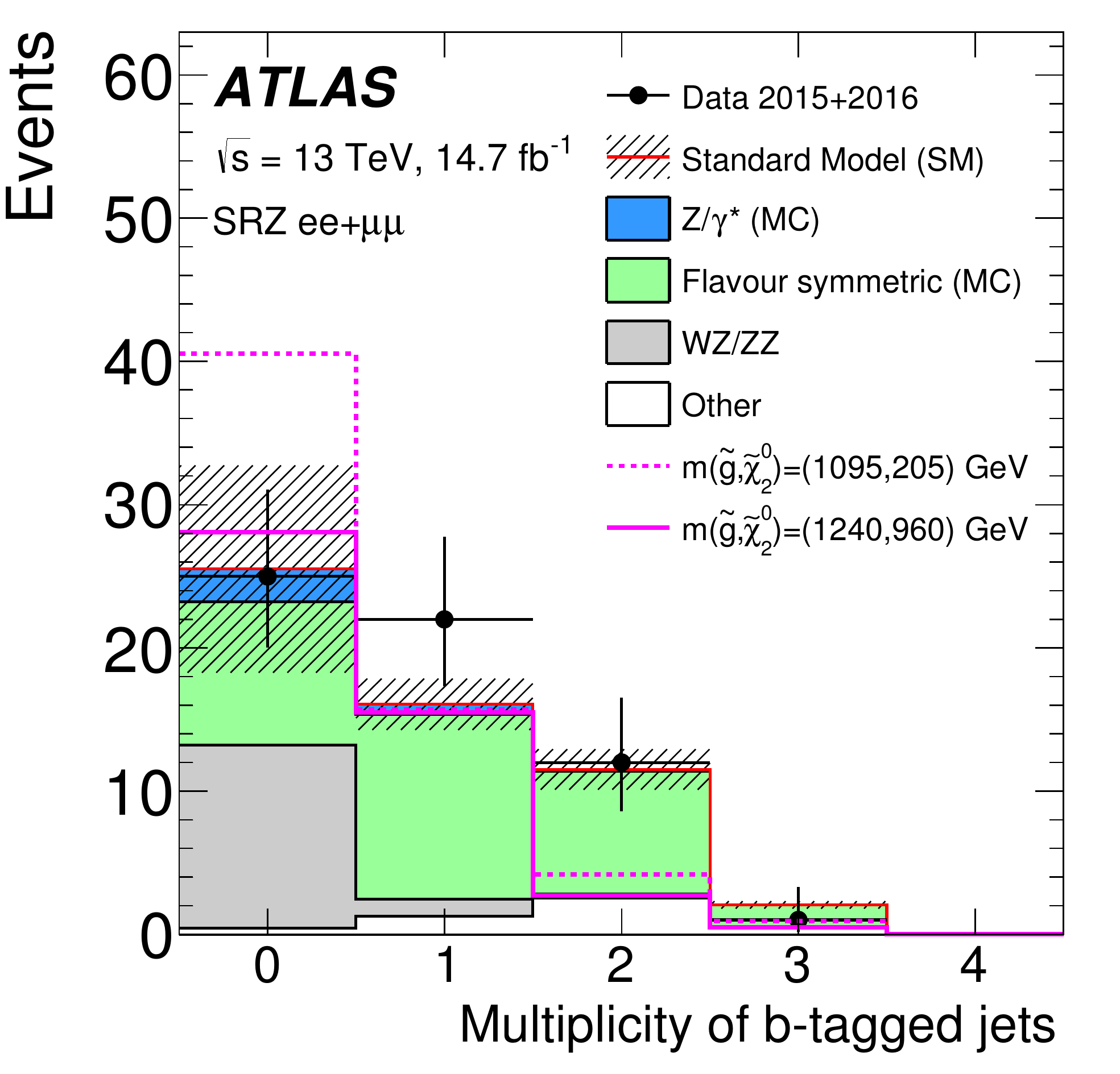}
\caption{
The \mll\ (top left), $\pt^{\ell\ell}$ (top right), \met\ (middle left), \htincl\ (middle right), 
jet multiplicity (bottom left) and $b$-tagged jet multiplicity (bottom right) distributions in SRZ. 
Two examples of signal models from the $\tilde{g}$--$\chitwozero$ on-shell grid, described in Section~\ref{sec:mc}, with $(m(\tilde{g}),m(\tilde{\chi}^{0}_{2}))=(1095, 205)$~\GeV\ 
and $(m(\tilde{g}),m(\tilde{\chi}^{0}_{2}))=(1240, 960)$~\GeV, are overlaid. 
In the case of the \met, \htincl\ and $\pt^{\ell\ell}$ distributions, the last bin contains the overflow. 
The flavour-symmetric and \dyjets\ backgrounds are taken from MC simulation and scaled to match their SRZ data-driven predictions.
The rare top and data-driven fake-lepton backgrounds are grouped under ``other'' backgrounds.
The hashed uncertainty bands include the statistical and systematic uncertainties in the background prediction.
}
\label{fig:extras}
\end{figure*}

\begin{figure*}[!htbp]
\centering
\includegraphics[width=0.44\textwidth]{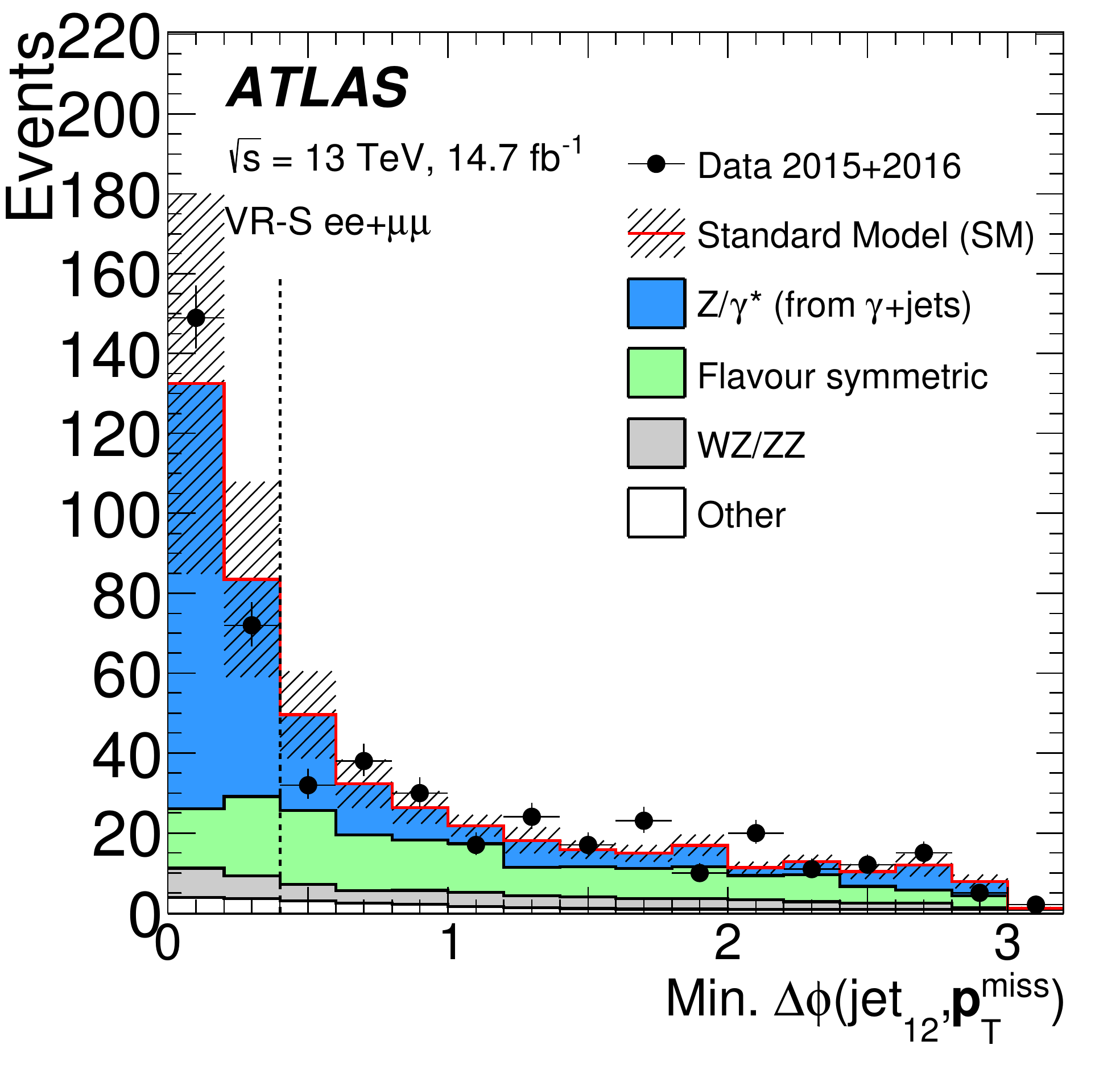}
\includegraphics[width=0.44\textwidth]{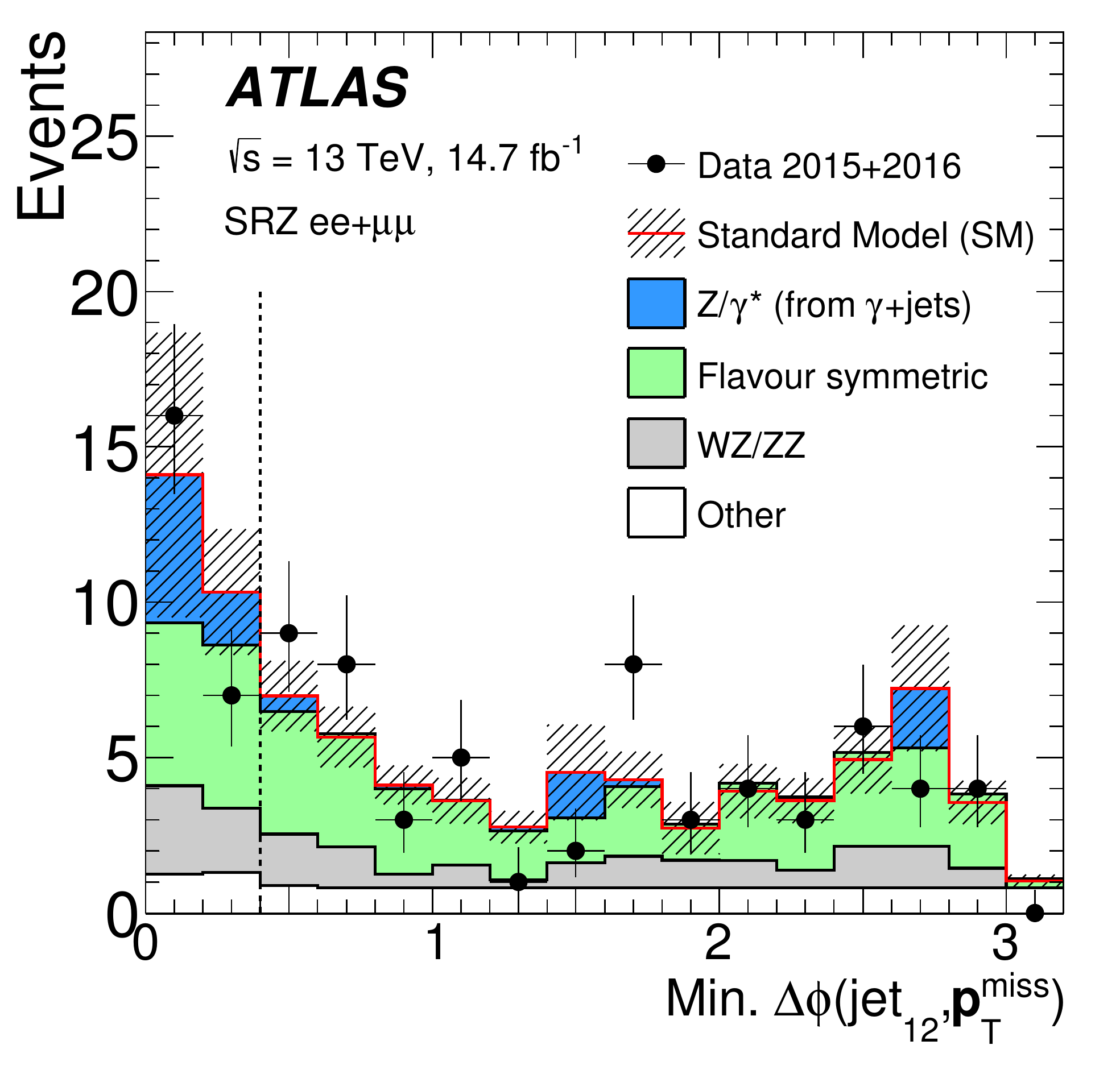}
\caption{
The min. $\Delta\phi(\text{jet}_{12},{\boldsymbol p}_{\mathrm{T}}^\mathrm{miss})$ distribution in (left) VR-S and (right) SRZ, 
where the min. $\Delta\phi(\text{jet}_{12},{\boldsymbol p}_{\mathrm{T}}^\mathrm{miss})>0.4$ requirement has been lifted.
The vertical dashed lines indicate the requirement in each region.
The flavour-symmetric and \dyjets\ distributions are taken completely from the data-driven estimate.
The rare top and data-driven fake-lepton backgrounds are grouped under ``other'' backgrounds.
The hashed uncertainty bands include the statistical and systematic uncertainties in the background prediction.
}
\label{fig:dphi}
\end{figure*}

The dilepton invariant-mass distribution for the $ee+\mu\mu$ and $e\mu$ channels with the kinematic requirements of SRZ, 
but over the full \mll\ range, is shown in Figure~\ref{fig:mll_n-1}.
Here the data are consistent with the expected background over the full \mll\ range.
The dilepton invariant-mass, jet and $b$-tagged jet multiplicity, \met, \htincl\ and $\pt^{\ell\ell}$ distributions in SRZ are shown in Figure~\ref{fig:extras}.
The shapes of the background distributions in these figures are obtained from MC simulation, 
where the MC simulation is normalised according to the data-driven estimates in the SR.
Here two representative examples of $\tilde{g}$--$\chitwozero$ on-shell signal models, with $(m(\tilde{g}),m(\tilde{\chi}^{0}_{2}))=(1095, 205)$~\GeV\
and $(m(\tilde{g}),m(\tilde{\chi}^{0}_{2}))=(1240, 960)$~\GeV, are overlaid. 
To demonstrate the modelling of the \dyjets\ background in VR-S and SRZ, 
Figure~\ref{fig:dphi} shows the minimum $\Delta\phi(\text{jet}_{12},{\boldsymbol p}_{\mathrm{T}}^\mathrm{miss})$ distribution over the full range, 
where $\Delta\phi(\text{jet}_{12},{\boldsymbol p}_{\mathrm{T}}^\mathrm{miss})>0.4$ is required in VR-S and SRZ. 
Here the \dyjets\ distribution is modelled using the full data-driven prediction from \gjets.
Two of the events in the SR contain a third signal lepton. 

\clearpage

\subsection{Results in the edge SRs}

\begin{figure}[h]
\centering
\includegraphics[width=.45\textwidth]{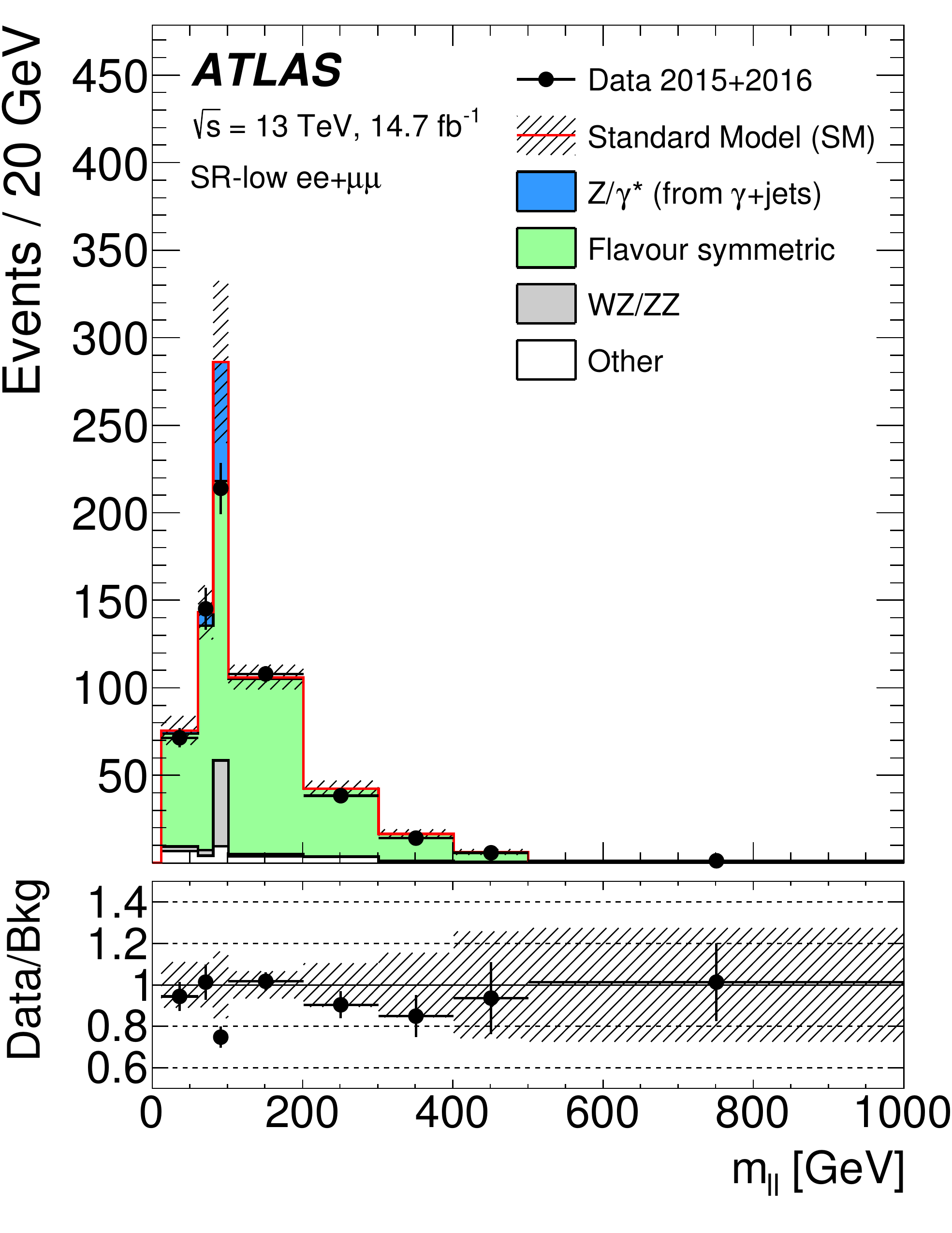}
\includegraphics[width=.45\textwidth]{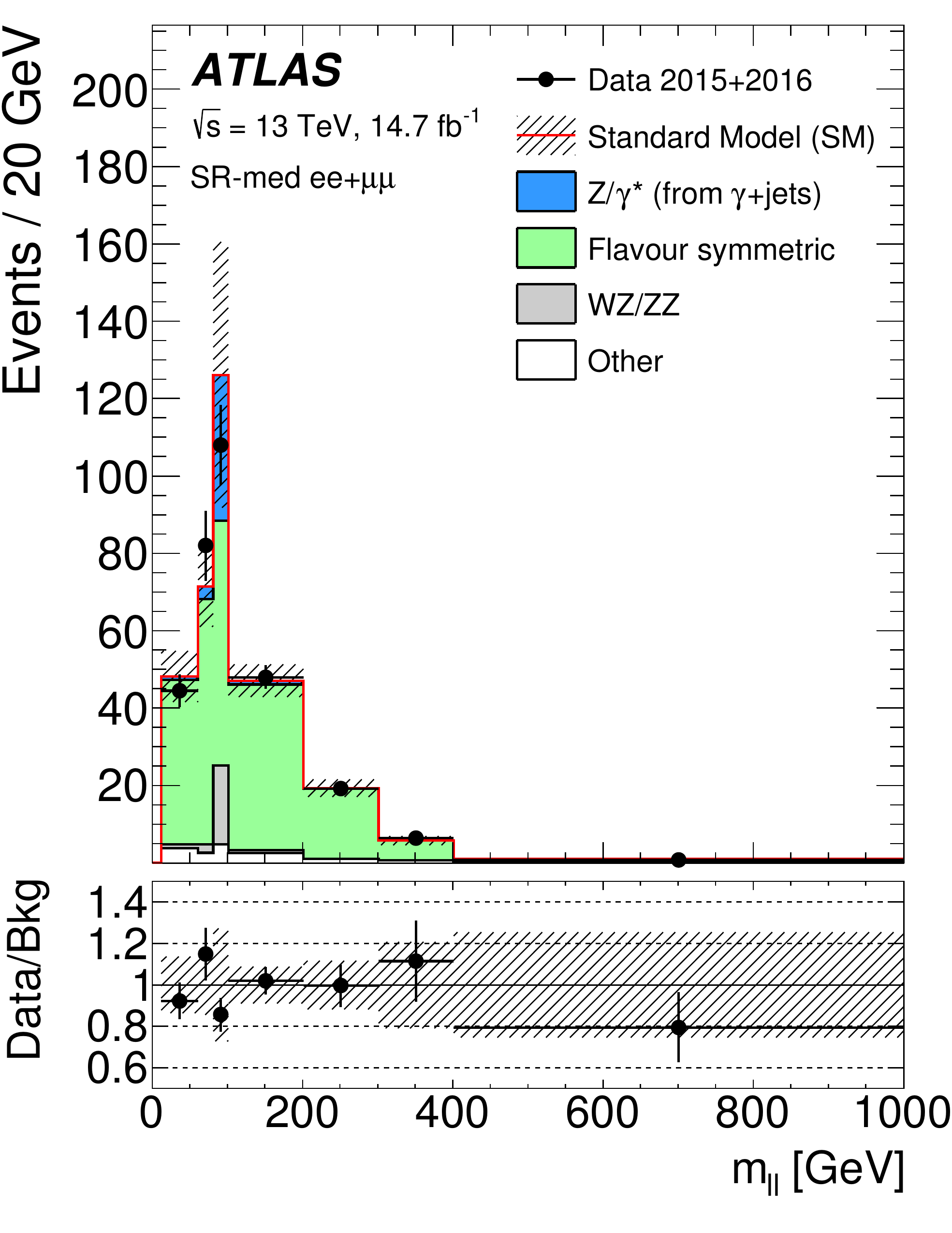} \\
\includegraphics[width=.45\textwidth]{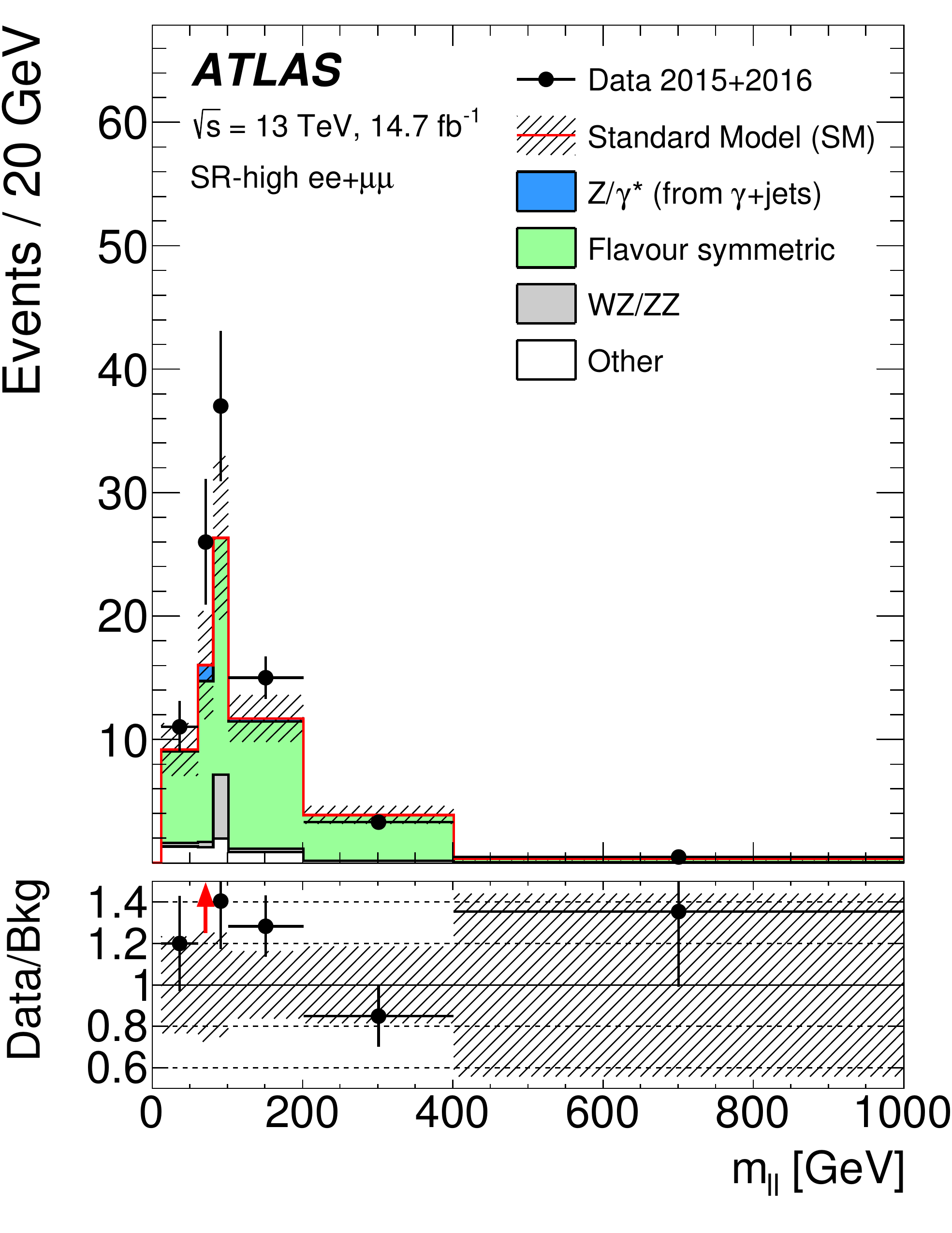}
\caption{ Expected and observed dilepton mass distributions, with the bin boundaries considered for the interpretation, in (top left) SR-low, (top-right) SR-medium, and (bottom) SR-high of the edge search.
These bins, and sets of neighbouring bins, make up the mll windows used for the interpretation.
The flavour-symmetric and \dyjets\ distributions are taken completely from the data-driven estimate.
The rare top and data-driven fake-lepton backgrounds are grouped under ``other'' backgrounds.
All statistical and systematic uncertainties are included in the hashed bands. 
The ratio of data to predicted background is shown in the bottom panels.
In cases where the data point is not accommodated by the scale of this panel,
a red arrow indicates the direction in which the point is out of range.
}
\label{fig:edgemll}
\end{figure}

\begin{figure}[h]
\centering
\includegraphics[width=.44\textwidth]{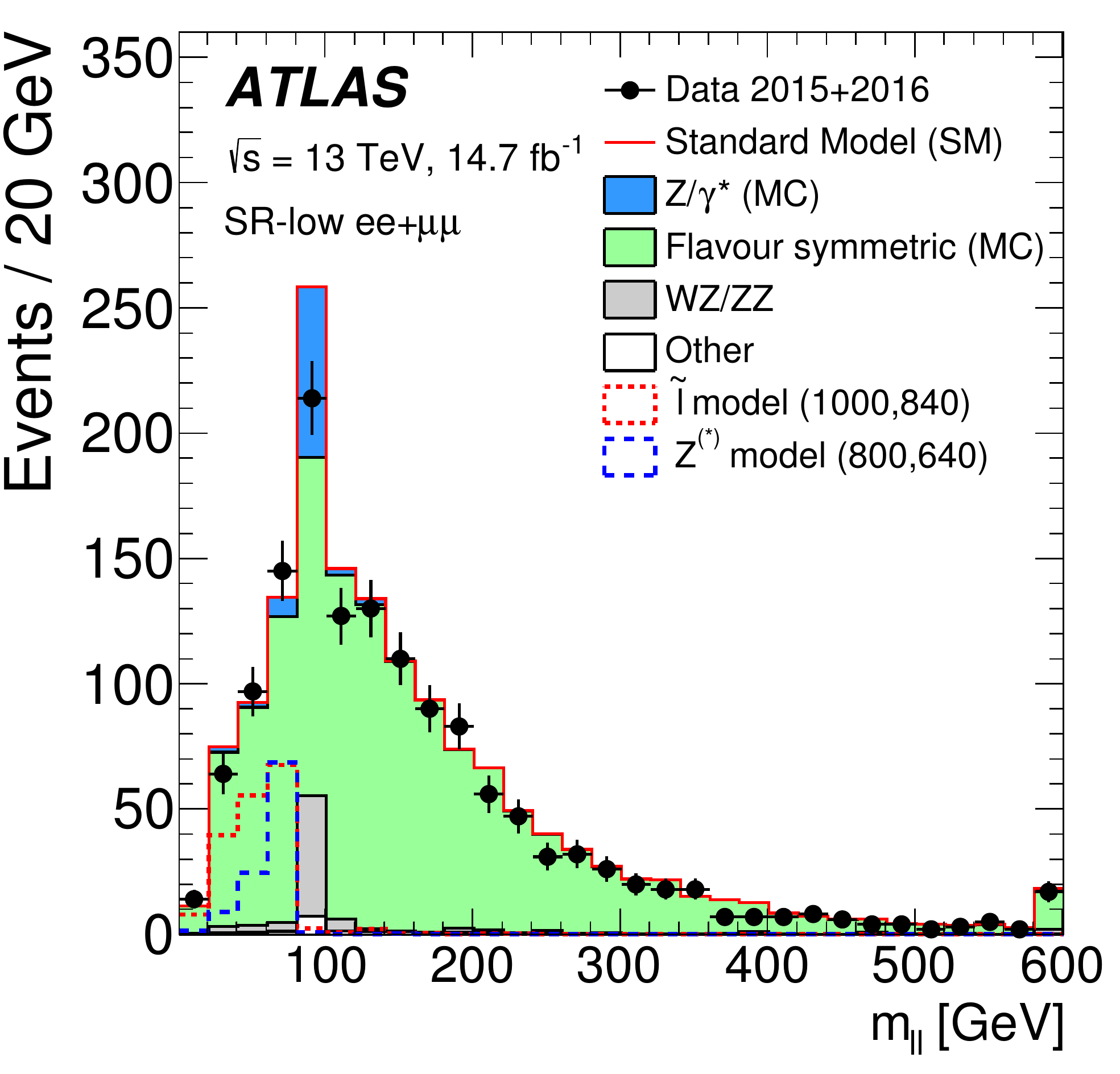} 
\includegraphics[width=.44\textwidth]{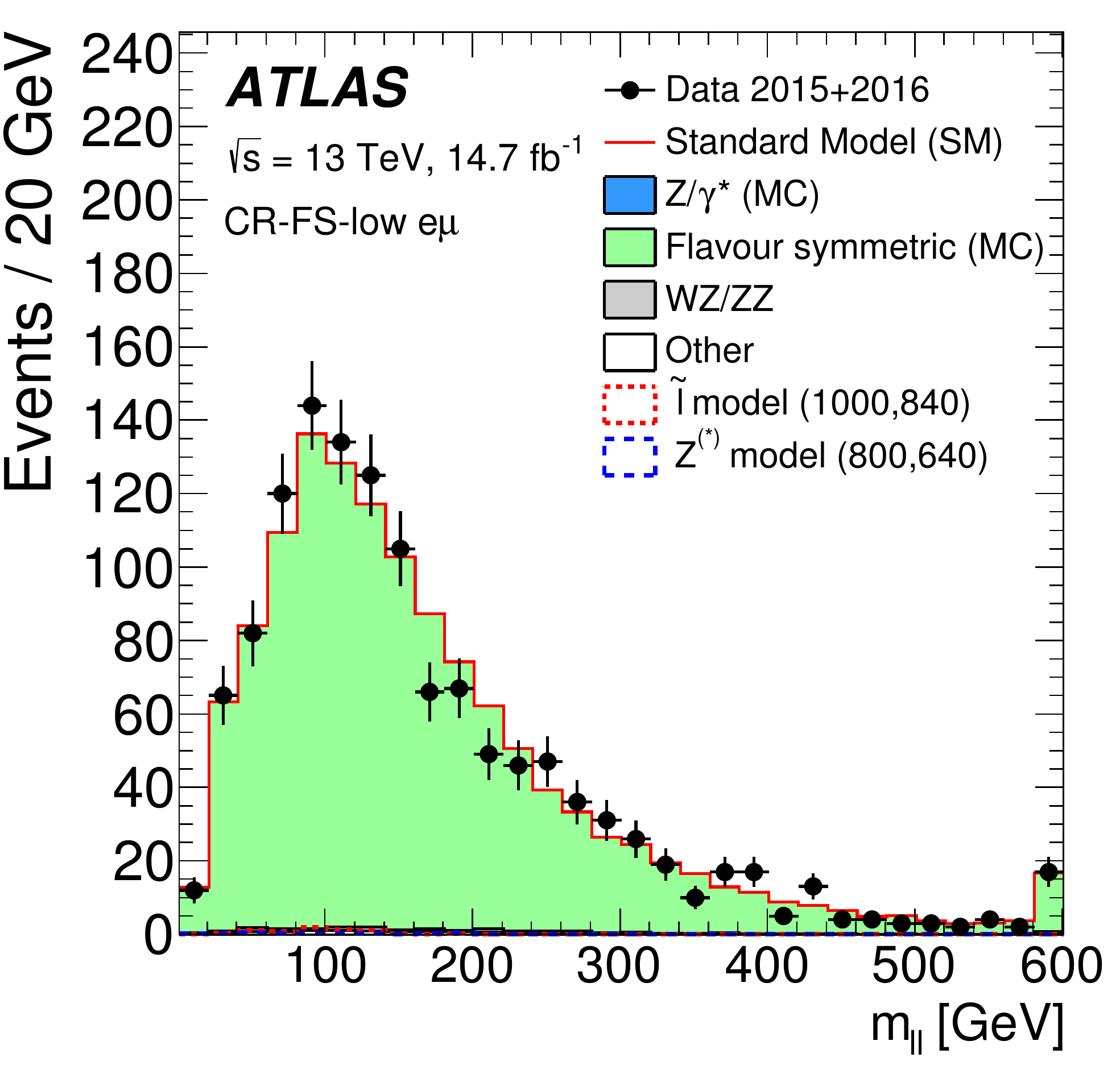} \\
\includegraphics[width=.44\textwidth]{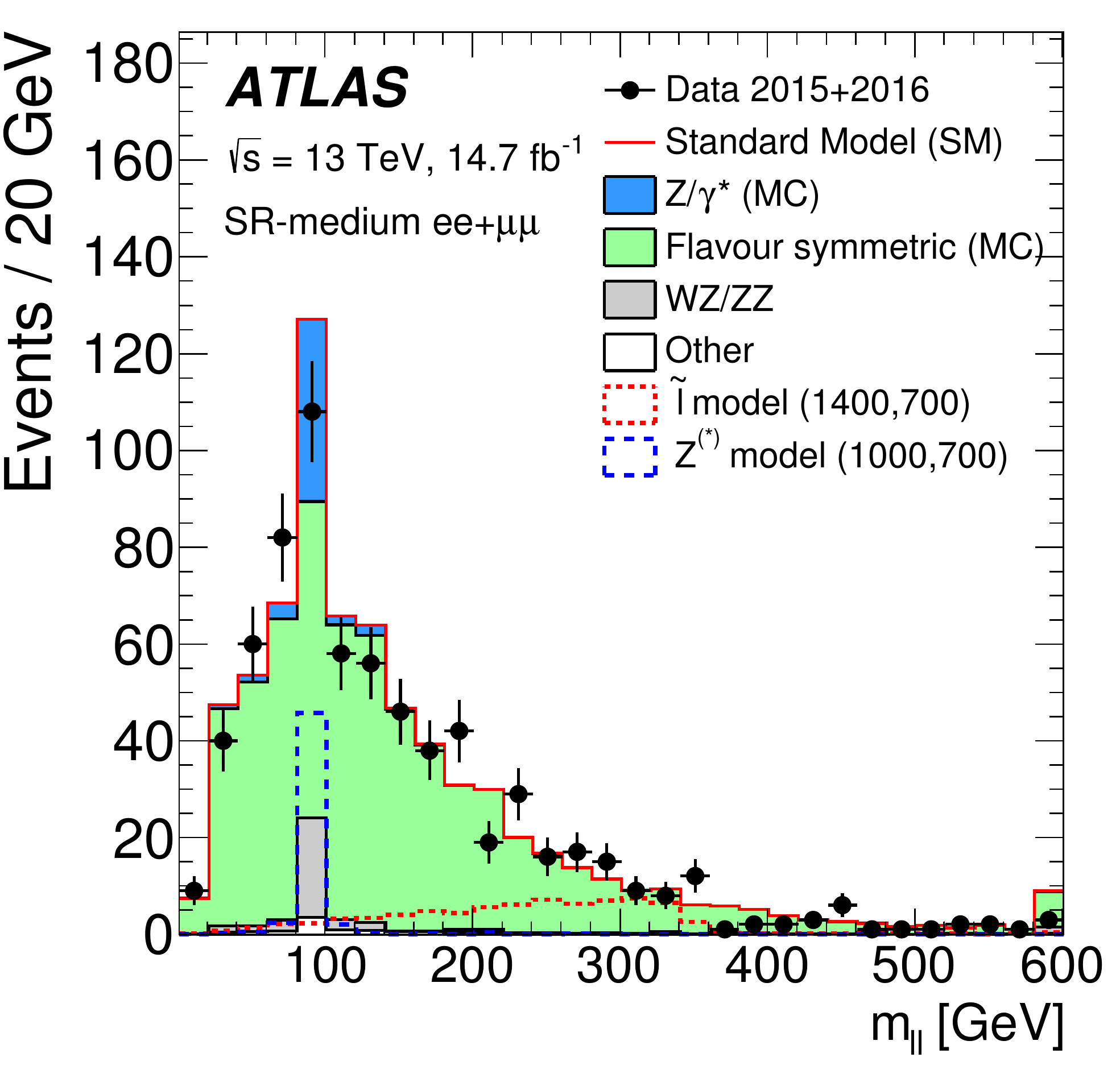} 
\includegraphics[width=.44\textwidth]{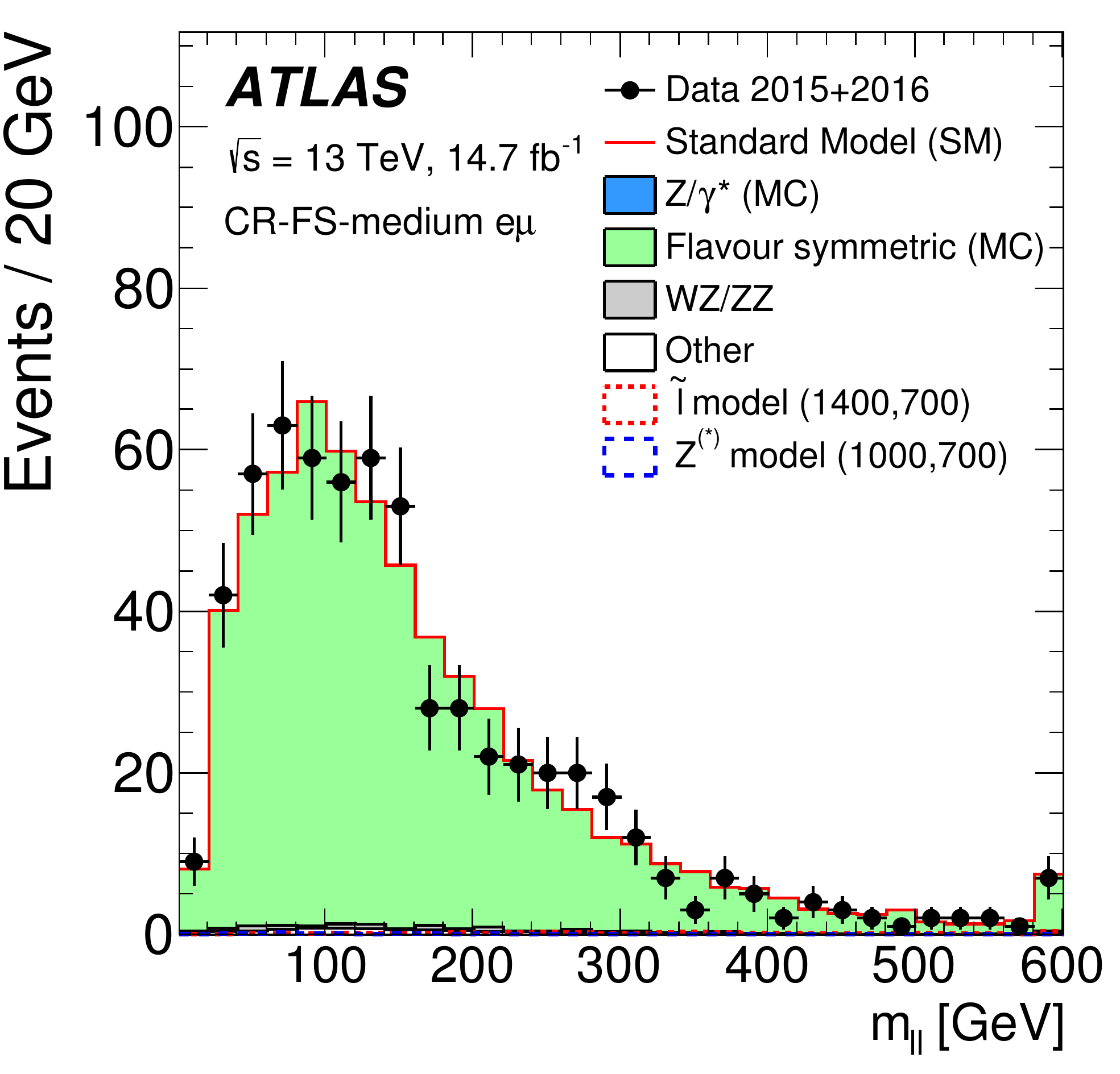} \\
\includegraphics[width=.44\textwidth]{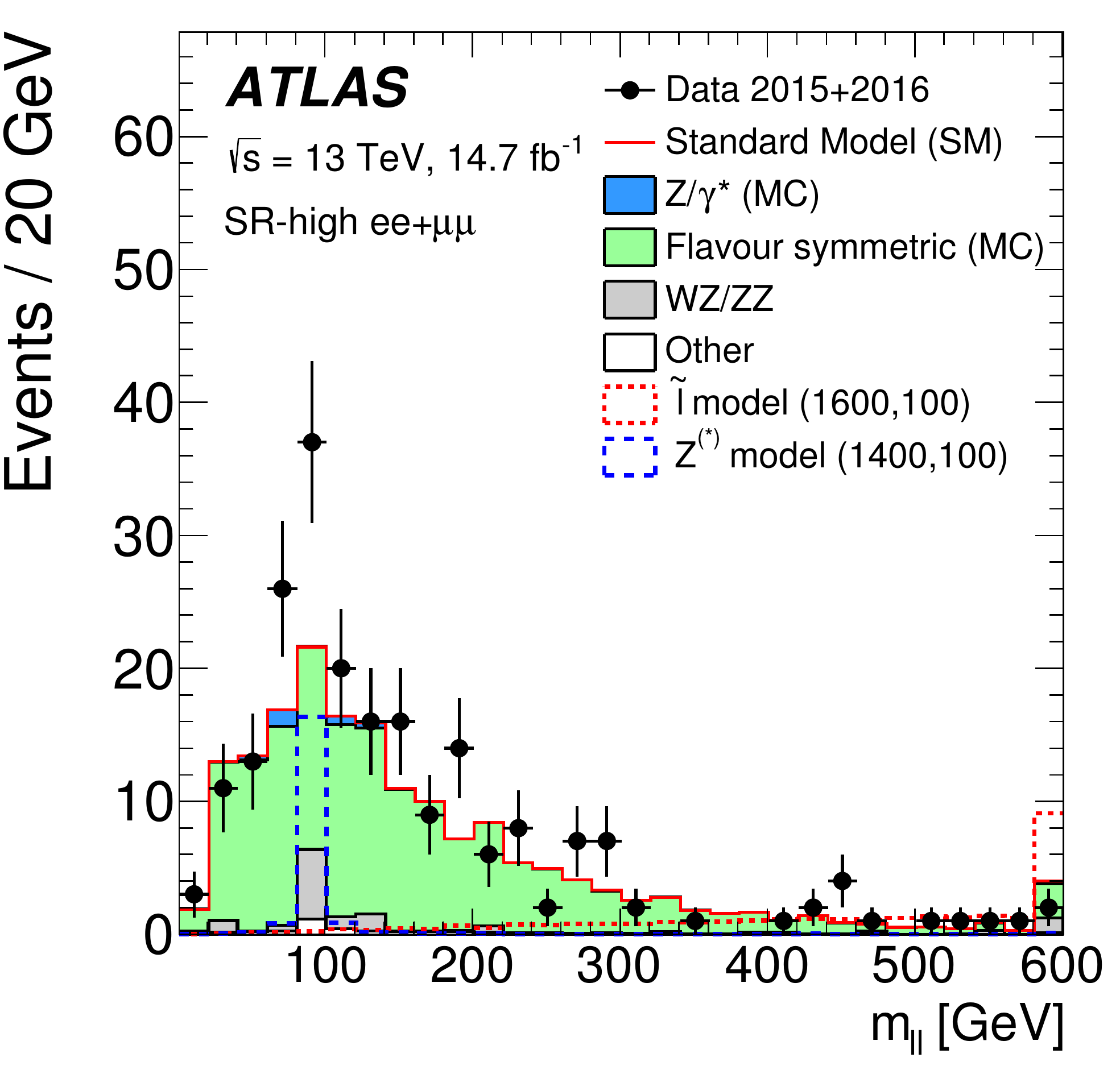} 
\includegraphics[width=.44\textwidth]{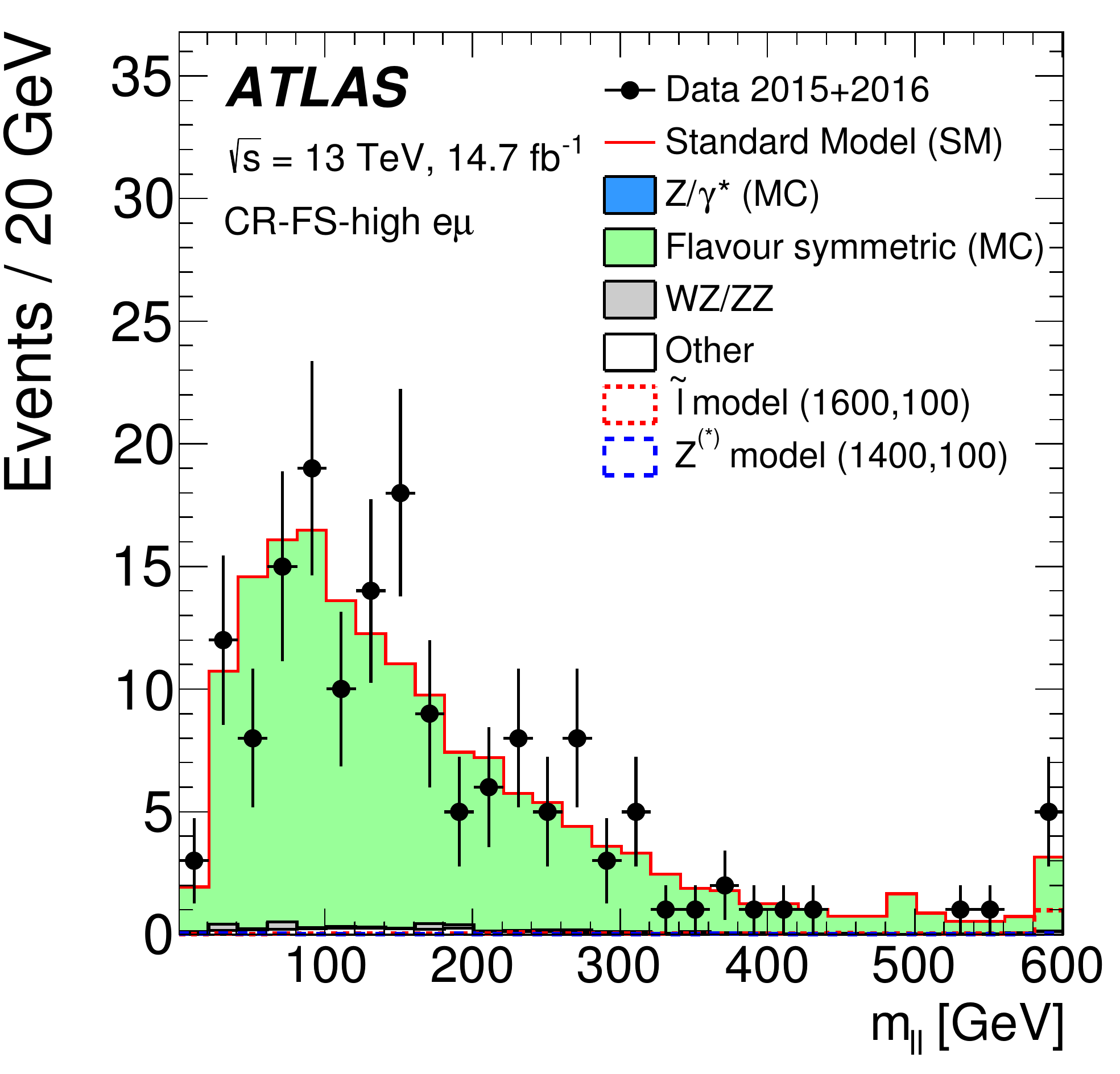}
\caption{ The dilepton mass distributions in the (top) SR-low (left) and CR-FS-low (right), (middle) SR-medium (left) and CR-FS-medium (right), and (bottom) SR-high (left) and CR-FS-high (right) regions
of the edge search. 
The \ttbar MC sample is normalised such that the total MC prediction matches data in the $e\mu$ channel for each region.
The \mll\ shape and normalisation for the \dyjets background is taken from the \gjets method. 
The rare top and data-driven fake-lepton backgrounds are grouped under ``other'' backgrounds.
Example signal benchmarks from the slepton and $Z^{(*)}$ models are overlaid on the distributions. The first (second) number
in parentheses is the gluino (LSP) mass.
The overflow is included in the last bin.}
\label{fig:edgemll2}
\end{figure}

The integrated yields in the edge signal regions are compared to the expected background in Table~\ref{tab:EdgeIntegratedYields}.
To allow for the visualisation of a potential edge, the full \mll\ distributions in the three search regions are compared to the expected background in Figure~\ref{fig:edgemll}. 
In addition, the observed \mll\ distributions are compared to the predictions from MC simulation in Figure~\ref{fig:edgemll2}, in which the \ttbar
background is scaled such that the total MC expected yield matches the data in the $e\mu$ CR.
The \ttbar normalisation factors are $\mu_{\ttbar}=0.85\pm0.03$, $0.75\pm0.04$, and $0.57\pm0.07$ in SR-low, SR-medium, and SR-high, respectively,
where the uncertainty is the data statistical uncertainty.
The data-driven flavour-symmetry prediction is used for the quantitative results of the analysis. This prediction does not rely on the \ttbar normalisation
scale factors discussed above.
The MC-based cross-check method is used to examine the \mll\ distribution in finer bins than can be achieved with the flavour-symmetry method, due to
the limited statistical precision of the $e\mu$ CR. 

\begin{table}
\begin{center}
\caption{
Breakdown of the expected background and observed data yields for SR-low, SR-medium and SR-high, integrated over the \mll\ spectrum.
The flavour-symmetric, \dyjets\ and fake-lepton components are all derived using data-driven estimates described in Section~\ref{sec:bg}.
All remaining backgrounds are taken from MC simulation.
The quoted uncertainties include statistical and systematic contributions.
\label{tab:EdgeIntegratedYields}}
\setlength{\tabcolsep}{0.0pc}

\begin{tabular*}{\textwidth}{@{\extracolsep{\fill}}lccc}
\noalign{\smallskip}\hline\noalign{\smallskip}
                                                                            & SR-low                 & SR-medium            & SR-high  \\[-0.05cm]
\noalign{\smallskip}\hline\noalign{\smallskip}

Observed events                                                             & $1394$                 & $689$                &   $212$ \\
\noalign{\smallskip}\hline\noalign{\smallskip}

Total expected background events                                            & $1500 \pm 100$         & $700 \pm 60$         &   $171 \pm 18$ \\
\noalign{\smallskip}\hline\noalign{\smallskip}

  Flavour-symmetric (\ttbar, $Wt$, $WW$ and $Z\rightarrow\tau\tau$) events  & $1270 \pm \makebox[3.5ex]{70\hfill}$          & $584 \pm 32$         &   $148 \pm 14$       \\
  \dyjets\ events                                                           & $\makebox[3ex]{\hfill 90} \pm \makebox[2ex]{50\hfill}$            & $\makebox[3ex]{\hfill 50} \pm \makebox[2ex]{40\hfill}$          &   $3^{+7}_{-3}$     \\
  $WZ/ZZ$ events                                                            & $\makebox[3ex]{\hfill 68} \pm \makebox[2ex]{31\hfill}$            & $\makebox[3ex]{\hfill 26} \pm \makebox[2ex]{11\hfill}$          &   $\makebox[3ex]{\hfill 7} \pm \makebox[2ex]{4\hfill}$        \\
  Rare top events                                                           & $19 \pm 5$             & $11.3 \pm 3.2$       &   $\makebox[3ex]{\hfill 4.2} \pm \makebox[2ex]{1.4\hfill}$  \\
 Fake-lepton events                                                         & $\makebox[3ex]{\hfill 59} \pm \makebox[2ex]{34\hfill}$            & $\makebox[3ex]{\hfill 32} \pm \makebox[2ex]{19\hfill}$          &   $10 \pm 8$    \\
\noalign{\smallskip}\hline\noalign{\smallskip}
\end{tabular*}
\end{center}
\end{table}

\begin{figure}[!t]
\centering
\includegraphics[width=1.0\textwidth]{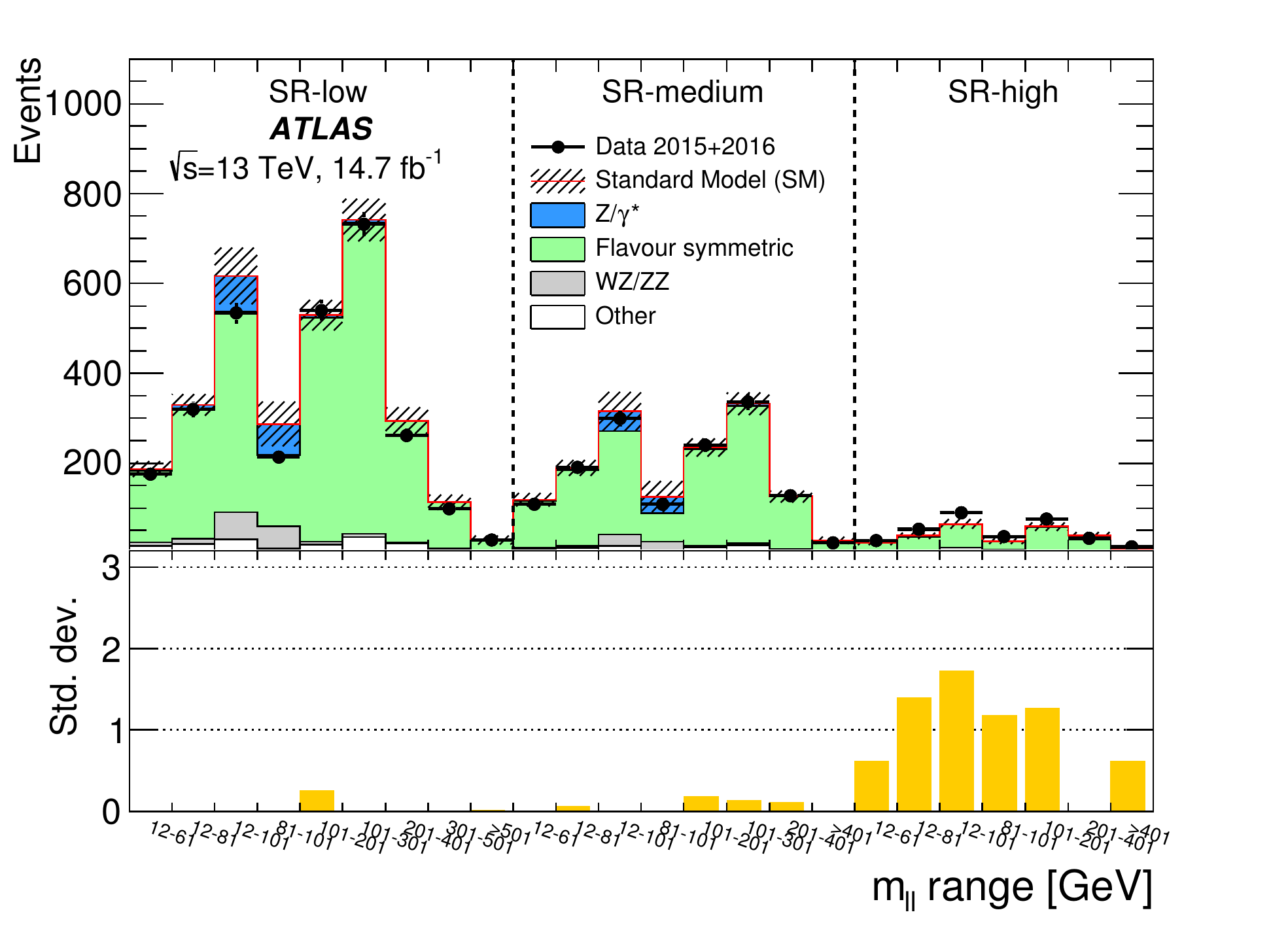}
\caption{ The expected and observed yields in the 24 (overlapping) \mll\ ranges of SR-low, SR-medium, and SR-high.
The data are compared to the sum of the expected backgrounds. 
The rare top and data-driven fake-lepton backgrounds are grouped under ``other'' backgrounds.
The significance of the difference between the data and the expected background (see text for details) is shown in the bottom plots; 
for regions in which the data yield is less than expected, the significance is set to zero.
The hashed uncertainty bands include the statistical and systematic uncertainties in the background prediction.
\label{fig:summary2}}
\end{figure}

As signal models may produce kinematic endpoints at any value of \mll, any excess must be searched for across the \mll\ distribution. 
To do this a ``sliding window'' approach is used.  
The binning in the SRs, shown in Figure~\ref{fig:edgemll}, defines many possible dilepton mass windows.
The 24 \mll\ ranges (9 for SR-low, 8 for SR-medium, and 7 for SR-high) are chosen because they are the most sensitive for at least one
grid point in the signal model parameter space. Some of the ranges overlap.  
The results in these regions are summarised in Figure~\ref{fig:summary2}, 
and the expected and observed yields in the combined $ee+\mu\mu$ channel for all 24 \mll\ ranges
are presented in Table~\ref{tab:edge_results}.
In SR-low and SR-medium, the data are consistent with the expected background across the full \mll\ range.
In SR-high the data show a slight excess above the background at low \mll.
Of these 24 \mll\ ranges, the largest excess is observed in SR-high with $12<\mll<101$~\GeV.
Here a total of 90 events are observed in data, compared to an expectation of $65\pm10$ events,
corresponding to a local significance of $1.7\sigma$.

\FloatBarrier
\input{edge_results}

%% file: edge_results.tex
\begin{table}
\begin{center}
\caption{Breakdown of the expected background and observed data yields in the edge signal regions.
The results are given for SR-low, SR-medium and SR-high in all 24 \mll\ ranges.
The \mll\ range in units of \GeV\ is indicated in the leftmost column of the table.
Left to right: the total expected background, with combined statistical and systematic uncertainties, observed data, 95\% CL upper limits on the visible cross section
($\langle\epsilon\sigma\rangle_{\rm obs}^{95}$) and on the number of
signal events ($S_{\rm obs}^{95}$).  The sixth column
($S_{\rm exp}^{95}$) shows the expected 95\% CL upper limit on the number of
signal events, given the expected number (and $\pm 1\sigma$ excursions) of background events.
The last two columns
indicate the discovery $p$-value ($p(s = 0)$)~\cite{Baak:2014wma}, and the Gaussian significance ($Z(s=0)$).
For an observed number of events lower than expected, the discovery $p$-value is truncated at 0.5 and the significance is set to zero.
\label{tab:edge_results}}
\resizebox{\textwidth}{!}{
\setlength{\tabcolsep}{0.0pc}
\small
\begin{tabular*}{\textwidth}{@{\extracolsep{\fill}}lcccccccc}
\noalign{\smallskip}\hline\noalign{\smallskip}
{\bf Signal Region}   & Total Bkg. & Data & $\langle\epsilon{\rm \sigma}\rangle_{\rm obs}^{95}$[fb]  &  $S_{\rm obs}^{95}$  & $S_{\rm exp}^{95}$ & $p(s=0)$ & $Z(s=0)$ \\
\noalign{\smallskip}\hline\noalign{\smallskip}
\noalign{\smallskip}\hline\noalign{\smallskip}
SR-low & & & & & &&&\\
\noalign{\smallskip}\hline\noalign{\smallskip}
12--61     & $187\pm18  $  &  $175$   & $2.68$ &  $39.4$ & $ { 48 }^{ +23 }_{ -14 }$ & $ 0.50$ & $0.00$ \\[.5ex]%
12--81     & $330\pm24  $  &  $320$   & $3.88$ &  $57.1$ & $ { 64 }^{ +30 }_{ -19 }$ & $ 0.50$ & $0.00$ \\[.5ex]%
12--101    & $617\pm63  $  &  $534$   & $4.64$ &  $68.2$ & $ { 98 }^{ +36 }_{ -26 }$ & $ 0.50$ & $0.00$ \\[.5ex]%
81--101    & $287\pm50  $  &  $214$   & $2.73$ &  $40.2$ & $ { 62 }^{ +22 }_{ -16 }$ & $ 0.50$ & $0.00$ \\[.5ex]%
101--201   & $529\pm34  $  &  $540$   & $6.80$ &  $99.9$ & $ { 91 }^{ +52 }_{ -29 }$ & $ 0.40$ & $0.26$ \\[.5ex]%
101--301   & $741\pm48  $  &  $732$   & $7.28$ &  $107$& $ { 113 }^{ +53 }_{ -33 }$& $ 0.50$ & $0.00$ \\[.5ex]%
201--401   & $295\pm30  $  &  $262$   & $3.43$ &  $50.5$ & $ { 70 }^{ +37 }_{ -21 }$ & $ 0.50$ & $0.00$ \\[.5ex]%
301--501   & $113\pm17  $  &  $99 $   & $2.37$ &  $34.8$ & $ { 46 }^{ +41 }_{ -16 }$ & $ 0.50$ & $0.00$ \\[.5ex]%
> 501      & $29\pm10   $  &  $29 $   & $1.88$ &  $27.7$ & $ { 27 }^{ +34 }_{ -10 }$ & $ 0.50$ & $0.01$ \\[.5ex]%
\noalign{\smallskip}\hline\noalign{\smallskip}
SR-medium & & & & &&  & \\
\noalign{\smallskip}\hline\noalign{\smallskip}
12--61     & $119\pm 15  $  &  $109 $   & $2.38$ &  $35.1$ & $ { 43 }^{ +29 }_{ -14 }$ & $ 0.50$ & $0.00$ \\[.5ex]%
12--81     & $190\pm 18  $  &  $191 $   & $3.57$ &  $52.5$ & $ { 51 }^{ +31 }_{ -15 }$ & $ 0.48$ & $0.06$ \\[.5ex]%
12--101    & $315\pm 43  $  &  $299$    & $5.12$ &  $75.3$ & $ { 81 }^{ +29 }_{ -20 }$ & $ 0.50$ & $0.00$ \\[.5ex]%
81--101    & $125\pm 35  $  &  $108 $   & $3.18$ &  $46.7$ & $ { 51 }^{ +17 }_{ -12 }$ & $ 0.50$ & $0.00$ \\[.5ex]%
101--201   & $235\pm 20  $  &  $240$    & $4.26$ &  $62.6$ & $ { 58 }^{ +37 }_{ -19 }$ & $ 0.42$ & $0.19$ \\[.5ex]%
101--301   & $332\pm 25  $  &  $336$    & $4.92$ &  $72.3$ & $ { 69 }^{ +39 }_{ -22 }$ & $ 0.45$ & $0.14$ \\[.5ex]%
201--401   & $126\pm 13  $  &  $128 $   & $3.27$ &  $48.0$ & $ { 46 }^{ +52 }_{ -16 }$ & $ 0.46$ & $0.11$ \\[.5ex]%
> 401      & $28\pm  8   $  &  $22 $    & $1.09$ &  $16.1$ & $ { 21 }^{ +19 }_{ -7 }$  & $ 0.50$ & $0.00$ \\[.5ex]% 
\noalign{\smallskip}\hline\noalign{\smallskip}
SR-high & & & & & & & \\
\noalign{\smallskip}\hline\noalign{\smallskip}
12--61    & $23\pm5     $  &  $27$   & $1.84$ &  $27.0$ & $ { 20 }^{ +31 }_{ -8 }$ & $ 0.27$ & $0.62$ \\[.5ex]%
12--81    & $39\pm7     $  &  $53$   & $3.32$ &  $48.9$ & $ { 26 }^{ +28 }_{ -10 }$ & $ 0.08$ & $1.40$ \\[.5ex]%
12--101   & $\makebox[3.3ex]{\hfill 65}\pm10    $  &  $90$   & $4.00$ &  $58.8$ & $ { 31 }^{ +17 }_{ -10 }$ & $ 0.04$ & $1.73$ \\[.5ex]% 
81--101   & $26\pm6     $  &  $37$   & $2.17$ &  $31.9$ & $ { 20 }^{ +13 }_{ -7 }$ & $ 0.12$ & $1.18$ \\[.5ex]%
101--201  & $59\pm9     $  &  $75$   & $3.68$ &  $54.1$ & $ { 31 }^{ +29 }_{ -11 }$& $ 0.10$ & $1.27$ \\[.5ex]% 
201--401  & $39\pm7     $  &  $33$   & $1.82$ &  $26.7$ & $ { 28 }^{ +14 }_{ -7 }$ & $ 0.50$ & $0.00$ \\[.5ex]%
> 401     & $10\pm 5    $  &  $14 $  & $2.04$ &  $30.0$ & $ { 21 }^{ +79 }_{ -10 }$& $ 0.27$ & $0.62$ \\[.5ex]%
\noalign{\smallskip}\hline\noalign{\smallskip}
\end{tabular*}
}
\end{center}
\end{table}

%% file: sec-interpretation.tex
In this section, exclusion limits are shown for the SUSY models detailed in Section~\ref{sec:susy}. 
The asymptotic $CL_{\text{S}}$ prescription~\cite{statforumlimits,clsread}, implemented in the HistFitter program~\cite{Baak:2014wma}, 
is used to determine cross-section upper limits at $95\%$ confidence level (CL) for the on-$Z$ search.
For the edge search, pseudo-experiments are used to evaluate the cross-section upper limits.
A Gaussian model for nuisance parameters is used for all signal and background uncertainties. 
Exceptions are the statistical uncertainties of the flavour-symmetry method, \gjets\ method and MC-based backgrounds, 
all of which are treated as Poissonian nuisance parameters.
The different experimental uncertainties are treated as correlated between signal and background events.
The theoretical uncertainty of the signal cross section is not accounted for in the limit-setting procedure.
Instead, following the initial limit determination, 
the impact of varying the signal cross section within its uncertainty is evaluated separately and indicated in the exclusion results. 
Limits are based on the combined $ee+\mu\mu$ results.
Possible signal contamination in the CRs is neglected in the limit-setting procedure;
the contamination is found to be negligible for signal points near the exclusion boundaries. 
Far from the exclusion boundary, although the signal contamination can be significant, 
the number of events appearing in the signal region is large enough that the points are still excluded, 
due to the relative branching fractions for the signal in the CR and SR. 
For example, for models with signal contamination of 50\% in CR-FS the signal-to-background ratio in SRZ is $\sim 10$.

\begin{figure}[htbp]
\centering
\includegraphics[width=.8\textwidth]{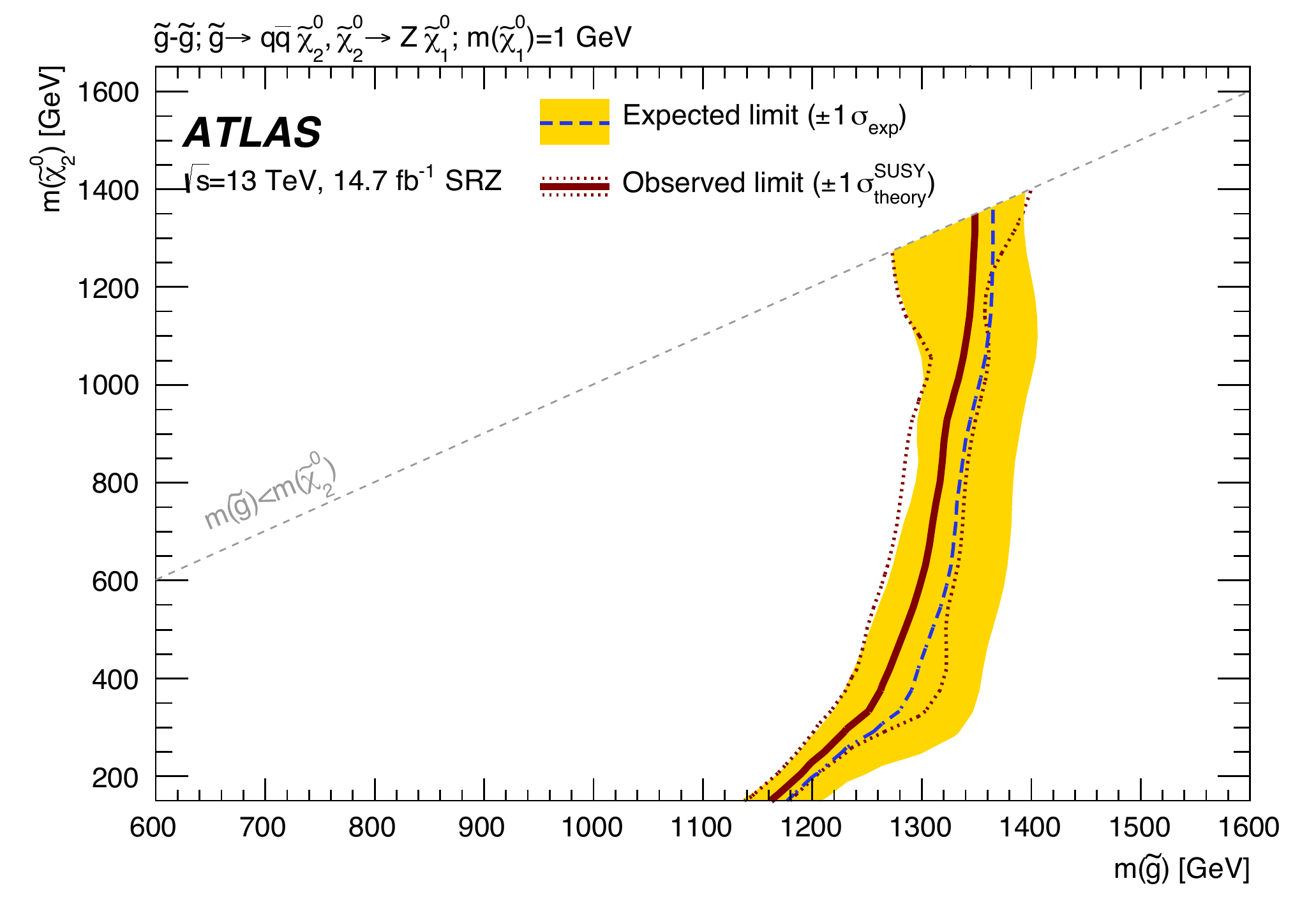}
\includegraphics[width=.8\textwidth]{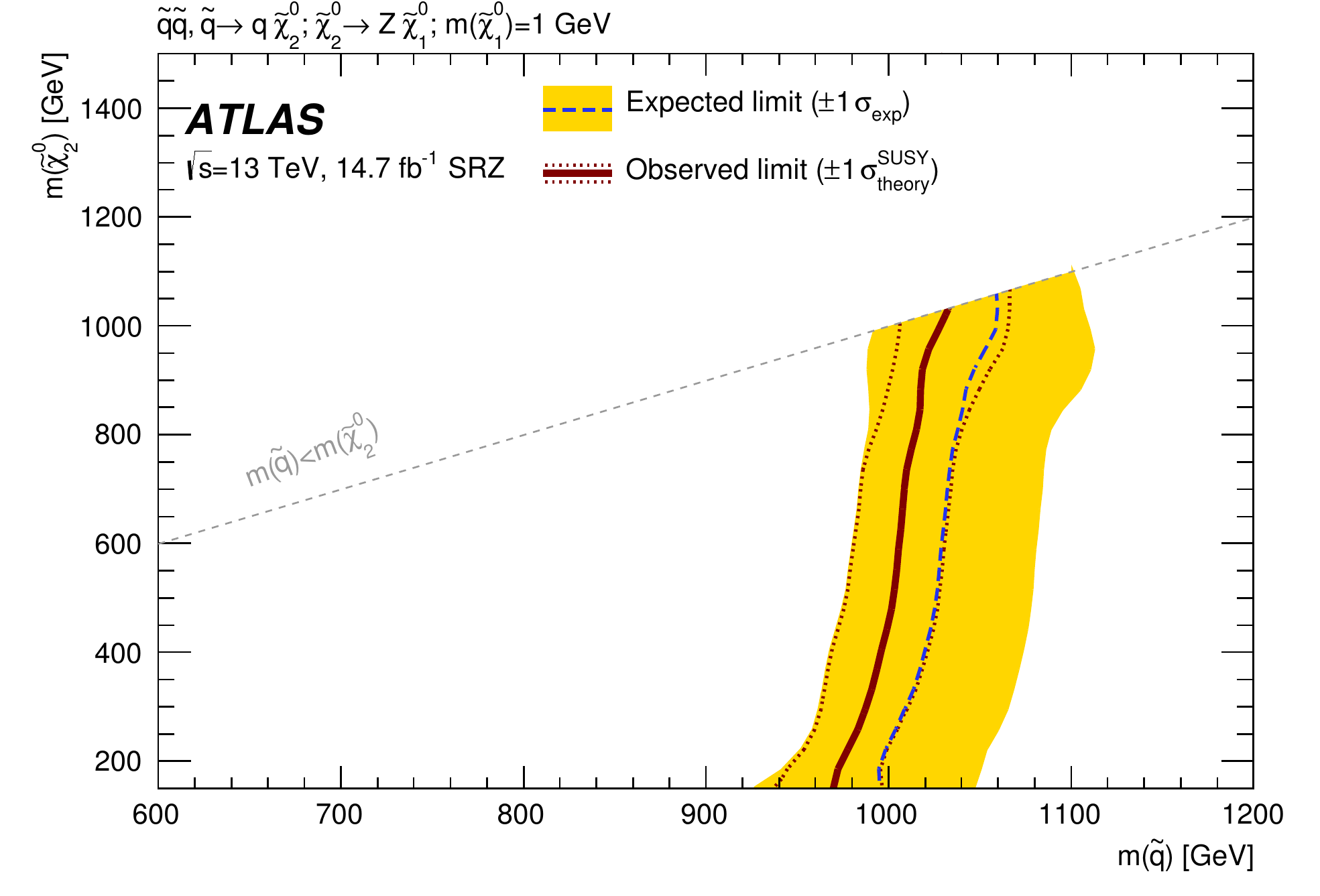}  
\caption{
Expected and observed exclusion contours derived from the results in SRZ for the (top) $\tilde{g}$--$\chitwozero$ on-shell grid and (bottom) $\tilde{q}$--$\chitwozero$ on-shell grid. 
The dashed blue line indicates the expected limits at $95\%$ CL and the yellow band shows the $1\sigma$ variation of the expected limit as a consequence of the uncertainties in the background prediction and the experimental uncertainties in the signal ($\pm1\sigma_\text{exp}$). 
The observed limits are shown by the solid red line, with the dotted red lines indicating the variation resulting from changing the signal cross section within its uncertainty ($\pm1\sigma^\text{SUSY}_\text{theory}$). \label{fig:excl_SMGGN2_1}
}

\end{figure}

The results of the on-shell $Z$ search are interpreted in a simplified model with gluino-pair production,
where each gluino decays as 
$\tilde{g} \rightarrow q\bar{q} \tilde{\chi}^{0}_{2}, \tilde{\chi}^{0}_{2} \rightarrow Z \tilde{\chi}^{0}_{1}$ and the $\tilde{\chi}^{0}_{1}$ mass is set to 1~\GeV. 
The expected and observed exclusion contours for this $\tilde{g}$--$\chitwozero$ on-shell grid are shown in the $m(\tilde{g})$--$m(\tilde{\chi}^{0}_{2})$ plane in Figure~\ref{fig:excl_SMGGN2_1}.
The expected (observed) lower limit on the gluino mass is about 1.35~\TeV\ (1.30~\TeV) for a $\tilde{\chi}_2^0$ with a mass of 1.1~\TeV\ in this model.
The impact of the systematic uncertainties in the background and the experimental uncertainties in the signal, 
shown with a coloured band, is about 100~\GeV\ on the gluino mass limit.
The systematic uncertainty of the signal cross section, shown as dotted lines around the observed contour, has an impact of about 40~\GeV.
Figure~\ref{fig:excl_SMGGN2_1} also shows the expected and observed exclusion limits for the $\tilde{q}$--$\chitwozero$ on-shell model. 
This is a simplified model with squark-pair production, where each squark decays to a quark and a neutralino, 
with the neutralino subsequently decaying to a $Z$ boson and an LSP with a mass of 1~\GeV.
In this model, exclusion is expected (observed) for squarks with masses below 1040~\GeV\ (980~\GeV) for a $\tilde{\chi}^{0}_{2}$ mass of 600~\GeV.

\begin{figure}[htbp]
\centering
\includegraphics[width=.8\textwidth]{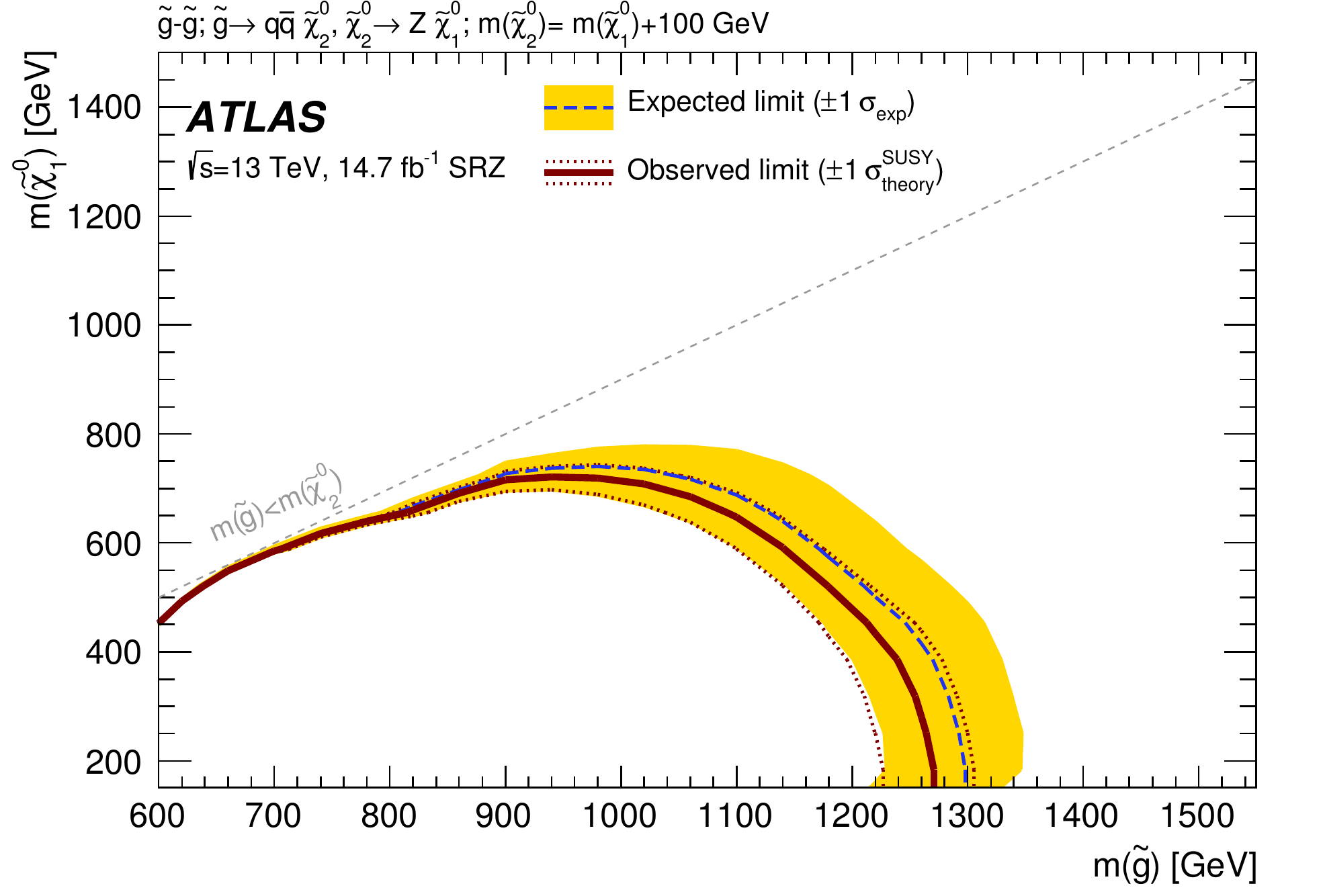}
\caption{
Expected and observed exclusion contours derived from the results in SRZ for the $\tilde{g}$--$\chionezero$ on-shell grid. 
The dashed blue line indicates the expected limits at $95\%$ CL and the yellow band shows the $1\sigma$ variation of the expected limit as a consequence of the uncertainties in the background prediction and the experimental uncertainties in the signal ($\pm1\sigma_\text{exp}$). 
The observed limits are shown by the solid red line, with the dotted red lines indicating the variation resulting from changing the signal cross section within its uncertainty ($\pm1\sigma^\text{SUSY}_\text{theory}$).
\label{fig:excl_SMGGN2}
}
\end{figure}

Figure~\ref{fig:excl_SMGGN2} shows the expected and observed exclusion contours for the $\tilde{g}$--$\chionezero$ on-shell model, 
in which the produced gluinos follow the same decay chain as in the model above.
In this case the mass difference $\Delta m = m(\tilde{\chi}^{0}_{2})-m(\tilde{\chi}^{0}_{1})$ is set to 100~\GeV.

The results of the edge search are interpreted in two simplified models with gluino-pair production, in which each gluino decays as
$\tilde{g} \rightarrow q\bar{q} \tilde{\chi}^{0}_{2}$. 
For each point in the signal-model parameter space, limits on the signal 
strength are calculated using the \mll\ window with the best expected sensitivity.
Details of the windows are described in Section~\ref{sec:result}. 

\begin{figure}[htbp]
\centering
\includegraphics[width=0.8\textwidth]{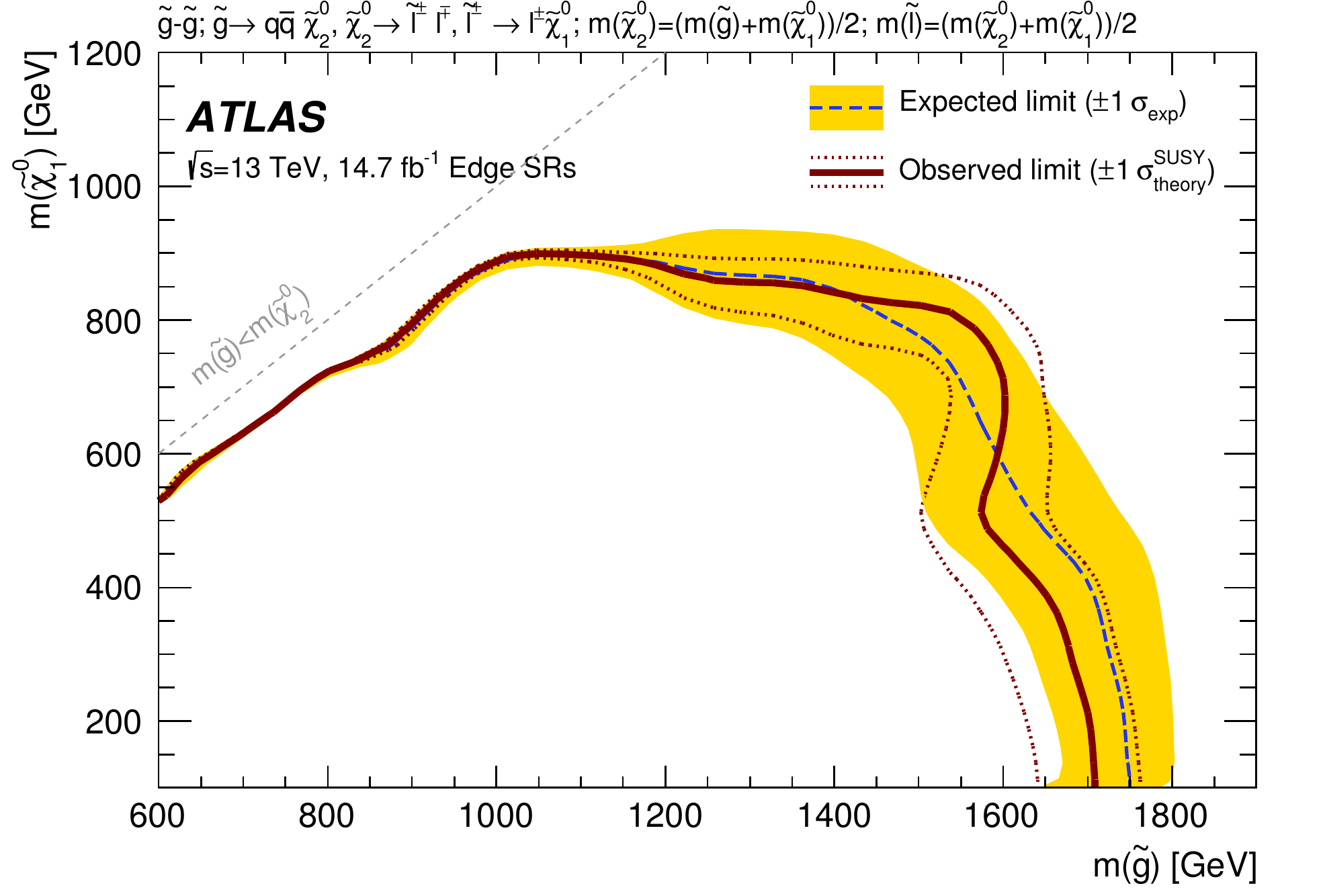}
\caption{
Expected and observed exclusion contours derived from the results in the edge search SRs for the slepton signal model.
The dashed blue line indicates the expected limits at $95\%$ CL and the yellow band shows the $1\sigma$ variation of the expected limit as a consequence of the uncertainties in the background prediction and the experimental uncertainties in the signal ($\pm1\sigma_\text{exp}$).
The observed limits are shown by the solid red lines, with the dotted red lines indicating the variation resulting from changing the signal cross section within its uncertainty ($\pm1\sigma^\text{SUSY}_\text{theory}$).
\label{fig:limit_slepton}}
\end{figure}

The excluded regions in the 
$m(\tilde{g})$--$m(\tilde{\chi}^{0}_{1})$ plane are presented in Figure~\ref{fig:limit_slepton} for the slepton model.
In this model, pair-produced gluinos each decay as $\tilde{g}\rightarrow q \bar{q} \tilde{\chi}^{0}_2, \tilde{\chi}^{0}_2 \rightarrow \ell^\pm \tilde{\ell}^\mp, \tilde{\ell}^\mp \rightarrow \ell^\mp \tilde{\chi}^{0}_{1}$.
Here, the results exclude gluinos with masses as large as 1.7~\TeV, with an expected limit of 1.75~\TeV\ for small $m(\tilde{\chi}^{0}_{1})$.
The results probe kinematic endpoints as small as
$m_{\ell\ell}^{\text{max}} = m(\tilde{\chi}^{0}_{2})-m(\tilde{\chi}^{0}_{1}) = 1/2(m(\tilde{g})-m(\tilde{\chi}^{0}_{1}) ) = 50$~\GeV.

\begin{figure}[htbp]
\centering
\includegraphics[width=0.8\textwidth]{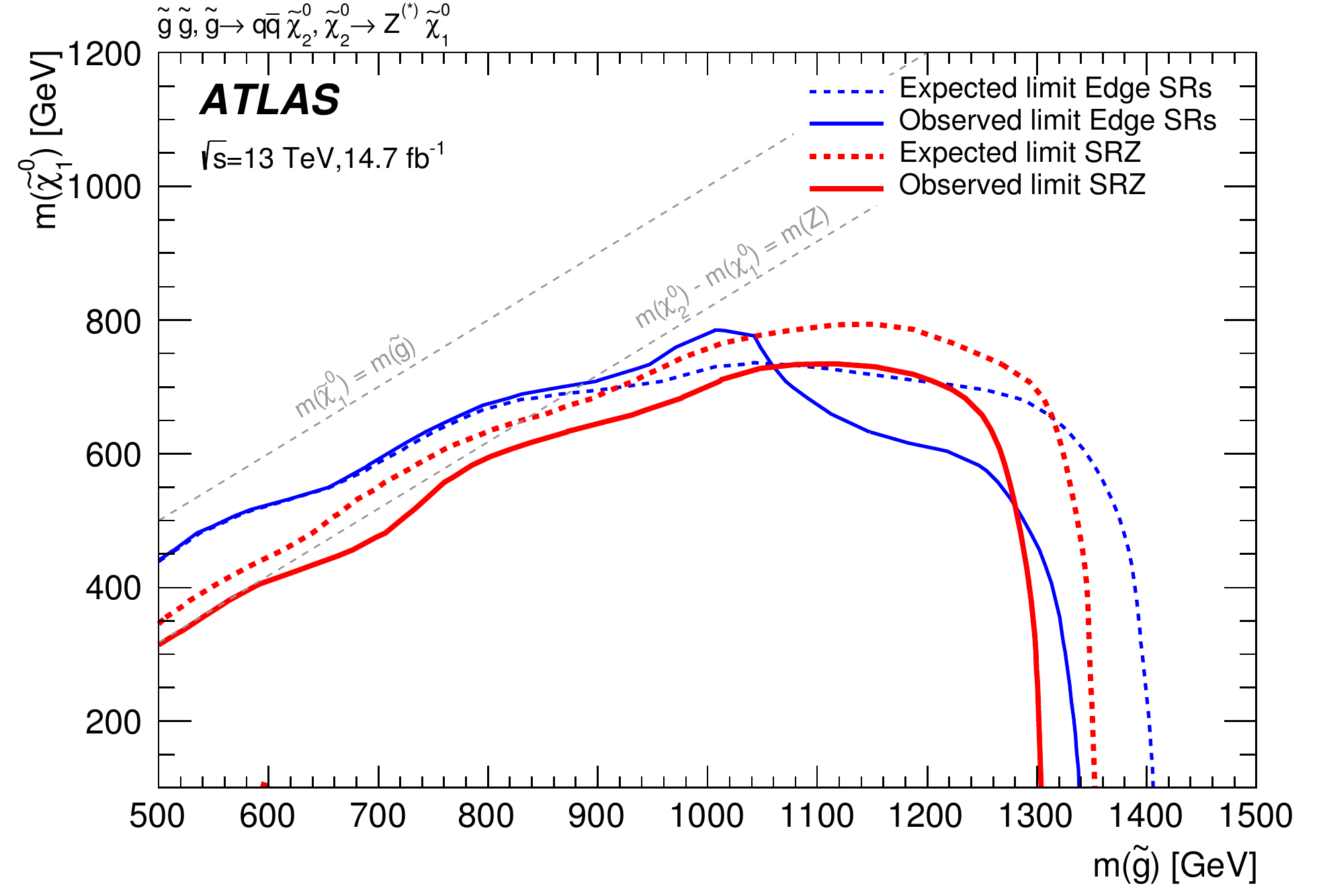}
\caption{
Expected and observed exclusion contours derived from the results in the edge search SRs and SRZ for the \zstar\ model.
The dashed and solid blue lines indicate the expected and observed limits at $95\%$ CL from the results in the edge SRs, 
while the thick dashed and solid red lines indicate the expected and observed limits at $95\%$ CL from the results in SRZ.
\label{fig:comb_limit_zstar}}
\end{figure}

The \zstar\ exclusion limits from the results in the edge SRs are compared with the same limits derived using the results in SRZ in Figure~\ref{fig:comb_limit_zstar}.
In this model, pair-produced gluinos each decay as $\tilde{g} \rightarrow q\bar{q} \tilde{\chi}^{0}_{2}, \tilde{\chi}^{0}_{2} \rightarrow \zstar \tilde{\chi}^{0}_{1}$, and the mass splitting between the $\tilde{\chi}^{0}_{2}$ and the $\tilde{\chi}^{0}_{1}$ determines whether the $Z$ boson is produced on-shell.
Here the edge limits extend into the more compressed region, whereas the expected SRZ exclusion probes higher $\tilde{\chi}^{0}_{1}$ masses in the on-shell regime.
At high gluino masses, the edge SRs provide stronger limits.
For the \zstar model, the expected and observed gluino mass limits are 1.4~\TeV\ and 1.34~\TeV\ (1.35 and 1.3~\TeV\ for the on-Z signal region), respectively, 
for \chionezero masses below 400~\GeV.
The sensitivity in the \zstar model is smaller than that of the slepton model because the leptonic branching fraction of the $Z$ boson suppresses the signal production rate.

Model-independent upper limits at 95\% CL on the number of events that could be attributed to non-SM sources ($S^{95}$) for SRZ are derived using the 
$CL_{\text{S}}$ prescription and neglecting possible signal contamination in the CRs.
For these upper limits, pseudo-experiments are used rather than the asymptotic approximation.
The expected and observed upper limits are given in Table~\ref{tab:results}.
The same information is given for the 24 \mll\ ranges of the edge search in Table~\ref{tab:edge_results}.

%% file: sec-conclusion.tex
This paper presents two searches for new phenomena in final states containing a \SF\ \OS\ lepton (electron or muon) pair,
jets, and large missing transverse momentum using \lumi\ of ATLAS data collected during 2015 and 2016 at the LHC at $\sqrt{s}=13$~\TeV.
The first search (on-shell $Z$ search) targets lepton pairs consistent with $Z$ boson decay, 
while the second search (edge search) targets a kinematic endpoint feature in the dilepton mass distribution.
For the edge search, a set of 24 mass ranges are considered, with different requirements on \MET and $\HT$, and different kinematic endpoint values in the dilepton
invariant-mass distribution.
The data in both searches are found to be consistent with the Standard Model prediction.
The results are interpreted in simplified models of gluino-pair production and squark-pair production, and exclude gluinos (squarks) with masses as large as 1.7~\TeV\ (980~\GeV).

%% file: Acknowledgements.tex
% Acknowledgements for papers with collision data
% Version 18-Jan-2017

% Standard acknowledgements start here
%----------------------------------------------
We thank CERN for the very successful operation of the LHC, as well as the
support staff from our institutions without whom ATLAS could not be
operated efficiently.

We acknowledge the support of ANPCyT, Argentina; YerPhI, Armenia; ARC, Australia; BMWFW and FWF, Austria; ANAS, Azerbaijan; SSTC, Belarus; CNPq and FAPESP, Brazil; NSERC, NRC and CFI, Canada; CERN; CONICYT, Chile; CAS, MOST and NSFC, China; COLCIENCIAS, Colombia; MSMT CR, MPO CR and VSC CR, Czech Republic; DNRF and DNSRC, Denmark; IN2P3-CNRS, CEA-DSM/IRFU, France; GNSF, Georgia; BMBF, HGF, and MPG, Germany; GSRT, Greece; RGC, Hong Kong SAR, China; ISF, I-CORE and Benoziyo Center, Israel; INFN, Italy; MEXT and JSPS, Japan; CNRST, Morocco; FOM and NWO, Netherlands; RCN, Norway; MNiSW and NCN, Poland; FCT, Portugal; MNE/IFA, Romania; MES of Russia and NRC KI, Russian Federation; JINR; MESTD, Serbia; MSSR, Slovakia; ARRS and MIZ\v{S}, Slovenia; DST/NRF, South Africa; MINECO, Spain; SRC and Wallenberg Foundation, Sweden; SERI, SNSF and Cantons of Bern and Geneva, Switzerland; MOST, Taiwan; TAEK, Turkey; STFC, United Kingdom; DOE and NSF, United States of America. In addition, individual groups and members have received support from BCKDF, the Canada Council, CANARIE, CRC, Compute Canada, FQRNT, and the Ontario Innovation Trust, Canada; EPLANET, ERC, ERDF, FP7, Horizon 2020 and Marie Sk{\l}odowska-Curie Actions, European Union; Investissements d'Avenir Labex and Idex, ANR, R{\'e}gion Auvergne and Fondation Partager le Savoir, France; DFG and AvH Foundation, Germany; Herakleitos, Thales and Aristeia programmes co-financed by EU-ESF and the Greek NSRF; BSF, GIF and Minerva, Israel; BRF, Norway; CERCA Programme Generalitat de Catalunya, Generalitat Valenciana, Spain; the Royal Society and Leverhulme Trust, United Kingdom.

The crucial computing support from all WLCG partners is acknowledged gratefully, in particular from CERN, the ATLAS Tier-1 facilities at TRIUMF (Canada), NDGF (Denmark, Norway, Sweden), CC-IN2P3 (France), KIT/GridKA (Germany), INFN-CNAF (Italy), NL-T1 (Netherlands), PIC (Spain), ASGC (Taiwan), RAL (UK) and BNL (USA), the Tier-2 facilities worldwide and large non-WLCG resource providers. Major contributors of computing resources are listed in Ref.~\cite{ATL-GEN-PUB-2016-002}.
%----------------------------------------------

%% file: atlas_authlist.tex
% ATLAS Collaboration author list
% Data extracted on 24-Oct-2016 for paper reference SUSY-2016-05
%\documentclass[11pt]{article}
%\usepackage{a4wide}%\begin{document}
\begin{flushleft}
{\Large The ATLAS Collaboration}

\bigskip

M.~Aaboud$^\textrm{\scriptsize 137d}$,
G.~Aad$^\textrm{\scriptsize 88}$,
B.~Abbott$^\textrm{\scriptsize 115}$,
J.~Abdallah$^\textrm{\scriptsize 8}$,
O.~Abdinov$^\textrm{\scriptsize 12}$,
B.~Abeloos$^\textrm{\scriptsize 119}$,
O.S.~AbouZeid$^\textrm{\scriptsize 139}$,
N.L.~Abraham$^\textrm{\scriptsize 151}$,
H.~Abramowicz$^\textrm{\scriptsize 155}$,
H.~Abreu$^\textrm{\scriptsize 154}$
% ...
\bigskip
\\
% ...
$^{8}$ Department of Physics, The University of Texas at Arlington, Arlington TX, United States of America\\
% ...
$^{12}$ Institute of Physics, Azerbaijan Academy of Sciences, Baku, Azerbaijan\\
% ...
$^{88}$ CPPM, Aix-Marseille Universit{\'e} and CNRS/IN2P3, Marseille, France\\
% ...
$^{115}$ Homer L. Dodge Department of Physics and Astronomy, University of Oklahoma, Norman OK, United States of America\\
% ...
$^{119}$ LAL, Univ. Paris-Sud, CNRS/IN2P3, Universit{\'e} Paris-Saclay, Orsay, France\\
% ...
$^{137}$ $^{(a)}$ Facult{\'e} des Sciences Ain Chock, R{\'e}seau Universitaire de Physique des Hautes Energies - Universit{\'e} Hassan II, Casablanca; $^{(b)}$ Centre National de l'Energie des Sciences Techniques Nucleaires, Rabat; $^{(c)}$ Facult{\'e} des Sciences Semlalia, Universit{\'e} Cadi Ayyad, LPHEA-Marrakech; $^{(d)}$ Facult{\'e} des Sciences, Universit{\'e} Mohamed Premier and LPTPM, Oujda; $^{(e)}$ Facult{\'e} des sciences, Universit{\'e} Mohammed V, Rabat, Morocco\\
% ...
$^{139}$ Santa Cruz Institute for Particle Physics, University of California Santa Cruz, Santa Cruz CA, United States of America\\
% ...
$^{151}$ Department of Physics and Astronomy, University of Sussex, Brighton, United Kingdom\\
% ...
$^{154}$ Department of Physics, Technion: Israel Institute of Technology, Haifa, Israel\\
$^{155}$ Raymond and Beverly Sackler School of Physics and Astronomy, Tel Aviv University, Tel Aviv, Israel\\
% ...
$^{a}$ Also at Department of Physics, King's College London, London, United Kingdom\\
$^{b}$ Also at Institute of Physics, Azerbaijan Academy of Sciences, Baku, Azerbaijan\\
$^{c}$ Also at Novosibirsk State University, Novosibirsk, Russia\\
$^{d}$ Also at TRIUMF, Vancouver BC, Canada\\
$^{e}$ Also at Department of Physics {\&} Astronomy, University of Louisville, Louisville, KY, United States of America\\
% ...
\end{flushleft}

%\end{document}
% Created with xml2latex.py